\documentclass[10pt, aps, prc, superscriptaddress, nofootinbib, 
               amsmath, 
               twocolumn, preprintnumbers, 
               showpacs, 
               raggedbottom, 
               floatfix, 
               longbibliography, 
               colorlinks=true, linkcolor=blue, citecolor=blue, urlcolor=blue, 
              ]{revtex4-1}
\pdfoutput=1
\usepackage[T1]{fontenc}
\usepackage[utf8]{inputenc}

\usepackage[english]{babel}     
\usepackage{adjustbox}          
\usepackage{colortbl}
\usepackage{xparse}             
\usepackage[caption=false]{subfig}
\usepackage{wrapfig,bm,bbm}
\usepackage[percent]{overpic} 

\usepackage[
            tightpage]{preview}

\newcommand{\beq}{\begin{equation}}
\newcommand{\eeq}{\end{equation}}
\newcommand{\bea}{\begin{eqnarray}}
\newcommand{\eea}{\end{eqnarray}}

\begin{document}

\newlength{\mybaselineskip}
\setlength{\mybaselineskip}{\baselineskip}

\title{Nuclear Fission Dynamics: Past, Present, Needs, and Future} 

\author{Aurel Bulgac}%
\affiliation{Department of Physics,  University of Washington, Seattle, Washington 98195--1560, USA}

\author{Shi Jin}%
\affiliation{Department of Physics,  University of Washington, Seattle, Washington 98195--1560, USA}

\author{Ionel Stetcu}%
\affiliation{Theoretical Division, Los Alamos National Laboratory, Los Alamos, NM 87545, USA}

\date{\today}

\begin{abstract}

Significant progress in the understanding of the fission process within a microscopic framework has been recently reported. Even though the complete description of this important nuclear reaction remains a computationally demanding task, recent developments in theoretical modeling and computational power have brought current microscopic simulations to the point where they can provide guidance and constraints to phenomenological models, without making recourse to parameters.  An accurate treatment compatible with our understanding of the inter-nucleon interactions should be able to describe the real-time dynamics of the 
fissioning system and could justify or rule out assumptions and approximations incompatible with the underlying universally accepted quantum-mechanical framework. Of particular importance are applications to observables that cannot be directly measured in experimental setups (such as the angular momentum distribution of the fission fragments, or the excitation energy sharing between the fission fragments, or fission of nuclei formed during the $r$-process), and their dependence of the excitation energy in the fissioning system. Even if accurate predictions are not within reach, being able to extract the trends with increasing excitation energy is important in various applications. The most advanced microscopic simulations of the fission process do not support the widely used assumption of adiabaticity of the large amplitude collective motion in fission, in particular for trajectories from the outer saddle towards the scission configuration. Hence, the  collective potential energy surface and inertia tensor, which are the essential elements of many simplified microscopic  theoretical approaches, become irrelevant. In reality, the dynamics of the fissioning system is slower than in the case of pure adiabatic motion by a factor of three to four times and is strongly overdamped. The fission fragment properties are defined only after the full separation, while in most of the current approaches no full separation can be achieved, which increases the uncertainties in describing fission-related observables in such methods.

\end{abstract}

\preprint{NT@UW-19-18,LA-UR-19-32211}

\maketitle


\section{The past}\label{sec:I}

In a matter of days after \textcite{Hahn:1939} communicated their yet unpublished results to 
Lise Meitner, she and her nephew Otto Frisch~\cite{Meitner:1939} understood that an unexpected
and qualitatively new type of nuclear reaction has been put in evidence and they dubbed it nuclear 
fission, in analogy to cell divisions in biology. Until that moment in time nuclear fission was considered 
a totally unthinkable process~\cite{Stuewer:2010,Pearson:2015}, ``as excluded by the small 
penetrability of the Coulomb barrier~\cite{Fermi:1934}, indicated by the Gamov's theory 
of alpha-decay''~\cite{Meitner:1939}. \textcite{Meitner:1939} also gave the correct physical 
interpretation of the nuclear fission mechanism. They understood that Bohr's compound 
nucleus~\cite{Bohr:1936} is formed after the absorption of a neutron, which eventually slowly 
evolves in shape, while the volume remains constant, and that the competition between the surface energy 
of a nucleus and its Coulomb energy leads to the eventual scission. \textcite{Meitner:1939} also
correctly estimated the total energy released in this process to be about 200 MeV. A few months later 
\textcite{Bohr:1939} filled in all the technical details and the long  road to developing a microscopic theory
of nuclear fission ensued. In the years since, a few more crucial theoretical results have been firmly established:
i) the defining role of  quantum shell effects~\cite{Strutinsky:1967,Brack:1972} and in 
particular the special role played by the pairing type of nucleon-nucleon interaction in shape 
evolution~\cite{Bertsch:1980,Bertsch:1997}; ii) the fact that the subsequent emission of neutrons and 
gammas can be described quite accurately using statistical methods~\cite{Weisskopf:1937,Hauser:1952}; iii)
and that the non-relativistic Schr{\" o}dinger equation should be adequate as well, as no genuine relativistic
effects, such a retardation, are expected to play any noticeable role in fission dynamics. 

Whether a fissioning nucleus undergoes either spontaneous fission or induced fission the time 
it takes the nucleus to evolve from its ground state shape until outside the barrier or past the outer 
saddle is very long in case of neutron induced fission $\approx{\cal O}(10^{-15})$ sec. 
in comparison with the time the nucleus slides downhill 
until scission, which is estimated to be ${\cal O}(10^{-20})$ sec. Therefore, the saddle-to-scission stage of
nuclear shape evolution is the fastest and arguably the most non-equilibrium stage of the nucleus dynamics 
from the moment a neutron has been captured and the stage which plays a crucial role in
in determining the FFs properties. 
The many intricacies of the fission process, the multitude of aspects, which required a deep theoretical 
understanding, ultimately rooted in the quantum nature of this phenomenon defied the efforts of many 
generations of theorists, and a huge plethora of mostly phenomenological models have been put forward, 
often based on contradictory assumptions between models. The extensive range 
of assumptions, range from adiabatic evolution on top of which one adds (relatively weak) dissipation 
and fluctuations, to strongly overdamped motion, when the role of the collective inertia becomes irrelevant, 
to full statistical equilibrium near the scission configuration, and to mixing quantum and classical descriptions.
 
Microscopically inspired models are typically based on the (ill suited choice of words, as the our analysis shows)
adiabatic approximation, which is often conflated with slow motion. The class of adiabatic transformations, during
which only mechanical work is performed and no heat transfer or entropy production occurs, are only a subclass
of slow motion or quasistatic processes.
Theorists believed that the nuclear shape evolution until the moment of scission was so 
slow that individual nucleons had a sufficient time to adapt to avoided single-particle level 
crossings~\cite{Hill:1953} and the entire nucleus would follow the lowest ``molecular term,'' using the
Born-Oppenheimer chemical terminology~\cite{Born:1927}, see Fig. \ref{fig:esp_q}. 
Following this assumption at first the 
generator coordinate method (GCM) has been introduced by \textcite{Hill:1953,Griffin:1957} and later on 
a related alternative approach, the adiabatic time-dependent Hartree-Fock (ATDHF) 
method~\cite{Baranger:1972,Baranger:1978,Villars:1978,Ring:2004}. GCM is still one of the most 
popular tools still in use in the microscopic theories of fission~\cite{Pomorski:2012,Schunck:2016,
Goutte:2005,Regnier:2016,Zdeb:2017,Schunck:2019,Younes:2019,Ring:2004}.  
The GCM and ATDHF method have been shown to be basically 
equivalent~\cite{Goeke:1980}, when GCM is defined with complex generator coordinates~\cite{Peierls:1962}.
As however \textcite{Goeke:1980} succinctly state: ``Usually the $|q\rangle$ 
is obtained by an educated guess using the preconceived knowledge of the process.'' 
(Typically $|q\rangle$  stands for a generalized Slater determinant, aka Hartree-Fock-Bogoliubov
many-nucleon wave function.) Even though many 
efforts have been dedicated to find a better way to choose the collective or generator coordinates the 
methods proved to be too difficult to implement in practice and the quality of the decoupling between 
collective and intrinsic degrees of freedom either not very good or difficult to assess ~\cite{Dang:2000}. 
An exact separation between collective and 
intrinsic degrees of freedom (DoF) it is equivalent to an adiabatic evolution of the set of collective 
DoF. Then the collective DoF would always follow the lowest ``molecular orbital'' and only work would be 
performed on the intrinsic DoF, and thus with no heat transfer, and the intrinsic 
system would remain ``cold'' during the entire evolution. 

\section{The present}\label{sec:II}

\begin{figure}
\includegraphics[clip, width=1.0\columnwidth]{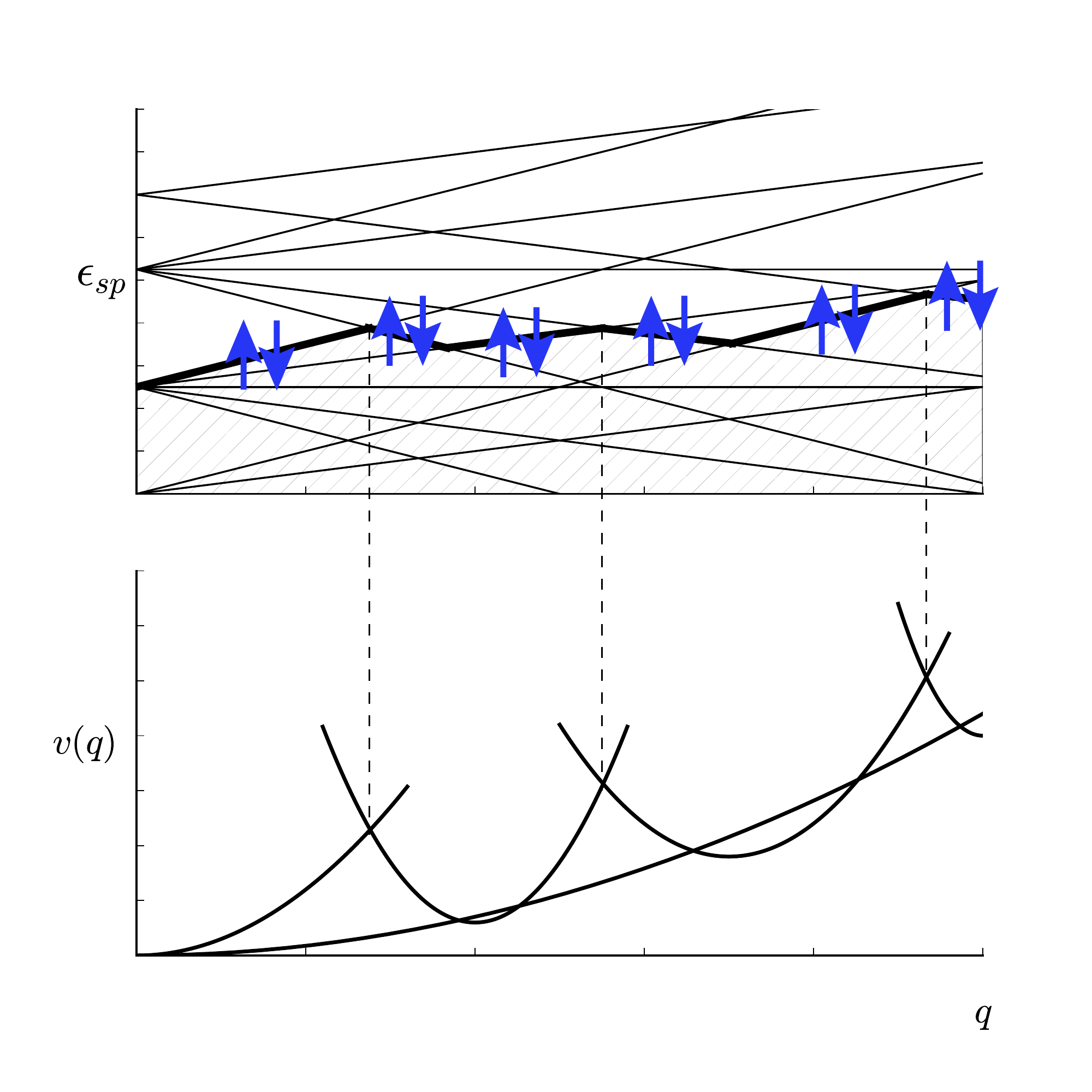}
\caption{\label{fig:esp_q} 
The schematic evolution of the single-particle nucleons levels
(upper panel) and of the total nuclear energy (lower panel)  as a function of 
deformation parameter $q$ \cite{Bertsch:1980,Barranco:1990,Bulgac:2019b}. The thick line 
represents the Fermi level and the up/down arrows depict the Cooper pairs of 
nucleons on the Fermi level only, in time-reversed orbits $(m, -m)$. }
\end{figure}

Unfortunately until recently this crucial
aspect of large amplitude collective nuclear dynamics, whether the large amplitude collective 
motion in fission is indeed adiabatic was never tested and, as our results 
unequivocally show, the adiabatic assumption is strongly violated. Surprisingly at first glance,
we have recently shown that the evolution of the nuclear shape is in reality significantly slower that the adiabatic
assumption would predict. One would naively expect that if the motion is even slower then at 
an avoided level crossing, see Fig. \ref{fig:esp_q}, the probability that the system would follow the 
lower ``molecular term'' is even greater and thus the adiabatic assumption would be even more likely
be valid. An analogy with a classical system can help and demonstrate just the opposite. If a railroad car is released 
on top of a hill, it will convert basically all the gravitational potential energy difference when reaching 
the bottom of the hill into kinetic energy. Thus only pure mechanical work on the intrinsic DoF
of the railroad car will occur with essentially no heat transfer or entropy production. 
This is an adiabatic process. However, if one were instead to block the wheels of the railroad car, 
the friction will slow down the car and almost the entire gravitational potential energy difference will 
be converted into heat, the wheels will get red hot, thus increasing the ``intrinsic energy'' of the car, 
and the speed of the car at the bottom of the hill would be rather small.

The only practical theoretical framework to consider in the treatment of the 
dynamics of large nuclei is the (Time-Dependent) 
Density Function Theory ((TD)DFT), which has been formulated a long time 
ago~\cite{Hohenberg:1964,Kohn:1965fk,Kohn:1999fk,Dreizler:1990lr,Gross:2006,Gross:2012}.
One of the main difficulties consist in constructing the energy density functional, for which no rigorous
recipes exist, DFT and TDDFT arrive at the mathematical conclusion that the  
stationary or time-dependent solution of the many-body Schr\"odinger equation is in a one-to-one-to-one
correspondence with the number density, an (arbitrary) applied one-body external potential, and that the 
number density can be obtained by solving the much simpler DFT or TDDFT equations. There is a 
continual debate in nuclear physics that DFT is not applicable to nuclei, which are self-bound
isolated systems. At the same time however, no one would argue that DFT cannot describe
neutrons and protons in the neutron star crust and deeper into the star.
Neutrons are delocalized in the neutron star crust and below it.
In the rod, slab, and tube phases in the neutron star crust and below 
both protons and neutrons are delocalized and 
in this respect they are similar to electrons in solids. 
One can imagine that in the future one might produce a nuclear trap using some kind of 
$\gamma$-lasers, similarly to
what nowadays experiments are made with cold atoms. Until then one can mentally imagine that one can 
put an isolated nucleus in a spherical infinite square well potential with a radius about 3...5x the nuclear 
radius (or even a harmonic potential) and compare the results of such a DFT treatment of the nucleus with 
the widely accepted DFT {\it alter ego}, the NEDF approach, a tool of choice in 
theoretical nuclear calculations. The results of the these two approaches are numerically indistinguishable under 
these conditions, and therefore the debate alluded above is merely pedantic.
One deficiency of a pure DFT approach is that the number density alone cannot disentangle between a 
normal and a superfluid system, and one needs an order parameter, as one does in the case of magnetization. 
The practical  local density approximation (LDA)~\cite{Kohn:1965fk}, which is the local formulation of DFT, 
has to be augmented with the anomalous density~\cite{Bulgac:2013a,Bulgac:2019}, and it was
dubbed the superfluid LDA (SLDA). In Refs.~\cite{Bulgac:2013a,Bulgac:2019} one can find detailed
reviews of the developments, verification, and validation of TDSLDA for a variety of physical systems,
ranging from cold atoms, nuclei, and  to neutron star crust.

A (TD)DFT framework for nuclear structure and dynamics should satisfy several requirements 
(in this order of importance):
i) the DFT and the Schr\"odinger description of observables should be identical, as both in ultimate instance rely
on the same inter particle interactions; ii) both DFT and Schr\"odinger 
equations should describe correctly Nature, thus we need accurate interactions between nucleons; iii) the 
numerical implementation of the (TD)DFT  should faithfully reproduce the theory. At present we definitely do not have
acceptable answers to the requirements i) and ii) and rely instead to a significant amount of phenomenology. 
The numerical  implementation of DFT without any drastic physical restrictions became possible only 
relatively recently, with the advent of supercomputers~\cite{Bulgac:2013a,Bulgac:2019}.

The TDSLDA is formulated in terms of Bogoliubov quasi-particle wave functions (qpwfs). 
The evolution of the qpwfs  is governed by the equations:
\begin{align} \label{eq:tdslda}
i\hbar \frac{\partial}{\partial t}
\begin{pmatrix}
u_{k\uparrow}  \\
u_{k\downarrow} \\
v_{k\uparrow} \\
v_{k\downarrow}
\end{pmatrix}
=
\begin{pmatrix}
h_{\uparrow \uparrow}  & h_{\uparrow \downarrow} & 0 & \Delta \\
h_{\downarrow \uparrow} & h_{\downarrow \downarrow} & -\Delta & 0 \\
0 & -\Delta^* &  -h^*_{\uparrow \uparrow}  & -h^*_{\uparrow \downarrow} \\
\Delta^* & 0 & -h^*_{\downarrow \uparrow} & -h^*_{\downarrow \downarrow} 
\end{pmatrix}
\begin{pmatrix}
u_{k\uparrow} \\
u_{k\downarrow} \\
v_{k\uparrow} \\
v_{k\downarrow}
\end{pmatrix},
\end{align}
where we have suppressed the spatial $\bm{r}$ and time coordinate $t$, and $k$  
labels the qpwfs (including the isospin) $[u_{k\sigma}(\bm{r},t), v_{k\sigma}(\bm{r},t)]$, 
with $\sigma = \uparrow, \downarrow$ the z-projection of the nucleon spin. 
The single-particle (sp) Hamiltonian $h_{\sigma\sigma'}(\bm{r}, t)$, and the 
pairing field $\Delta(\bm{r}, t)$ are functionals of various neutron and proton 
densities, which are computed from the qpwfs, see 
Ref.~\cite{Jin:2017} for technical details. No proton-neutron pairing is assumed 
in the present study, and the pairing field is singlet in character. 
A TDSLDA extension to a more complex pairing mechanisms is straightforward. 

\begin{figure}
\includegraphics[width=\columnwidth]{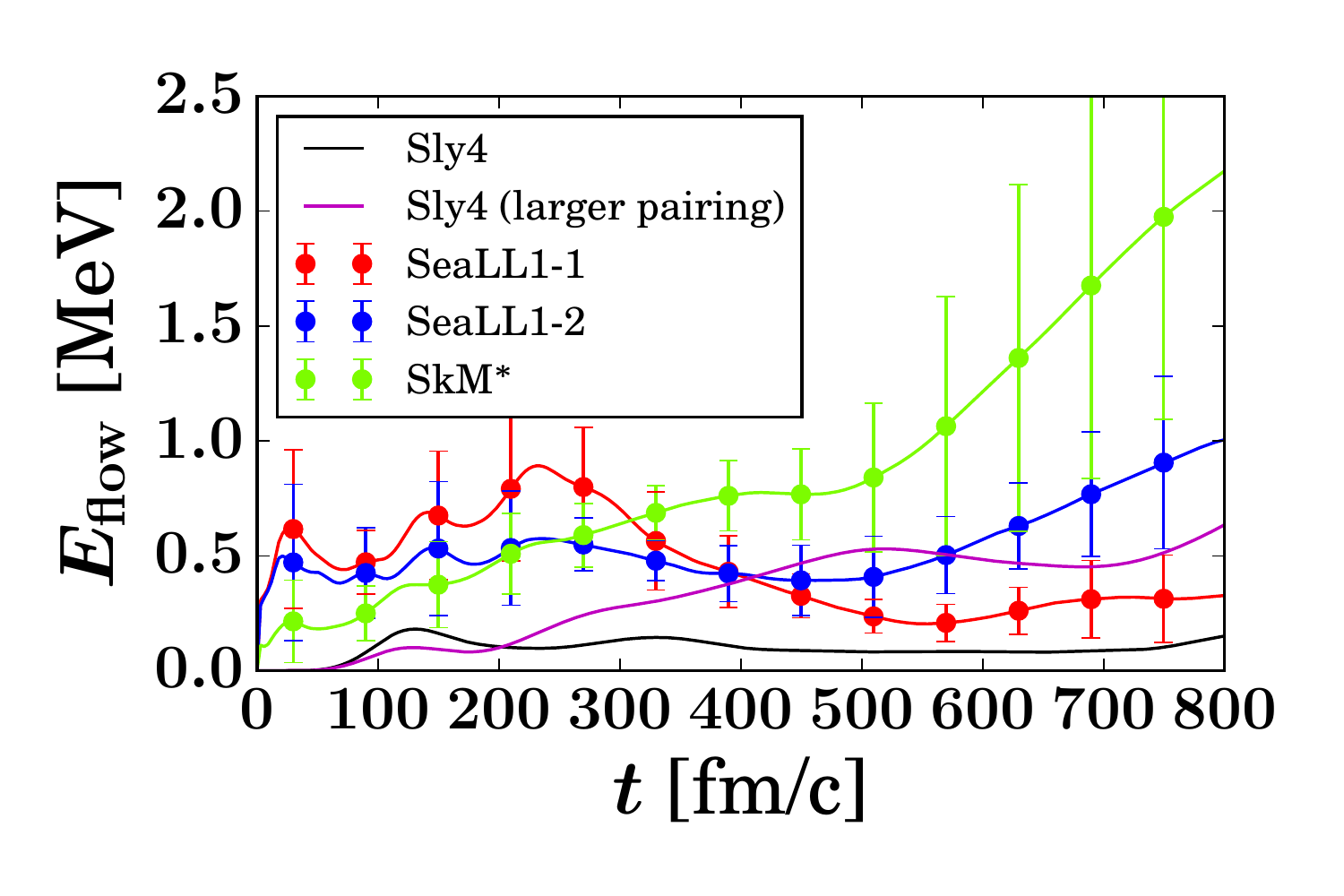}
\caption{
\label{fig:ecoll} 
The collective flow energy evaluated for NEDFs~\cite{Bulgac:2016}
realistic pairing SLy4 (dash-dot line), enhanced pairing SLy4* (dash line), and
for SkM*(dotted and dash-dot lines with error bars), and SeaLL1 
(solid and dashed lines with errors bars) sets~\cite{Bulgac:2019b}.  The error bars illustrate the
size of the variations due to different initial conditions in case of various 
SeaLL1-1,2 and SkM*-1,2 NEDFs used.  In the case of realistic pairing 
NEDF Sly4 (larger pairing) the time has been scaled by a factor of 1/10. }
\end{figure}

While a nuclear system evolves in time one can uniquely separate the energy into collective 
kinetic energy and intrinsic energy contributions~\cite{Bulgac:2019b} using the 
nuclear energy density functional (NEDF)
${\cal E}\left (\tau({\bf r},t), n({\bf r},t),...\right )$, in a similar manner as in hydrodynamics:
\bea
&& E_\text{tot} \!\!=
E_\text{coll}(t)+E_\text{int}(t)
\equiv \int \!\!d{\bf r}\frac{ mn({\bf r},t){\bf v}^2({\bf r},t)}{2} \nonumber \\
&& +\int \!\!d{\bf r}\, {\cal E}\left (\tau({\bf r},t)-n({\bf r},t)m^2 {\bf v}^2({\bf r},t), n({\bf r},t),...\right ).\label{eq:etot}
 \eea
 Above $n({\bf r},t)$ is the number density, $\tau({\bf r},t)$ is the kinetic density energy, and 
 ${\bf p}({\bf r},t)=mn({\bf r},t){\bf v}({\bf r},t)$ are linear 
 momentum and local collective/hydrodynamic velocity
 densities, and ellipses stand for various other densities.  
${{\bf p}({\bf r},t)}/{n({\bf r},t)}$ is the position of the center of the local Fermi sphere in momentum space. 
 The first term in Eq.~\eqref{eq:etot} is the collective/hydrodynamic energy
 flow $E_\text{coll}$ and the second term is the intrinsic energy
 $E_\text{int}$ in the local rest frame.  For the sake of simplicity
 we have suppressed the spin and isospin DoF, even though they are included
 in all the actual calculations.
 The collective energy $E_\text{coll}(t)$ is not vanishing only in 
 the presence of currents and vanishes exactly for stationary states. 
 The inertia tensor in $E_\text{coll}(t)$ in the case of irrotational collective motion 
 is fully equivalent to the Werner-Wheeler inertial tensor~\cite{Pomorski:2012}.
 The intrinsic energy $E_\text{int}(t)$  is determined only by the fermionic matter distribution.
 The qualitative new result established in Ref.~\cite{Bulgac:2019b} and also
 illustrated here in Fig.~\ref{fig:ecoll} is that in an fully unrestricted TDSLDA the collective 
 flow energy is almost negligible until scission, in total discrepancy with what one would have 
 naively expected if the adiabatic assumptions would be satisfied.  

Our simulations point to an
unexpectedly small $E_\text{coll}$ from saddle-to-scission,
corresponding to a collective speed ${v_\text{coll}}/{c}\approx
0.002\cdots 0.004$, significantly smaller than the Fermi velocity
${v_F}/{c}\approx 0.25$, see Fig. \ref{fig:ecoll}.  
Since in TDSLDA one simulates the one-body
dynamics exactly, it is natural to discuss adiabaticity at the mean-field
level.  The transition rate between sp states is suppressed if the
time to cross an avoided level-crossing configuration satisfies the
restriction $ \Delta t\ll {\hbar}/{\Delta \epsilon} \approx 400$ fm/c,
where $\Delta \epsilon = {1}/{\rho_{\rm sp}(\epsilon_F)}$ is the average sp
energy level spacing at the Fermi level. Since on the way from
saddle-to-scission the time required is $1\dots 3\times10^3$ fm/c and 
several dozen of avoided level crossings
occur~\cite{Barranco:1990,Bertsch:2018}, this condition is clearly
violated.  Somewhat surprisingly, the adiabatic assumption is also violated even in the case of SLy4 NEDF, 
see Ref.~\cite{Bulgac:2016} and Fig.~\ref{fig:ecoll}, when the saddle-to-scission time is 
${\cal O}(10^4)$ fm/c as well.
The collective motion is thus expected to be strongly
overdamped. From saddle-to-scission the nucleus behaves as a very
viscous fluid, the role of collective inertia is strongly suppressed,
and the trajectories follow predominantly the direction of the
steepest descent with the terminal velocity determined by the balance
between the friction and the driving conservative forces, see Fig.~\ref{fig:ecoll}.  

This result serves as the first microscopic
justification for the assumption  of the overdamped Brownian motion
model~\cite{Weidenmuller:1984,Randrup:2011,Randrup:2011a,Randrup:2013,Ward:2017,Albertsson:2020} 
and partially to the scission-point
model~\cite{Wilkins:1972,Wilkins:1976,Lemaitre:2015,Lemaitre:2019}.  
In both these phenomenological models it is assumed that the preformed FFs 
are in statistical equilibrium and that the collective energy flow is either vanishing or very small. 
The main difference is that in the scission-point model there is no mechanism to ensure that 
all equilibrium scission configurations could be reached dynamically, 
while the nucleus evolves from the saddle-to-scission.  Moreover, the relaxed FF properties
are defined only after the FFs become sufficiently well separated, see below.
It is equally unexpected
that in the case of enhanced pairing, when the pairing condensates
retain their long-range order throughout the entire saddle-to-scission
evolution, the collective dynamics remains strongly overdamped.

\begin{figure}
\includegraphics[clip, width=.8\columnwidth, angle = 90]{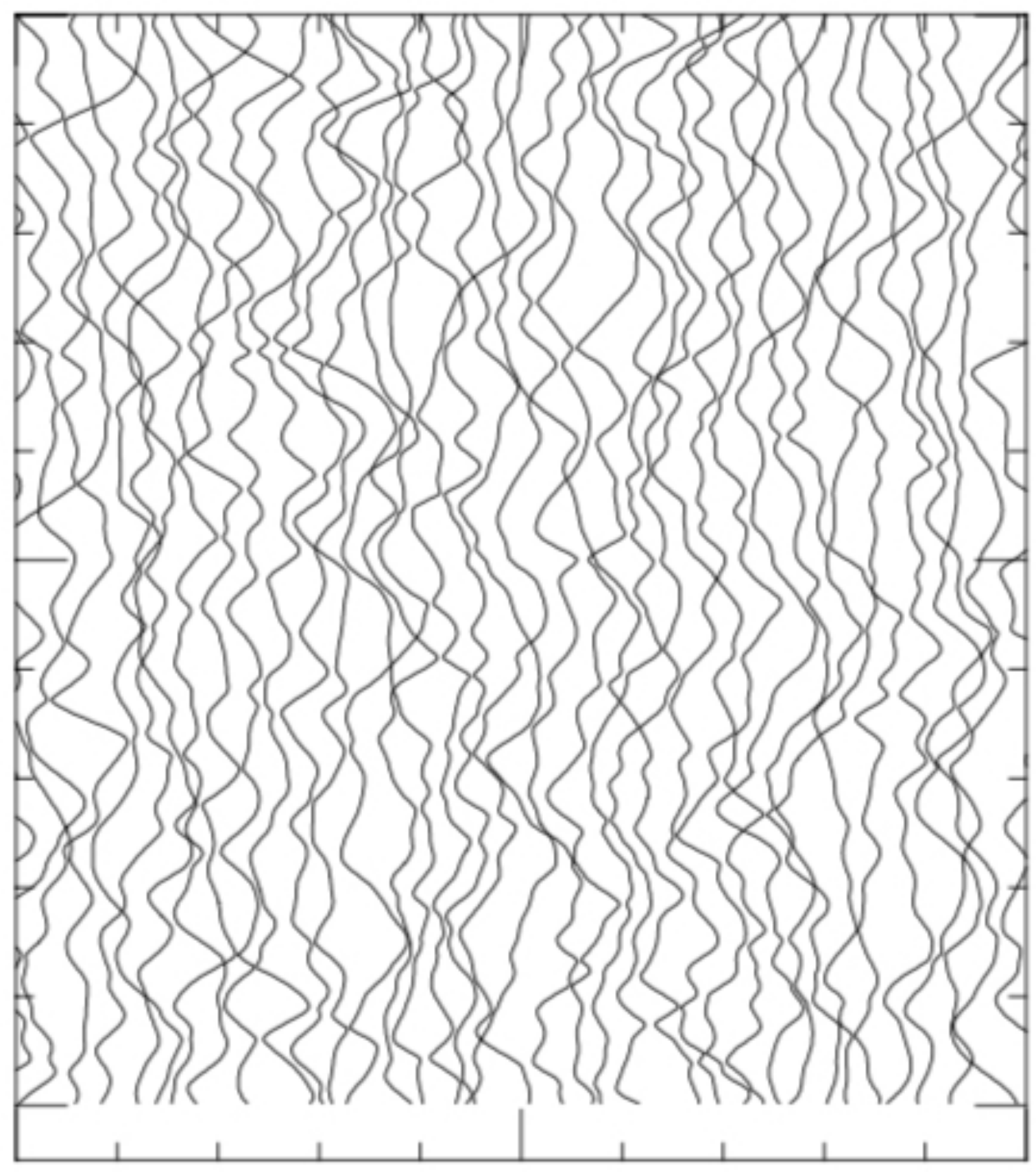}
\caption{\label{fig:grab23} 
In nuclei the level density increases with the excitation energy quite fast, practically exponentially
at energies of the order of the neutron separation energy, when 
$\rho(E^*)\propto \exp \left (\sqrt{2aE^*}\right ) $~\cite{Bethe:1936,Bohr:1969},  and it reaches values of 
${\cal O}(10^5)$ MeV$^{-1}$ and various potential energy surfaces, 
corresponding to different ``molecular terms'' display a large number of avoided level crossings, 
see Ref.~\cite{Bulgac:1996}. }
\end{figure}

While evolving from saddle-to-scission a nucleus encounters a large number of avoided level crossings
and instead of following the lowest potential energy surface, as would happen in an adiabatic evolution,
many transitions to higher excited levels occur. Similarly to what is known for decades in 
chemistry~\cite{Tully:1971,Tully:1990,Tully:1994}, one should consider not only the lowest potential energy 
surface, when initially the system could be found on the  lowest with unit probability, but all ``molecular terms,'' 
which  become populated during the evolution, see Fig. \ref{fig:grab23}. The separation between 
such potential energy surfaces reaches values of order ${\cal O}(10)$ keV or even less, 
and only if the system traverses a level crossing in a time much longer than
$\approx {\hbar}/{10 \;\textrm{keV}}\approx 20,000$ fm/c or longer the nucleus will remain on the lowest 
potential energy surface. On the other hand the saddle-to-scission time is ${\cal O}(10^3)$ fm/c and 
this is why the adiabatic assumption is in the final analysis strongly violated.
 
 \begin{figure}
\includegraphics[clip, width=1.0\columnwidth]{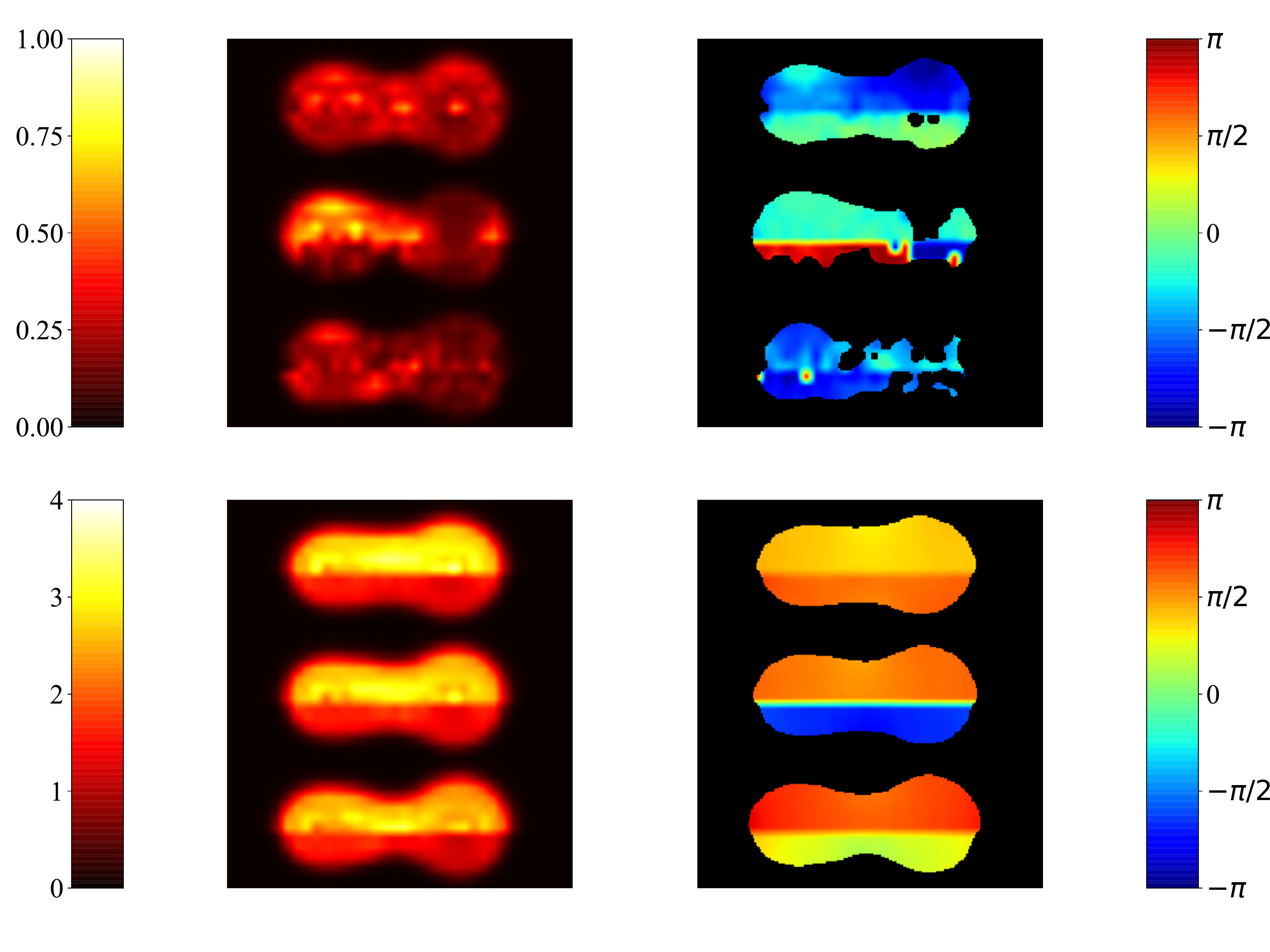}
\caption{\label{fig:snapshots} 
These are some selected snapshots of the neutron/proton magnitudes (left) and phases (right) of pairing gaps 
(upper half/lower/half of each frame) in the case of realistic pairing strengths (upper panels) and enhanced
pairing strength (lower panels)~\cite{Bulgac:2019b}, obtained in the case of SLy4 NEDF.  }
\end{figure}

Why do pairing correlations play an important role in fission dynamics? In a static nuclear configuration
each single-particle level is doubly-degenerate, due to the Kramers degeneracy. At the level crossing both 
nucleons on the highest occupied level (homo - highest occupied molecular orbital) would have to transition 
simultaneously  to the lowest down-sloping level (lumo - lowest unoccupied molecular orbital)  
to ensure that the local momentum distribution remains
approximately spherical, as otherwise it would acquire an oblate 
shape~\cite{Hill:1953,Bertsch:1980,Bertsch:1997}, while the shape of the nucleus becomes more prolate.
Nucleon-nucleon interactions at low momentum transfer can be modeled with a reasonable accuracy 
with a zero-range $\delta$-interaction, which favors the transitions between pairs of time-reversed 
orbitals, exactly as the Kramers degenerate orbitals, see Fig.~\ref{fig:esp_q}.  
The up-sloping levels are characterized by larger projections of the angular momentum on the fission axis, 
$|m|\approx  k_Fr_0A^{1/3}$, and these levels should be depopulated, 
since in a FF the largest angular momenta are smaller, $\approx k_Fr_0(A/2)^{1/3}$.
While evolving from one level crossing to 
the next, the entire evolution is likely rather well reproduced by a simple one-body dynamics, 
as each single-particle level occupation probability changes little. What one-body dynamics 
lacks is the contributions arising from the Boltzmann collision integral. However, at each level 
crossing the two-correlated nucleon pairs will undergo a collision, and at low 
energies transitions between pairs of time-reversed orbitals
expected to dominate the collision rate. One should take with a grain of salt this simplistic picture 
of ``collisions'' and jumps between sp levels, as nothing happens instantaneously or at one point 
in space in quantum mechanics. In the presence of a Bose-Einstein condensate of nucleon  
Cooper pairs the nucleus has a superfluid component and pair transfers are enhanced
due to the Bose enhancement factor. The dynamics of the nuclear systems then approaches 
the evolution of classical inviscid (no viscosity) or perfect fluid. An illustration of this behavior was exemplified in 
Fig. 4 in Ref.~\cite{Bulgac:2019} and in Fig. \ref{fig:snapshots}. 
When the magnitude of the pairing field was artificially increased 
from a realistic value to a value 3...4x larger the evolution time from saddle-to-scission 
decreased by a factor of $\approx 10$ and at the same time the long range coherence of the pairing field across 
the entire nucleus survived. For realistic values of pairing strengths during the descent from the 
saddle-to-scission both proton and neutron pairing fields fluctuate both in space and time,
long range order basically vanishes, but quite often it is revived. 

\begin{figure}
\includegraphics[clip, width=1.0\columnwidth]{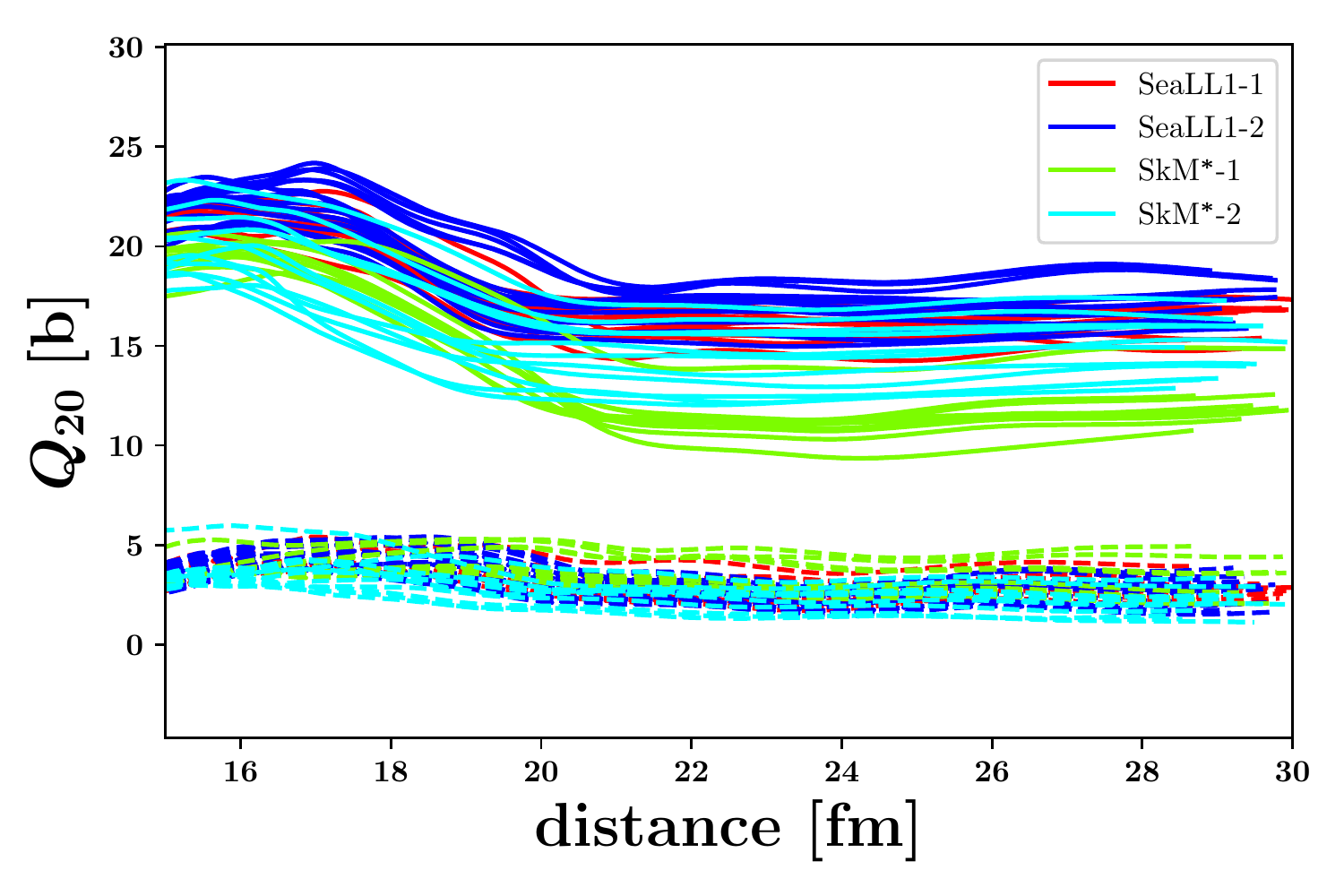}
\includegraphics[clip, width=1.0\columnwidth]{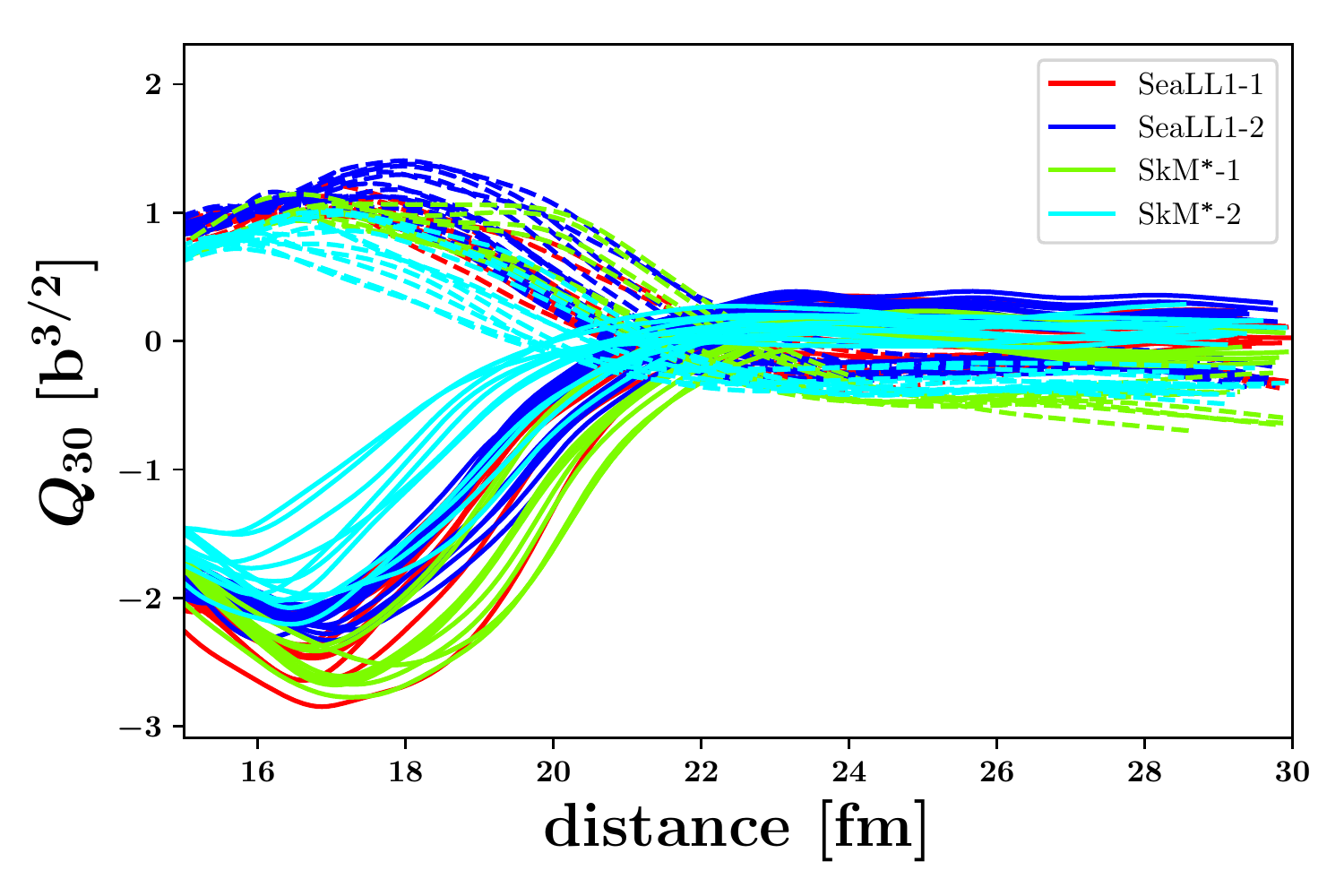}
\caption{\label{fig:q2q3} 
The evolution of the quadrupole and octupole moments of the FFs 
as a functions of their separation after scission.~\cite{Bulgac:2019b}. 
Different lines correspond to different initial conditions and different NEDFs, 
see Ref.~\cite{Bulgac:2019b} more details. The solid/dashed lines are for the light/heavy 
FFs respectively. }
\end{figure}

Another important aspect which emerged from our TDSLDA fission simulations~\cite{Bulgac:2016,Bulgac:2019b},
which is of significant importance in the implementation of various phenomenological models, concerns
the shapes of FFs at scission and when they reach their relaxed shapes, see Fig.~\ref{fig:q2q3}. 
This behavior is apparently confirmed indirectly by experiments.  In  Langevin or
Fokker-Planck~\cite{Grange:1983,Frobrich:1998,Sierk:2017,Ishizuka:2017,Sadhukhan:2016,Sadhukhan:2017}, 
TDGCM~\cite{Regnier:2016,Regnier:2019}, and 
scission-point~\cite{Wilkins:1972,Wilkins:1976,Lemaitre:2015,Lemaitre:2019} models 
the calculation of the FFs yields 
consider only a very limited range of nuclear shapes. In particular in such simulations one never introduces the 
octupole FF moments. Our results, as well recent analysis by \textcite{Scamps:2018}, clearly 
demonstrate that the FFs emerge at scission octupole deformed and also with a 
significantly larger quadrupole deformation than the relaxed values. 
Moreover, even after scission the FFs there is a significant Coulomb interaction between them, which leads
to the excitation of both low energy and giant resonances in FF~\cite{Mustafa:1971,Simenel:2014}. 
This interaction enables additional excitation energy exchange between 
the FFs after scission, and it also affects their total kinetic energy, a behavior also seen in our simulation, 
but yet not documented. In statistical scission-point 
models there is no dynamics, and only the competition between FFs configurations at the scission point are
considered~\cite{Wilkins:1972,Wilkins:1976,Lemaitre:2015,Lemaitre:2019}, a model to which our
results lend partial support.  However, the only shapes considered are quadrupole
deformation of the relaxed FFs, which clearly is not what our dynamical 
simulation demonstrate. 

\begin{figure}
\includegraphics[clip,width=\columnwidth]{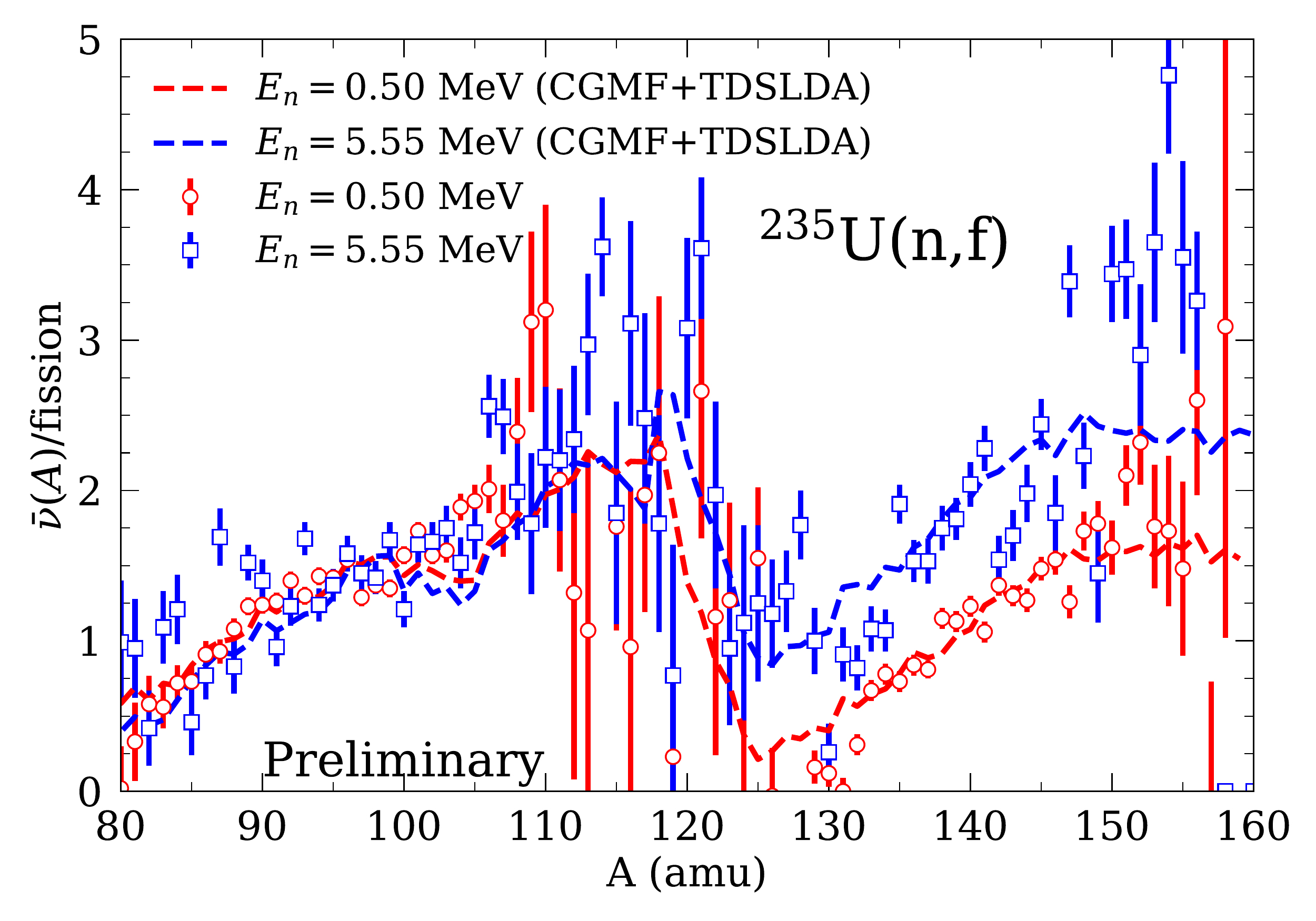}
\caption{\label{fig:nubar}We compare here the average neutron 
multiplicity $\bar{\nu}(A)$ emitted by FFs using a
CGMF simulation~\cite{Becker:2013}, which assumes an $E_n$ dependence for the energy sharing 
extracted using the the excitation  energy sharing between the FFs in our calculation with NEDF SeaLL1, 
as a function of the equivalent incident neutron energy in 
$^{235}$U(n,f) reaction along with available experimental data~\cite{Muller:1984}. 
Note that in this figure the parametrization was based on $^{240}$Pu calculations, 
while $^{236}$U calculations are in progress.}
\end{figure}

Our simulations put in evidence another very important aspect, the mechanism of the 
excitation energy sharing between the FFs. As a rule the heavy FF emerges 
in the end cooler than the light FF, even though they have been in contact for 
quite a long time before scission. Moreover, when increasing the initial energy excitation of 
the fissioning nucleus we have established that only the heavy FF becomes hotter 
and that is reflected in the average neutron multiplicity number of emitted, and results which is 
in apparent agreement with experimental findings, see Fig.~\ref{fig:nubar}. Experimentally it is extremely difficult 
to infer the excitation energies of the FFs, which are a crucial input in various statistical 
codes~\cite{Randrup:2009,Becker:2013,Randrup:2019}. \textcite{Bertsch:2019} argue that the FFs spin distribution, 
which determines the prompt gamma angular distribution, can be used to infer information of the 
excitation energy sharing between FFs. \textcite{Randrup:2019} point to a pronounced anti-correlation 
between ${\bar \nu}(A)$ and mean total kinetic energy $\bar{TKE}$, which can be used reduce uncertainties
in data analysis. \textcite{Schmidt:2010,Schmidt:2011} suggested a phenomenological
model,  the ``energy-sorting'' mechanism based on the empirical constant temperature parametrization  
of the nuclear level densities due to \textcite{Gilbert:1965}. In this ``energy-sorting'' model the FFs before 
scission have different temperatures, with a lower temperature of the heavy fragment, which would 
generate an energy  flow from the light/hotter to the heavy/cooler fragment. Our simulations 
demonstrate however that the near scission the two FFs have properties quite different (shape, excitation 
energy, pairing correlations) from the properties of relaxed fragments. It is therefore problematic to relate 
the properties of excited isolated nuclei with the properties of FFs in contact before rupture.

\begin{figure}
\includegraphics[clip,width=\columnwidth]{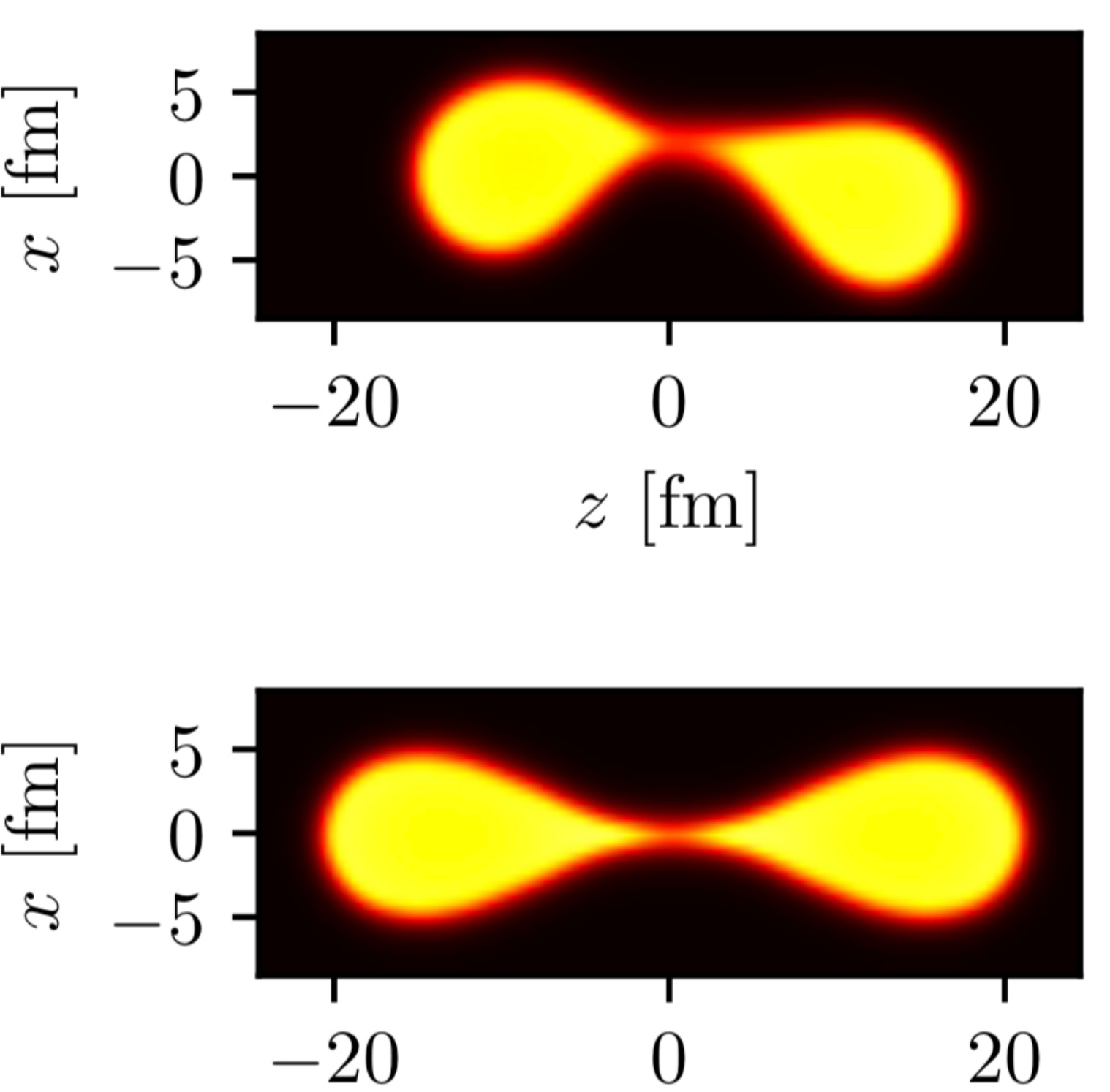}
\caption{\label{fig:bend}  In the upper panel we show a typical scission configuration when 
unlimited type of fluctuations are allowed as compared to a typical TDSLDA scission configuration, 
in which axial symmetry is assumed in this case for $^{258}$Fm~\cite{Bulgac:2019a}. }
\end{figure}
As we have mentioned above, the number of collective DoF
and their character is a long standing problem in microscopic inspired
theoretical and phenomenological models. The choice of collective DoF and their 
character is guided by the authors' intuition, their computational and other abilities,
educated guess, imagination, and/or available resources~\cite{Goeke:1980}. It was never proven or
demonstrated that a GCM representation of the nuclear wave function is ever accurate. 
The wave function of a many-fermion system in TDGCM is constructed according to the following prescription
\beq
\Psi_\text{GCM}({\bf x},t) =\int d{\bf q}  f({\bf q},t)\Phi({\bf x}|{\bf q}) ,\label{eq:GCM}
\eeq
where $\Phi({\bf x}|{\bf q})$ are (generalized) Slater determinants depending on nucleon spatial 
coordinates, spin, and isospin ${\bf x} =(x_1,\ldots,x_A),\; x_k=( {\bf r}_k, \sigma_k,\tau_k)$ and
parameterized by the collective coordinates ${\bf q}=(q_1,\ldots,q_n)$, and where $f({\bf q},t)$    
is the collective wave function. There is no criterion or small parameter which controls the accuracy
of such a representation, that is
\beq
1 - | \langle \Psi_\text{GCM}|\Psi_\text{exact}\rangle| \ll \epsilon.
\eeq
We have presented rather simple arguments that most likely such an accurate representation
does not exist in general, particularly in fission large amplitude collective motion~\cite{Bulgac:2019b}.
A rather simple estimate of the possible number of different shapes, and therefore of independent terms in 
Eq.~\eqref{eq:GCM}, shows that it is basically infinitely small in comparison with the number of possible
terms in an exact multi-configurational Slater determinant representation of $\Psi_\text{exact}$. 
Can we check whether GCM is a good approximation for the saddle-to-scission evolution? Now
we definitely can and we have the answer, and since
\beq
E_\text{total}=E_\text{coll}(t)+E_\text{int}(t)\approx E_\text{int}(q,T)\approx V(q,T)\approx const.
\eeq
where  $T$ is the temperature of the intrinsic system, 
we know that the collective motion is strongly overdamped, that there is an irreversible 
energy flow from the collective/shape DoF to the intrinsic DoF, 
and that during the saddle-to-scission evolution the temperature of the intrinsic system increases, 
as does its (entanglement) entropy as well. We should also remember, that the evolution of the
fissioning nucleus from saddle-to-scission is a truly non-equilibrium one, and the notion of a 
``slowly evolving'' nuclear temperature $T$ might be questionable.
This suggest that in phenomenological calculations a 
more physically motivated choice for the potential energy surface would be one in which with
increasing deformation one would increase the ``temperature $T$,'' so as to keep the 
``collective potential'' energy $V(q,T)$ essentially
equal to the initial excitation energy of the compound nucleus. Basically this is the prescription that 
Randrup et al.~\cite{Ward:2017,Albertsson:2020} have implemented lately.

There were indirect indication in microscopic approaches
that the number of collective DoF can vary along the fission path.  In the overwhelming majority of 
fission studies the only deformation of the nucleus close to the ground state is the axial symmetric 
quadrupole  moment $Q_{20}$, in spite of the fact that the Bohr-Mottelson five-dimensional collective 
Hamiltonian is one tool of choice to describe the low energy structure of open-shell 
nuclei~\cite{Bohr:1969,Bertsch:2007,Delaroche:2010}.
\textcite{Ryssens:2015} have shown in a beyond mean field calculation of the 
1-dimensional fission path the nucleus
$^{240}$Pu is axially deformed around the ground state configuration, it becomes triaxially deformed 
when it reaches the fission isomer region, and before it reaches the outer fission barrier it breaks the axial 
symmetry as well. We have recently developed a framework into which one can include fluctuations
and dissipation in a quantum approach~\cite{Bulgac:2019a}, which allows for shape fluctuations
of any kind. What we observed that by allowing a bending mode to become active, see Fig.~\ref{fig:bend},
the mass and TKE distributions acquire shapes in quite nice agreement with observations. It was 
discussed quite some time ago that such bending modes might be responsible for the angular distributions
of the FFs~\cite{Moretto:1989} and this type of distributions can be also 
extracted from TDSLDA calculations in the near future~\cite{Bertsch:2019,Bulgac:2019x}. 
Phenomenological or even TDGCM approaches do not consider so far such bending modes, which 
are definitely physically relevant for a large number of observables.

\section{What are the needs?}\label{sec:III}

Nuclear fission is a complex process, in which a heavy nucleus evolves from a compact shape to a 
configuration in which two or more fragments are produced, and is accompanied by emission of prompt 
neutrons and gamma rays (and, eventually, electrons and antineutrinos).  Most of the energy is released 
in form of kinetic energy of the fragments, while prompt neutrons emitted before beta decays 
play the main role in applications like energy production.

The dynamics of the nuclear system from the formation of compound nucleus until after the acceleration 
of FFs and prompt particle emission is too rapid to be experimentally resolved. On the 
other hand, the time scale of the weak interaction, which governs the decay of FFs in experiments towards 
stability ranges from seconds to minutes, and thus the dynamics involving beta decay towards stability, 
including delayed neutron and gamma emissions, can be decoupled from the initial more rapid part that 
is only governed by the strong interactions. Consequently, the fragment properties directly influence 
the properties of prompt fission and gamma rays, which have been the subject of comprehensive, albeit not 
exhaustive, experimental investigations over the years. Thus, at Los Alamos National Laboratory, experimental campaigns have investigated the prompt fission neutron spectra for neutron-induced fission of major actinides ($^{235,238}$U, $^{239}$Pu) for a large range of incident neutron energies using the ChiNu experimental setup \cite{Haight:2015aa,Kelly-Keegan-J.:2018aa}. On the other hand, experiments that measure the average neutron multiplicity as a function of pre-fission neutron mass \cite{Maslin:1967aa,Nishio:1998aa,Tsuchiya:2000,Batenkov:2005,Gook:2014aa,Gook:2018aa}, which can be used to guide energy sharing in fission fragments, are more scarce, and for a limited number of reactions (usually spontaneous fission or neutron-induced fission with thermal neutrons). The surrogate reaction technique is used by Lawrence Livermore National Lab and Texas A\&M to measure the $^{239}$Pu and $^{241}$Pu prompt fission neutron multiplicity (average and distribution) as a function of equivalent incident neutron energy \cite{Akindele:2019aa,Wang:2019aa}. Complementary, significant resources have been devoted toward measuring the prompt fission gamma rays produced in the decay of fission fragments, after the neutron emission, at Los Alamos, using the DANCE calorimeter \cite{Ullmann:2013,Chyzh:2013,Chyzh:2014,Jandel:2014aa}, and worldwide employing high-resolution detectors 
\cite{Billnert:2013,Oberstedt:2013,Lebois:2015aa,Gatera:2017,Makii:2019aa}. Such measurements complement existing measurements of prompt-gamma rays by  \textcite{Verbinski:1973}, and many others \cite{Peelle1971,Pleasonton:1972,Pleasonton:1973}. Furthermore, data on total gamma production can be useful in evaluating the prompt fission gamma properties, if no other data exist, fission dominates, and the other gamma-producing channels can be modeled with reasonable accuracy \cite{Stetcu:2020aa}.

In addition to detecting and measuring the products of decay of FFs, a concerted effort has been directed towards the direct measurement of the FF distributions. While simpler, the ``2E'' measurements \cite{Straede:1987aa,Vives:2000aa,Birgersson:2009aa,Duke:2016} have the shortcoming that additional information is required to identify the mass of the fragment, hence the poor 4-6 amu resolution. Reducing such large uncertainties was the main argument for the construction of the SPIDER “2E-2v” experimental setup \cite{Tovesson:2013,Meierbachtol:2016} at the Los Alamos Neutron Science Center, which should start taking data in the near future. In this setup the
two fragment kinetic energies and velocities are measured simultaneously and can achieve one mass unit resolution, as also demonstrated by similar setups, COSI FAN
TUTTE at ILL, France and VERDI at JRC-Geel, Belgium. The Sofia experiment \cite{Pellereau:2017aa} at GSI has produced an extensive amount of experimental data with accuracies in mass and charge less than one mass unit, but suffer from the fact that a large range of excitation energies that cannot be disentangled is produced in Coulomb excitations. LLNL and LANL activation measurements at TUNL  have
provided invaluable data on the incident energy dependence of cumulative fission product yields \cite{Gooden:2016aa,Wang:2019aa}, and ongoing efforts are directed toward testing the Bohr hypothesis of independence between entrance and
outgoing channels in a compound reaction by comparing fission yields in the $^{239}$Pu(n,f) and
$^{240}$Pu($\gamma$,f) reactions \cite{Silano:2020aa} that lead to the same fissioning system, at least in mass and charge. At CARIBOU, two projects are underway: (i) measurement of fission product properties (isomeric yield ratios, gamma-ray decay branching ratios, and $\beta$-delayed neutron emission properties) and (ii) improvements of the antineutrino spectrum simulations by performing measurements of the $\beta$-decay data using GAMASPHERE \cite{Savard:2019aa}.
Other experiments related to the properties of FFs have concentrated on measuring the average total kinetic energy (TKE) of the fission fragments  at the LANSCE WNR facility for energy from a few hundred keVs to 200 MeV \cite{Meierbachtol:2016,Duke:2016}.
The High Rigidity Spectrometer~\cite{HRS} at FRIB~\cite{FRIB} will allow to explore the fissioning of very neutron-rich  nuclei, 
study their shell structure, the presence of superheavy, the limits of stability and the equation of state of neutron-rich nuclear matter, the neutron skins, and shed light on the nature of the $r$-process,  so far an unknown territory. 

Treating the evolution of a heavy nucleus from a compact configuration until the start of beta decays 
in the FFs, including prompt neutron and gamma emissions, is a complicated task, 
computationally unfeasible within a unified microscopic approach. Within this reality, one has to consider a mixture 
of approaches, in which the initial part of the fission process is treated within a microscopic framework 
that can inform more phenomenological treatments that model the emission of prompt neutrons and 
gamma rays, and whose results can be directly compared against a large set of experimental data. And the 
input from microscopic models does not necessarily need to be restricted to the fission process. The 
systematics of several physical quantities used in the phenomenological models of neutron and gamma 
emission is based on data for stable nuclei. Since the FFs are nuclei far from stability, it is 
important to investigate within more microscopic models the validity of various systematics far from 
stability, as prompt neutron and gamma observables are sensitive to ingredients like level densities, 
optical models, or gamma strength functions.

Several models have been proposed to describe the shape dynamics, and to some extent many of them 
are able to reproduce experimental quantities like the pre-neutron emission mass distributions, irrespective 
of the approximations involved. But, as noted above, the direct measurement of the FFs 
before neutron emission is not possible. Therefore, even before one goes into details regarding the validity 
of the approximations involved in theoretical models, one needs to consider the corrections involved 
in the analysis of the experimental data. The post-neutron emission FF mass distributions are 
obtained using information on average prompt neutron multiplicity as a function of the fragment mass, 
$\bar\nu(A)$, either from measurements, where available, or from phenomenological models. Given the 
scarcity of the data, and the fact that these phenomenological models have been built in a systematics based on 
little data, the uncertainties arising from $\bar\nu(A)$ could be considerable. Another assumption is that no 
neutrons are emitted during scission or at neck rupture. It is conceivable that the number of neutrons emitted 
during the scission dynamics is not necessarily large, but how large is large?  If the fraction of scission neutrons 
is significant, as some phenomenological models predict, citing the experimental prompt fission spectrum as 
evidence to support the claim~\cite{Carjan:2010,Carjan:2012,Rizea:2013,Carjan:2015,Capote:2016a,Carjan:2019}, what 
is the impact on the experimental analysis? While these models are not universally accepted, a microscopic 
approach should be able to answer such a question, and the assumption of small numbers of neutrons emitted 
during scission should not be implicitly included in the model. Moreover, it would be rather impossible to asses 
the fraction of scission neutrons if the model does not follow the dynamics of the process until full separation.

All the codes that model the prompt neutron and gamma emissions have been built in the assumption that 
the neutron emission proceeds after the full acceleration of the fission 
fragments ~\cite{Vogt:2013,Becker:2013,Litaize:2015,Schmidt:2016,Schmidt:2018,Talou:2018}. 
This assumption can have  significant consequences, as at full acceleration the neutrons are maximally boosted. 
If the emission occurs instead during the acceleration, then the prompt neutron spectrum in the lab frame 
can noticeably be altered.

In addition to validating assumptions in phenomenological codes, a reliable theoretical model should provide 
information on observables that cannot be measured, but are essential in modeling the neutron and gamma 
emission. Thus, while the total excitation energy available in the fragments can be determined from the Q-value and 
TKE, no additional constrains on how this energy is shared between the FFs are available. As the most 
efficient way to lower the energy in FFs is via neutron emission, $\bar\nu(A)$ can be used to 
parametrize the excitation energy sharing. However, such data are scarcely available, usually for a limited number of 
spontaneous and neutron-induced fission with thermal neutrons, and thus guidance from microscopic models on
 how the excitation energy is shared with increasing incident neutron energy is required not only for minor actinides,
 but even for induced fission reaction of major actinides $^{235, 238}$U and $^{239}$Pu isotopes.

Gamma emission competes with the neutron emission when the excitation energy in the nucleus is around the
neutron separation energy. The spin of the fragments, which cannot be directly measured experimentally,
determines the exact strength on the neutron-gamma competition. Only indirect information can be extracted
regarding the spin distribution of the FFs from properties of prompt gamma rays, like average 
multiplicity, prompt gamma fission spectra \cite{Stetcu:2014a}, or isomeric ratios \cite{Stetcu:2013}. The 
measurements of gamma rays are also scarce with increasing the incident neutron energy. Experimentally, one 
observes an increase of the total prompt fission gamma energy released  in fission \cite{Frehaut:1983}, which has 
been interpreted as an increase of the average spin with the incident energy. However, 
only little experimental data exist and no microscopic model has been able to reproduce such trends. 

The main message of this section is that experimentally one cannot isolate and measure properties of  post-scission 
FFs, and any full model will have to include simulations of prompt neutron and gamma emission, in 
order to compare and validate against experimental data. Given that many quantities used in the modeling of prompt
neutron and gamma emission are taken from data systematics, which in general is available only for nuclei close to stability, 
the reliability of the theoretical model near scission and beyond, until full separation, is very important. Such models 
should  allow for a full separation of the fragments, and employ approximations that are validated and under control, 
as to reduce any uncertainties regarding the FF properties.

\section{What lessons have we learned so far and what is the most likely path to the Future?}

While pairing is not the engine driving the fission dynamics, pairing provides the essential lubricant, 
without which the evolution may arrive rather quickly to a screeching 
halt~\cite{Goddard:2015,Goddard:2016,Tanimura:2015}. So far we have not considered whether 
proton-neutron pairing might have a role in fission. It is very unlikely that a condensate of proton-neutron 
pairs exists in heavy nuclei, but as we have learned from our simulations, proton-neutron transitions with
L=0 between single-particle orbitals could be important as the neutron-neutron and proton-proton transitions 
with (S,L,T)=(0,0,1) at low excitation energies, even in the absence of a condensate of such pairs, see the 
discussion concerning Figs.~\ref{fig:esp_q} and \ref{fig:snapshots}.

TDSLDA framework for fission dynamics, while it does not incorporate fluctuations, has provided a lot of 
insight into the real quantum dynamics, and it revealed extremely valuable information into nuclear 
processes and quantities, which are either not easy or impossible to obtain in laboratory or observations:
FFs excitation energies and angular momenta distributions prior  to neutron and gamma emission, element 
formation in astrophysical environments, as well as other nuclear reactions in a parameter free approach. 
In particular, the excitation energy sharing mechanism between FFs and its evolution with the initial excitation 
energy of the compound nucleus was not accessible until now within a dynamic approach. Fluctuations, 
which are essential in order to reproduce mass and charge yields for example, can be now incorporated 
into a pure quantum framework~\cite{Bulgac:2019a}.

The quality of the agreement with experimental data is surprisingly good,  especially taking into account 
the fact that no attempt was made to reproduce any fission data. Basically all phenomenological NEDFs
satisfy the most important requirements to describe the gross properties of nuclear fission: saturation, 
realistic surface tension and symmetry energy, Coulomb energy, realistic pairing and shell corrections
energy. Nevertheless, the quality of existing NEDFs needs improving. One can make a strong argument that
we have now a clear path  from more phenomenology and adjusted parameters to more fundamental 
theory and increased predictive power~\cite{Bulgac:2017a}.

Perhaps the most important aspect we have observed in all our simulations is the strong violation of the 
adiabatic assumption in fission large amplitude collective motion. Basically since the 1950's the adiabatic
assumption was the main simplification included in all microscopic frameworks, GCM, ATDHF and a large 
majority of phenomenological models as well, such as the Langevin  and the  Fokker-Planck equations, where 
one needs to introduce  a potential energy surface and an inertia tensor in the space of the collective variables. 
If the collective motion is overdamped, the inertia tensor becomes irrelevant, and moreover, considering only
the lowest potential energy surface is physically unacceptable,  see Figs.~\ref{fig:grab23} and \ref{fig:ecoll} 
and the corresponding discussion in section~\ref{sec:II}. As we have established in Ref.~\cite{Bulgac:2019a} 
the fluctuations, or equivalently the role of two-body collisions, does not affect this conclusion. 

A somewhat unexpected result was the character of the energy sharing mechanism between the fission 
fragments, the fact that the heavy fragment is cooler than the light fragment and it has 
less excitation energy~\cite{Bulgac:2016,Bulgac:2019b}.  And this conclusion is not 
a result of the fact that  the heavy FF is closer to the double magic $^{132}$Sn nucleus. 
In the original experiment of \textcite{Hahn:1939} the heavy fragment had a charge closer to 
$Z\approx 52-56$, a fact recognized also by \textcite{Meitner:1939,Meitner:1939a}, and explained by 
\textcite{Scamps:2018},  this is due to a stabilization of the octupole deformation in FFs, 
also observed in our simulations~\cite{Bulgac:2016,Bulgac:2019b}. With increasing excitation energy 
of the fissioning nucleus the heavy FF appears to be the only one who absorb the increase 
and emits more neutrons, see Fig.~\ref{fig:nubar} and the accompanying discussion. The character of 
the excitation energy sharing mechanism has major consequences on the predicted spectrum of emitted 
neutrons and gammas.

Another important outcome was the clear indications that many more collective DoF appear 
to be relevant in fission than have ever been considered in either phenomenological or microscopically 
inspired models such as GCM and ATDHF frameworks. At scission, both FFs are octupolly
deformed.  Moreover, the FFs attain their relaxed shapes only after the separation between 
them is $\approx 5...6$ fm. No phenomenological or microscopically inspired approach on the market 
follow the FFs to such large separations. As we stressed in section~\ref{sec:III} the spectrum of emitted 
neutrons is affected by the number of neutrons emitted before the full FFs acceleration. Another 
DoF, which appears to be relevant as well is the bending mode, see Fg.~\ref{fig:bend} 
and Ref.~\cite{Bulgac:2019a}, the inclusion of which likely is going to influence the angular momentum 
distributions of the FFs~\cite{Moretto:1989}. 

We have pointed to several directions into which phenomenological models and theoretical models such as 
GCM and ATDHF would have to be altered, in order to describe nuclear fission in a manner more consistent 
with theoretical expectations inferred from unrestricted quantum mechanical simulations, see section~\ref{sec:III}.
We did not cover or mention all phenomenological models on the market and not all microscopically 
inspired theoretical frameworks, as this is not  a review of such approaches, for which we
recommend Refs.~\cite{Schmidt:2018,Andreyev:2018}.

\begin{acknowledgments}

We thank many people with whom we had discussion over the years and for their input: 
I. Abdurrahman, G.F. Bertsch, P. Magierski, K.J. Roche, N. Schunck. We thank 
G.F. Bertsch and J. Randrup who made a number of suggestions on our manuscript, 
which has been released as a preprint~\cite{Bulgac:2020}.
 
The work of AB and SJ was supported by U.S. Department of Energy,
Office of Science, Grant No. DE-FG02-97ER41014 and in part by NNSA
cooperative agreement DE-NA0003841.  The work of IS was supported by the US Department of Energy through the 
Los Alamos National Laboratory. Los Alamos National Laboratory is operated 
by Triad National Security, LLC, for the National Nuclear Security Administration 
of U.S. Department of Energy (Contract No. 89233218CNA000001). 
The TDSLDA calculations have been
performed at the OLCF Summit and Titan, and CSCS Piz Daint, and for generating
initial configurations for direct input into the TDSLDA code at OLCF
Titan and Summit and NERSC Edison. This research used resources of the Oak Ridge
Leadership Computing Facility, which is a U.S. DOE Office of Science
User Facility supported under Contract No. DE- AC05-00OR22725 and of
the National Energy Research Scientific computing Center, which is
supported by the Office of Science of the U.S. Department of Energy
under Contract No. DE-AC02-05CH11231.  We acknowledge PRACE for
awarding us access to resource Piz Daint based at the Swiss National
Supercomputing Centre (CSCS), decision No. 2018194657.
This work is supported by "High Performance Computing 
Infrastructure" in Japan, Project ID: hp180048. A series of simulations 
were carried out on the Tsubame 3.0 supercomputer at Tokyo Institute of Technology.
This research used resources provided by the Los Alamos National Laboratory 
Institutional Computing Program.

  \end{acknowledgments}


\providecommand{\selectlanguage}[1]{}
\renewcommand{\selectlanguage}[1]{}

\bibliography{latest_fission-IS}

\begin{thebibliography}{140}%
\makeatletter
\providecommand \@ifxundefined [1]{%
 \@ifx{#1\undefined}
}%
\providecommand \@ifnum [1]{%
 \ifnum #1\expandafter \@firstoftwo
 \else \expandafter \@secondoftwo
 \fi
}%
\providecommand \@ifx [1]{%
 \ifx #1\expandafter \@firstoftwo
 \else \expandafter \@secondoftwo
 \fi
}%
\providecommand \natexlab [1]{#1}%
\providecommand \enquote  [1]{``#1''}%
\providecommand \bibnamefont  [1]{#1}%
\providecommand \bibfnamefont [1]{#1}%
\providecommand \citenamefont [1]{#1}%
\providecommand \href@noop [0]{\@secondoftwo}%
\providecommand \href [0]{\begingroup \@sanitize@url \@href}%
\providecommand \@href[1]{\@@startlink{#1}\@@href}%
\providecommand \@@href[1]{\endgroup#1\@@endlink}%
\providecommand \@sanitize@url [0]{\catcode `\\12\catcode `\$12\catcode
  `\&12\catcode `\#12\catcode `\^12\catcode `\_12\catcode `\%12\relax}%
\providecommand \@@startlink[1]{}%
\providecommand \@@endlink[0]{}%
\providecommand \url  [0]{\begingroup\@sanitize@url \@url }%
\providecommand \@url [1]{\endgroup\@href {#1}{\urlprefix }}%
\providecommand \urlprefix  [0]{URL }%
\providecommand \Eprint [0]{\href }%
\providecommand \doibase [0]{http://dx.doi.org/}%
\providecommand \selectlanguage [0]{\@gobble}%
\providecommand \bibinfo  [0]{\@secondoftwo}%
\providecommand \bibfield  [0]{\@secondoftwo}%
\providecommand \translation [1]{[#1]}%
\providecommand \BibitemOpen [0]{}%
\providecommand \bibitemStop [0]{}%
\providecommand \bibitemNoStop [0]{.\EOS\space}%
\providecommand \EOS [0]{\spacefactor3000\relax}%
\providecommand \BibitemShut  [1]{\csname bibitem#1\endcsname}%
\let\auto@bib@innerbib\@empty
\bibitem [{\citenamefont {Hahn}\ and\ \citenamefont
  {Strassmann}(1939)}]{Hahn:1939}%
  \BibitemOpen
  \bibfield  {author} {\bibinfo {author} {\bibfnamefont {O.}~\bibnamefont
  {Hahn}}\ and\ \bibinfo {author} {\bibfnamefont {F.}~\bibnamefont
  {Strassmann}},\ }\bibfield  {title} {\enquote {\bibinfo {title} {{\"Uber den
  Nachweis und das Verhalten der bei der Bestrahlung des Urans mittels
  Neutronen entstehenden Erdalkalimetalle}},}\ }\href {\doibase
  10.1007/BF01488241} {\bibfield  {journal} {\bibinfo  {journal}
  {Naturwissenschaften}\ }\textbf {\bibinfo {volume} {27}},\ \bibinfo {pages}
  {11} (\bibinfo {year} {1939})}\BibitemShut {NoStop}%
\bibitem [{\citenamefont {Meitner}\ and\ \citenamefont
  {Frisch}(1939{\natexlab{a}})}]{Meitner:1939}%
  \BibitemOpen
  \bibfield  {author} {\bibinfo {author} {\bibfnamefont {L.}~\bibnamefont
  {Meitner}}\ and\ \bibinfo {author} {\bibfnamefont {O.~R.}\ \bibnamefont
  {Frisch}},\ }\bibfield  {title} {\enquote {\bibinfo {title} {{Disintegration
  of Uranium by Neutrons: a New Type of Nuclear Reaction}},}\ }\href
  {http://dx.doi.org/10.1038/143239a0} {\bibfield  {journal} {\bibinfo
  {journal} {Nature}\ }\textbf {\bibinfo {volume} {143}},\ \bibinfo {pages}
  {239} (\bibinfo {year} {1939}{\natexlab{a}})}\BibitemShut {NoStop}%
\bibitem [{\citenamefont {Stuewer}(2010)}]{Stuewer:2010}%
  \BibitemOpen
  \bibfield  {author} {\bibinfo {author} {\bibfnamefont {R.}~\bibnamefont
  {Stuewer}},\ }\href@noop {} {\enquote {\bibinfo {title} {{An act of creation:
  The Meitner--Frisch interpretation of nuclear fission}},}\ } (\bibinfo {year}
  {2010}),\ \Eprint
  {http://arxiv.org/abs/http://edition-open-access.de/proceedings/5/11/index.html}
  {http://edition-open-access.de/proceedings/5/11/index.html} \BibitemShut
  {NoStop}%
\bibitem [{\citenamefont {Pearson}(2015)}]{Pearson:2015}%
  \BibitemOpen
  \bibfield  {author} {\bibinfo {author} {\bibfnamefont {J.~M.}\ \bibnamefont
  {Pearson}},\ }\bibfield  {title} {\enquote {\bibinfo {title} {{On the belated
  discovery of fission}},}\ }\href {\doibase 10.1063/PT.3.2817} {\bibfield
  {journal} {\bibinfo  {journal} {Physics Today}\ }\textbf {\bibinfo {volume}
  {68 (6)}},\ \bibinfo {pages} {40} (\bibinfo {year} {2015})}\BibitemShut
  {NoStop}%
\bibitem [{\citenamefont {Fermi}\ \emph {et~al.}(1934)\citenamefont {Fermi},
  \citenamefont {Amaldi}, \citenamefont {d'Agostino}, \citenamefont {Rasetti},\
  and\ \citenamefont {Segre}}]{Fermi:1934}%
  \BibitemOpen
  \bibfield  {author} {\bibinfo {author} {\bibfnamefont {E.}~\bibnamefont
  {Fermi}}, \bibinfo {author} {\bibfnamefont {E.}~\bibnamefont {Amaldi}},
  \bibinfo {author} {\bibfnamefont {O.}~\bibnamefont {d'Agostino}}, \bibinfo
  {author} {\bibfnamefont {F.}~\bibnamefont {Rasetti}}, \ and\ \bibinfo
  {author} {\bibfnamefont {E.}~\bibnamefont {Segre}},\ }\bibfield  {title}
  {\enquote {\bibinfo {title} {{Artificial radioactivity produced by neutron
  bombardment}},}\ }\href {\doibase 10.1098/rspa.1934.0168} {\bibfield
  {journal} {\bibinfo  {journal} {Proc. Roy. Soc. A}\ }\textbf {\bibinfo
  {volume} {146}},\ \bibinfo {pages} {249} (\bibinfo {year}
  {1934})}\BibitemShut {NoStop}%
\bibitem [{\citenamefont {Bohr}(1936)}]{Bohr:1936}%
  \BibitemOpen
  \bibfield  {author} {\bibinfo {author} {\bibfnamefont {N.}~\bibnamefont
  {Bohr}},\ }\bibfield  {title} {\enquote {\bibinfo {title} {{Neutron Capture
  and Nuclear Constitution}},}\ }\href {\doibase 10.1038/137344a0} {\bibfield
  {journal} {\bibinfo  {journal} {Nature}\ }\textbf {\bibinfo {volume} {137}},\
  \bibinfo {pages} {344 and 351} (\bibinfo {year} {1936})}\BibitemShut
  {NoStop}%
\bibitem [{\citenamefont {Bohr}\ and\ \citenamefont
  {Wheeler}(1939)}]{Bohr:1939}%
  \BibitemOpen
  \bibfield  {author} {\bibinfo {author} {\bibfnamefont {N.}~\bibnamefont
  {Bohr}}\ and\ \bibinfo {author} {\bibfnamefont {J.~A.}\ \bibnamefont
  {Wheeler}},\ }\bibfield  {title} {\enquote {\bibinfo {title} {{The Mechanism
  of Nuclear Fission}},}\ }\href {\doibase 10.1103/PhysRev.56.426} {\bibfield
  {journal} {\bibinfo  {journal} {Phys. Rev.}\ }\textbf {\bibinfo {volume}
  {56}},\ \bibinfo {pages} {426} (\bibinfo {year} {1939})}\BibitemShut
  {NoStop}%
\bibitem [{\citenamefont {Strutinsky}(1967)}]{Strutinsky:1967}%
  \BibitemOpen
  \bibfield  {author} {\bibinfo {author} {\bibfnamefont {V.M.}\ \bibnamefont
  {Strutinsky}},\ }\bibfield  {title} {\enquote {\bibinfo {title} {Shell
  effects in nuclear masses and deformation energies},}\ }\href {\doibase
  https://doi.org/10.1016/0375-9474(67)90510-6} {\bibfield  {journal} {\bibinfo
   {journal} {Nucl. Phys. A}\ }\textbf {\bibinfo {volume} {95}},\ \bibinfo
  {pages} {420} (\bibinfo {year} {1967})}\BibitemShut {NoStop}%
\bibitem [{\citenamefont {Brack}\ \emph {et~al.}(1972)\citenamefont {Brack},
  \citenamefont {Damgaard}, \citenamefont {Jensen}, \citenamefont {Pauli},
  \citenamefont {Strutinsky},\ and\ \citenamefont {Wong}}]{Brack:1972}%
  \BibitemOpen
  \bibfield  {author} {\bibinfo {author} {\bibfnamefont {M.}~\bibnamefont
  {Brack}}, \bibinfo {author} {\bibfnamefont {J.}~\bibnamefont {Damgaard}},
  \bibinfo {author} {\bibfnamefont {A.~S.}\ \bibnamefont {Jensen}}, \bibinfo
  {author} {\bibfnamefont {H.~C.}\ \bibnamefont {Pauli}}, \bibinfo {author}
  {\bibfnamefont {V.~M.}\ \bibnamefont {Strutinsky}}, \ and\ \bibinfo {author}
  {\bibfnamefont {C.~Y.}\ \bibnamefont {Wong}},\ }\bibfield  {title} {\enquote
  {\bibinfo {title} {{Funny Hills: The Shell-Correction Approach to Nuclear
  Shell Effects and Its Applications to the Fission Process}},}\ }\href
  {\doibase 10.1103/RevModPhys.44.320} {\bibfield  {journal} {\bibinfo
  {journal} {Rev. Mod. Phys.}\ }\textbf {\bibinfo {volume} {44}},\ \bibinfo
  {pages} {320} (\bibinfo {year} {1972})}\BibitemShut {NoStop}%
\bibitem [{\citenamefont {Bertsch}(1980)}]{Bertsch:1980}%
  \BibitemOpen
  \bibfield  {author} {\bibinfo {author} {\bibfnamefont {G.}~\bibnamefont
  {Bertsch}},\ }\bibfield  {title} {\enquote {\bibinfo {title} {The nuclear
  density of states in the space of nuclear shapes},}\ }\href {\doibase
  10.1016/0370-2693(80)90458-X} {\bibfield  {journal} {\bibinfo  {journal}
  {Phys. Lett. B}\ }\textbf {\bibinfo {volume} {95}},\ \bibinfo {pages} {157}
  (\bibinfo {year} {1980})}\BibitemShut {NoStop}%
\bibitem [{\citenamefont {Bertsch}\ and\ \citenamefont
  {Bulgac}(1997)}]{Bertsch:1997}%
  \BibitemOpen
  \bibfield  {author} {\bibinfo {author} {\bibfnamefont {G.~F.}\ \bibnamefont
  {Bertsch}}\ and\ \bibinfo {author} {\bibfnamefont {A.}~\bibnamefont
  {Bulgac}},\ }\bibfield  {title} {\enquote {\bibinfo {title} {{Comment on
  ``Spontaneous Fission: A Kinetic Approach''}},}\ }\href {\doibase
  10.1103/PhysRevLett.79.3539} {\bibfield  {journal} {\bibinfo  {journal}
  {Phys. Rev. Lett.}\ }\textbf {\bibinfo {volume} {79}},\ \bibinfo {pages}
  {3539} (\bibinfo {year} {1997})}\BibitemShut {NoStop}%
\bibitem [{\citenamefont {Weisskopf}(1937)}]{Weisskopf:1937}%
  \BibitemOpen
  \bibfield  {author} {\bibinfo {author} {\bibfnamefont {V.}~\bibnamefont
  {Weisskopf}},\ }\bibfield  {title} {\enquote {\bibinfo {title} {{Statistics
  and Nuclear Reactions}},}\ }\href {\doibase 10.1103/PhysRev.52.295}
  {\bibfield  {journal} {\bibinfo  {journal} {Phys. Rev.}\ }\textbf {\bibinfo
  {volume} {52}},\ \bibinfo {pages} {295} (\bibinfo {year} {1937})}\BibitemShut
  {NoStop}%
\bibitem [{\citenamefont {Hauser}\ and\ \citenamefont
  {Feshbach}(1952)}]{Hauser:1952}%
  \BibitemOpen
  \bibfield  {author} {\bibinfo {author} {\bibfnamefont {W.}~\bibnamefont
  {Hauser}}\ and\ \bibinfo {author} {\bibfnamefont {H.}~\bibnamefont
  {Feshbach}},\ }\bibfield  {title} {\enquote {\bibinfo {title} {The inelastic
  scattering of neutrons},}\ }\href {\doibase 10.1103/PhysRev.87.366}
  {\bibfield  {journal} {\bibinfo  {journal} {Phys. Rev.}\ }\textbf {\bibinfo
  {volume} {87}},\ \bibinfo {pages} {366} (\bibinfo {year} {1952})}\BibitemShut
  {NoStop}%
\bibitem [{\citenamefont {Hill}\ and\ \citenamefont
  {Wheeler}(1953)}]{Hill:1953}%
  \BibitemOpen
  \bibfield  {author} {\bibinfo {author} {\bibfnamefont {D.~L.}\ \bibnamefont
  {Hill}}\ and\ \bibinfo {author} {\bibfnamefont {J.~A.}\ \bibnamefont
  {Wheeler}},\ }\bibfield  {title} {\enquote {\bibinfo {title} {{Nuclear
  Constitution and the Interpretation of Fission Phenomena}},}\ }\href
  {\doibase 10.1103/PhysRev.89.1102} {\bibfield  {journal} {\bibinfo  {journal}
  {Phys. Rev.}\ }\textbf {\bibinfo {volume} {89}},\ \bibinfo {pages} {1102}
  (\bibinfo {year} {1953})}\BibitemShut {NoStop}%
\bibitem [{\citenamefont {Born}\ and\ \citenamefont
  {Oppenheimer}(1927)}]{Born:1927}%
  \BibitemOpen
  \bibfield  {author} {\bibinfo {author} {\bibfnamefont {M.}~\bibnamefont
  {Born}}\ and\ \bibinfo {author} {\bibfnamefont {J.R.}\ \bibnamefont
  {Oppenheimer}},\ }\bibfield  {title} {\enquote {\bibinfo {title} {{Zur
  Quantentheorie der Molekeln}},}\ }\href {\doibase 10.1002/andp.19273892002}
  {\bibfield  {journal} {\bibinfo  {journal} {Annalen der Physik}\ }\textbf
  {\bibinfo {volume} {389}},\ \bibinfo {pages} {457} (\bibinfo {year}
  {1927})}\BibitemShut {NoStop}%
\bibitem [{\citenamefont {Griffin}\ and\ \citenamefont
  {Wheeler}(1957)}]{Griffin:1957}%
  \BibitemOpen
  \bibfield  {author} {\bibinfo {author} {\bibfnamefont {J.~J.}\ \bibnamefont
  {Griffin}}\ and\ \bibinfo {author} {\bibfnamefont {J.~A.}\ \bibnamefont
  {Wheeler}},\ }\bibfield  {title} {\enquote {\bibinfo {title} {{Collective
  Motions in Nuclei by the Method of Generator Coordinates}},}\ }\href
  {\doibase 10.1103/PhysRev.108.311} {\bibfield  {journal} {\bibinfo  {journal}
  {Phys. Rev.}\ }\textbf {\bibinfo {volume} {108}},\ \bibinfo {pages} {311}
  (\bibinfo {year} {1957})}\BibitemShut {NoStop}%
\bibitem [{\citenamefont {Baranger}(1972)}]{Baranger:1972}%
  \BibitemOpen
  \bibfield  {author} {\bibinfo {author} {\bibfnamefont {M.}~\bibnamefont
  {Baranger}},\ }\bibfield  {title} {\enquote {\bibinfo {title} {Microscopic
  view of nuclear collective properties},}\ }\href {\doibase
  10.1051/jphyscol:1972506} {\bibfield  {journal} {\bibinfo  {journal} {J.
  Phys.}\ }\textbf {\bibinfo {volume} {33}},\ \bibinfo {pages} {C5} (\bibinfo
  {year} {1972})}\BibitemShut {NoStop}%
\bibitem [{\citenamefont {Baranger}\ and\ \citenamefont
  {V\'en\'eroni}(1978)}]{Baranger:1978}%
  \BibitemOpen
  \bibfield  {author} {\bibinfo {author} {\bibfnamefont {M.}~\bibnamefont
  {Baranger}}\ and\ \bibinfo {author} {\bibfnamefont {M.}~\bibnamefont
  {V\'en\'eroni}},\ }\bibfield  {title} {\enquote {\bibinfo {title} {{An
  adiabatic time-dependent Hartree-Fock theory of collective motion in finite
  systems}},}\ }\href {\doibase 10.1016/0003-4916(78)90265-8} {\bibfield
  {journal} {\bibinfo  {journal} {Ann. Phys.}\ }\textbf {\bibinfo {volume}
  {114}},\ \bibinfo {pages} {123} (\bibinfo {year} {1978})}\BibitemShut
  {NoStop}%
\bibitem [{\citenamefont {Villars}(1978)}]{Villars:1978}%
  \BibitemOpen
  \bibfield  {author} {\bibinfo {author} {\bibfnamefont {F.}~\bibnamefont
  {Villars}},\ }\bibfield  {title} {\enquote {\bibinfo {title} {{Adiabatic
  time-dependent Hartree-Fock theory in nuclear physics}},}\ }\href {\doibase
  10.1016/0375-9474(77)90253-6} {\bibfield  {journal} {\bibinfo  {journal}
  {Nucl. Phys. A}\ }\textbf {\bibinfo {volume} {285}},\ \bibinfo {pages} {269}
  (\bibinfo {year} {1978})}\BibitemShut {NoStop}%
\bibitem [{\citenamefont {Ring}\ and\ \citenamefont
  {Schuck}(2004)}]{Ring:2004}%
  \BibitemOpen
  \bibfield  {author} {\bibinfo {author} {\bibfnamefont {P.}~\bibnamefont
  {Ring}}\ and\ \bibinfo {author} {\bibfnamefont {P.}~\bibnamefont {Schuck}},\
  }\href@noop {} {\emph {\bibinfo {title} {{The Nuclear Many-Body Problem}}}},\
  \bibinfo {edition} {1st}\ ed.,\ \bibinfo {series} {Theoretical and
  Mathematical Physics Series}\ No.~\bibinfo {number} {17}\ (\bibinfo
  {publisher} {Springer-Verlag},\ \bibinfo {address} {Berlin Heidelberg New
  York},\ \bibinfo {year} {2004})\BibitemShut {NoStop}%
\bibitem [{\citenamefont {Krappe}\ and\ \citenamefont
  {Pomorski}(2012)}]{Pomorski:2012}%
  \BibitemOpen
  \bibfield  {author} {\bibinfo {author} {\bibfnamefont {J.~K.}\ \bibnamefont
  {Krappe}}\ and\ \bibinfo {author} {\bibfnamefont {K.}~\bibnamefont
  {Pomorski}},\ }\href {\doibase 10.1007/978-3-642-23515-3} {\emph {\bibinfo
  {title} {{Theory of Nuclear Fission}}}}\ (\bibinfo  {publisher} {Springer
  Heidelberg},\ \bibinfo {year} {2012})\BibitemShut {NoStop}%
\bibitem [{\citenamefont {Schunck}\ and\ \citenamefont
  {Robledo}(2016)}]{Schunck:2016}%
  \BibitemOpen
  \bibfield  {author} {\bibinfo {author} {\bibfnamefont {N.}~\bibnamefont
  {Schunck}}\ and\ \bibinfo {author} {\bibfnamefont {L.~M.}\ \bibnamefont
  {Robledo}},\ }\bibfield  {title} {\enquote {\bibinfo {title} {{Microscopic}
  theory of nuclear fission: a review},}\ }\href {\doibase
  10.1088/0034-4885/79/11/116301} {\bibfield  {journal} {\bibinfo  {journal}
  {{Rep. Prog. Phys.}}\ }\textbf {\bibinfo {volume} {79}},\ \bibinfo {pages}
  {116301} (\bibinfo {year} {2016})}\BibitemShut {NoStop}%
\bibitem [{\citenamefont {Goutte}\ \emph {et~al.}(2005)\citenamefont {Goutte},
  \citenamefont {Berger}, \citenamefont {Casoli},\ and\ \citenamefont
  {Gogny}}]{Goutte:2005}%
  \BibitemOpen
  \bibfield  {author} {\bibinfo {author} {\bibfnamefont {H.}~\bibnamefont
  {Goutte}}, \bibinfo {author} {\bibfnamefont {J.~F.}\ \bibnamefont {Berger}},
  \bibinfo {author} {\bibfnamefont {P.}~\bibnamefont {Casoli}}, \ and\ \bibinfo
  {author} {\bibfnamefont {D.}~\bibnamefont {Gogny}},\ }\bibfield  {title}
  {\enquote {\bibinfo {title} {Microscopic approach of fission dynamics applied
  to fragment kinetic energy and mass distributions in $^{238}\mathrm{U}$},}\
  }\href {\doibase 10.1103/PhysRevC.71.024316} {\bibfield  {journal} {\bibinfo
  {journal} {Phys. Rev. C}\ }\textbf {\bibinfo {volume} {71}},\ \bibinfo
  {pages} {024316} (\bibinfo {year} {2005})}\BibitemShut {NoStop}%
\bibitem [{\citenamefont {Regnier}\ \emph {et~al.}(2016)\citenamefont
  {Regnier}, \citenamefont {Dubray}, \citenamefont {Schunck},\ and\
  \citenamefont {Verri\`ere}}]{Regnier:2016}%
  \BibitemOpen
  \bibfield  {author} {\bibinfo {author} {\bibfnamefont {D.}~\bibnamefont
  {Regnier}}, \bibinfo {author} {\bibfnamefont {N.}~\bibnamefont {Dubray}},
  \bibinfo {author} {\bibfnamefont {N.}~\bibnamefont {Schunck}}, \ and\
  \bibinfo {author} {\bibfnamefont {M.}~\bibnamefont {Verri\`ere}},\ }\bibfield
   {title} {\enquote {\bibinfo {title} {Fission fragment charge and mass
  distributions in $^{239}\mathrm{Pu}(n,f)$ in the adiabatic nuclear energy
  density functional theory},}\ }\href {\doibase 10.1103/PhysRevC.93.054611}
  {\bibfield  {journal} {\bibinfo  {journal} {Phys. Rev. C}\ }\textbf {\bibinfo
  {volume} {93}},\ \bibinfo {pages} {054611} (\bibinfo {year}
  {2016})}\BibitemShut {NoStop}%
\bibitem [{\citenamefont {Zdeb}\ \emph {et~al.}(2017)\citenamefont {Zdeb},
  \citenamefont {Dobrowolski},\ and\ \citenamefont {Warda}}]{Zdeb:2017}%
  \BibitemOpen
  \bibfield  {author} {\bibinfo {author} {\bibfnamefont {A.}~\bibnamefont
  {Zdeb}}, \bibinfo {author} {\bibfnamefont {A.}~\bibnamefont {Dobrowolski}}, \
  and\ \bibinfo {author} {\bibfnamefont {M.}~\bibnamefont {Warda}},\ }\bibfield
   {title} {\enquote {\bibinfo {title} {Fission dynamics of
  $^{252}\mathbf{Cf}$},}\ }\href {\doibase 10.1103/PhysRevC.95.054608}
  {\bibfield  {journal} {\bibinfo  {journal} {Phys. Rev. C}\ }\textbf {\bibinfo
  {volume} {95}},\ \bibinfo {pages} {054608} (\bibinfo {year}
  {2017})}\BibitemShut {NoStop}%
\bibitem [{\citenamefont {{N. Schunck, editor}}(2019)}]{Schunck:2019}%
  \BibitemOpen
  \bibfield  {author} {\bibinfo {author} {\bibnamefont {{N. Schunck,
  editor}}},\ }\href {\doibase 10.1088/2053-2563/aae0ed} {\emph {\bibinfo
  {title} {{Energy Density Functional Methods for Atomic Nuclei}}}}\ (\bibinfo
  {publisher} {IOP Publishing, Bristol, UK},\ \bibinfo {year}
  {2019})\BibitemShut {NoStop}%
\bibitem [{\citenamefont {Younes}\ \emph {et~al.}(2019)\citenamefont {Younes},
  \citenamefont {Gogny},\ and\ \citenamefont {Berger}}]{Younes:2019}%
  \BibitemOpen
  \bibfield  {author} {\bibinfo {author} {\bibfnamefont {W.}~\bibnamefont
  {Younes}}, \bibinfo {author} {\bibfnamefont {D.~M.}\ \bibnamefont {Gogny}}, \
  and\ \bibinfo {author} {\bibfnamefont {J.~F.}\ \bibnamefont {Berger}},\
  }\href {\doibase 10.1007/978-3-030-04424-4} {\emph {\bibinfo {title} {{A
  Microscopic Theory of Fission Based on the Generator Coordinate Method}}}},\
  \bibinfo {series} {{Lectures Notes in Physics}}, Vol.\ \bibinfo {volume}
  {950}\ (\bibinfo  {publisher} {Springer, Berlin},\ \bibinfo {year}
  {2019})\BibitemShut {NoStop}%
\bibitem [{\citenamefont {Goeke}\ and\ \citenamefont
  {Reinhard}(1980)}]{Goeke:1980}%
  \BibitemOpen
  \bibfield  {author} {\bibinfo {author} {\bibfnamefont {K.}~\bibnamefont
  {Goeke}}\ and\ \bibinfo {author} {\bibfnamefont {P.-G.}\ \bibnamefont
  {Reinhard}},\ }\bibfield  {title} {\enquote {\bibinfo {title} {The
  generator-coordinate-method with conjugate parameters and the unification of
  microscopic theories for large amplitude collective motion},}\ }\href
  {\doibase 10.1016/0003-4916(80)90210-9} {\bibfield  {journal} {\bibinfo
  {journal} {Ann. Phys.}\ }\textbf {\bibinfo {volume} {124}},\ \bibinfo {pages}
  {249} (\bibinfo {year} {1980})}\BibitemShut {NoStop}%
\bibitem [{\citenamefont {Peierls}\ and\ \citenamefont
  {Thouless}(1962)}]{Peierls:1962}%
  \BibitemOpen
  \bibfield  {author} {\bibinfo {author} {\bibfnamefont {R.E.}\ \bibnamefont
  {Peierls}}\ and\ \bibinfo {author} {\bibfnamefont {D.J.}\ \bibnamefont
  {Thouless}},\ }\bibfield  {title} {\enquote {\bibinfo {title} {{Variational
  approach to collective motion}},}\ }\href {\doibase
  10.1016/0029-5582(62)91025-8} {\bibfield  {journal} {\bibinfo  {journal}
  {Nucl. Phys.}\ }\textbf {\bibinfo {volume} {38}},\ \bibinfo {pages} {154}
  (\bibinfo {year} {1962})}\BibitemShut {NoStop}%
\bibitem [{\citenamefont {Dang}\ \emph {et~al.}(2000)\citenamefont {Dang},
  \citenamefont {Klein},\ and\ \citenamefont {Walet}}]{Dang:2000}%
  \BibitemOpen
  \bibfield  {author} {\bibinfo {author} {\bibfnamefont {G.D.}\ \bibnamefont
  {Dang}}, \bibinfo {author} {\bibfnamefont {A.}~\bibnamefont {Klein}}, \ and\
  \bibinfo {author} {\bibfnamefont {N.R.}\ \bibnamefont {Walet}},\ }\bibfield
  {title} {\enquote {\bibinfo {title} {Self-consistent theory of
  large-amplitude collective motion: applications to approximate quantization
  of nonseparable systems and to nuclear physics},}\ }\href {\doibase
  10.1016/S0370-1573(99)00119-2} {\bibfield  {journal} {\bibinfo  {journal}
  {Phys. Rep.}\ }\textbf {\bibinfo {volume} {335}},\ \bibinfo {pages} {93}
  (\bibinfo {year} {2000})}\BibitemShut {NoStop}%
\bibitem [{\citenamefont {Barranco}\ \emph {et~al.}(1990)\citenamefont
  {Barranco}, \citenamefont {Bertsch}, \citenamefont {Broglia},\ and\
  \citenamefont {Vigezzi}}]{Barranco:1990}%
  \BibitemOpen
  \bibfield  {author} {\bibinfo {author} {\bibfnamefont {F.}~\bibnamefont
  {Barranco}}, \bibinfo {author} {\bibfnamefont {G.F.}\ \bibnamefont
  {Bertsch}}, \bibinfo {author} {\bibfnamefont {R.A.}\ \bibnamefont {Broglia}},
  \ and\ \bibinfo {author} {\bibfnamefont {E.}~\bibnamefont {Vigezzi}},\
  }\bibfield  {title} {\enquote {\bibinfo {title} {{Large-amplitude motion in
  superfluid Fermi droplets}},}\ }\href {\doibase
  https://doi.org/10.1016/0375-9474(90)93232-U} {\bibfield  {journal} {\bibinfo
   {journal} {Nucl. Data Sheets Phys. A}\ }\textbf {\bibinfo {volume} {512}},\
  \bibinfo {pages} {253} (\bibinfo {year} {1990})}\BibitemShut {NoStop}%
\bibitem [{\citenamefont {Bulgac}\ \emph
  {et~al.}(2019{\natexlab{a}})\citenamefont {Bulgac}, \citenamefont {Jin},
  \citenamefont {Roche}, \citenamefont {Schunck},\ and\ \citenamefont
  {Stetcu}}]{Bulgac:2019b}%
  \BibitemOpen
  \bibfield  {author} {\bibinfo {author} {\bibfnamefont {A.}~\bibnamefont
  {Bulgac}}, \bibinfo {author} {\bibfnamefont {S.}~\bibnamefont {Jin}},
  \bibinfo {author} {\bibfnamefont {K.~J.}\ \bibnamefont {Roche}}, \bibinfo
  {author} {\bibfnamefont {N.}~\bibnamefont {Schunck}}, \ and\ \bibinfo
  {author} {\bibfnamefont {I.}~\bibnamefont {Stetcu}},\ }\bibfield  {title}
  {\enquote {\bibinfo {title} {Fission dynamics of $^{240}\mathrm{Pu}$ from
  saddle to scission and beyond},}\ }\href {\doibase
  10.1103/PhysRevC.100.034615} {\bibfield  {journal} {\bibinfo  {journal}
  {Phys. Rev. C}\ }\textbf {\bibinfo {volume} {100}},\ \bibinfo {pages}
  {034615} (\bibinfo {year} {2019}{\natexlab{a}})}\BibitemShut {NoStop}%
\bibitem [{\citenamefont {Hohenberg}\ and\ \citenamefont
  {Kohn}(1964)}]{Hohenberg:1964}%
  \BibitemOpen
  \bibfield  {author} {\bibinfo {author} {\bibfnamefont {P.}~\bibnamefont
  {Hohenberg}}\ and\ \bibinfo {author} {\bibfnamefont {W.}~\bibnamefont
  {Kohn}},\ }\bibfield  {title} {\enquote {\bibinfo {title} {{Inhomogeneous
  Electron Gas}},}\ }\href {\doibase 10.1103/PhysRev.136.B864} {\bibfield
  {journal} {\bibinfo  {journal} {Phys. Rev.}\ }\textbf {\bibinfo {volume}
  {136}},\ \bibinfo {pages} {B864--B871} (\bibinfo {year} {1964})}\BibitemShut
  {NoStop}%
\bibitem [{\citenamefont {Kohn}\ and\ \citenamefont
  {Sham}(1965)}]{Kohn:1965fk}%
  \BibitemOpen
  \bibfield  {author} {\bibinfo {author} {\bibfnamefont {W.}~\bibnamefont
  {Kohn}}\ and\ \bibinfo {author} {\bibfnamefont {L.~J.}\ \bibnamefont
  {Sham}},\ }\bibfield  {title} {\enquote {\bibinfo {title} {Self-consistent
  equations including exchange and correlation effects},}\ }\href {\doibase
  10.1103/PhysRev.140.A1133} {\bibfield  {journal} {\bibinfo  {journal} {Phys.
  Rev.}\ }\textbf {\bibinfo {volume} {140}},\ \bibinfo {pages} {A1133--A1138}
  (\bibinfo {year} {1965})}\BibitemShut {NoStop}%
\bibitem [{\citenamefont {Kohn}(1999)}]{Kohn:1999fk}%
  \BibitemOpen
  \bibfield  {author} {\bibinfo {author} {\bibfnamefont {W.}~\bibnamefont
  {Kohn}},\ }\bibfield  {title} {\enquote {\bibinfo {title} {{Nobel Lecture:
  Electronic structure of matter---wave functions and density functionals}},}\
  }\href {\doibase 10.1103/RevModPhys.71.1253} {\bibfield  {journal} {\bibinfo
  {journal} {Rev. Mod. Phys.}\ }\textbf {\bibinfo {volume} {71}},\ \bibinfo
  {pages} {1253--1266} (\bibinfo {year} {1999})}\BibitemShut {NoStop}%
\bibitem [{\citenamefont {Dreizler}\ and\ \citenamefont
  {{Gross}}(1990)}]{Dreizler:1990lr}%
  \BibitemOpen
  \bibfield  {author} {\bibinfo {author} {\bibfnamefont {R.~M.}\ \bibnamefont
  {Dreizler}}\ and\ \bibinfo {author} {\bibfnamefont {E.~K.~U.}\ \bibnamefont
  {{Gross}}},\ }\href {\doibase 10.1007/978-3-642-86105-5} {\emph {\bibinfo
  {title} {{Density Functional Theory: An Approach to the Quantum Many--Body
  Problem}}}}\ (\bibinfo  {publisher} {Springer-Verlag},\ \bibinfo {address}
  {Berlin},\ \bibinfo {year} {1990})\BibitemShut {NoStop}%
\bibitem [{\citenamefont {Marques}\ \emph {et~al.}(2006)\citenamefont
  {Marques}, \citenamefont {Ullrich}, \citenamefont {Nogueira}, \citenamefont
  {Rubio}, \citenamefont {Burke},\ and\ \citenamefont {{Gross}}}]{Gross:2006}%
  \BibitemOpen
  \bibinfo {editor} {\bibfnamefont {M.~A.~L.}\ \bibnamefont {Marques}},
  \bibinfo {editor} {\bibfnamefont {C.~A.}\ \bibnamefont {Ullrich}}, \bibinfo
  {editor} {\bibfnamefont {F.}~\bibnamefont {Nogueira}}, \bibinfo {editor}
  {\bibfnamefont {A.}~\bibnamefont {Rubio}}, \bibinfo {editor} {\bibfnamefont
  {K.}~\bibnamefont {Burke}}, \ and\ \bibinfo {editor} {\bibfnamefont
  {E.~K.~U.}\ \bibnamefont {{Gross}}},\ eds.,\ \href {\doibase
  10.1007/b11767107} {\emph {\bibinfo {title} {Time-Dependent Density
  Functional Theory}}},\ \bibinfo {series} {Lecture Notes in Physics}, Vol.\
  \bibinfo {volume} {706}\ (\bibinfo  {publisher} {Springer-Verlag},\ \bibinfo
  {address} {Berlin},\ \bibinfo {year} {2006})\BibitemShut {NoStop}%
\bibitem [{\citenamefont {Marques}\ \emph {et~al.}(2012)\citenamefont
  {Marques}, \citenamefont {Maitra}, \citenamefont {Nogueira}, \citenamefont
  {{Gross}},\ and\ \citenamefont {Rubio}}]{Gross:2012}%
  \BibitemOpen
  \bibinfo {editor} {\bibfnamefont {M.~A.~L.}\ \bibnamefont {Marques}},
  \bibinfo {editor} {\bibfnamefont {N.~T.}\ \bibnamefont {Maitra}}, \bibinfo
  {editor} {\bibfnamefont {F.~M.~S.}\ \bibnamefont {Nogueira}}, \bibinfo
  {editor} {\bibfnamefont {E.~K.~U.}\ \bibnamefont {{Gross}}}, \ and\ \bibinfo
  {editor} {\bibfnamefont {A.}~\bibnamefont {Rubio}},\ eds.,\ \href {\doibase
  10.1007/978-3-642-23518-4} {\emph {\bibinfo {title} {Fundamentals of
  Time-Dependent Density Functional Theory}}},\ \bibinfo {series} {Lecture
  Notes in Physics}, Vol.\ \bibinfo {volume} {837}\ (\bibinfo  {publisher}
  {Springer},\ \bibinfo {address} {Heidelberg},\ \bibinfo {year}
  {2012})\BibitemShut {NoStop}%
\bibitem [{\citenamefont {Bulgac}(2013)}]{Bulgac:2013a}%
  \BibitemOpen
  \bibfield  {author} {\bibinfo {author} {\bibfnamefont {A.}~\bibnamefont
  {Bulgac}},\ }\bibfield  {title} {\enquote {\bibinfo {title} {{Time-Dependent
  Density Functional Theory and the Real-Time Dynamics of Fermi
  Superfluids}},}\ }\href {\doibase 10.1146/annurev-nucl-102212-170631}
  {\bibfield  {journal} {\bibinfo  {journal} {Ann. Rev. Nucl. and Part. Sci.}\
  }\textbf {\bibinfo {volume} {63}},\ \bibinfo {pages} {97} (\bibinfo {year}
  {2013})}\BibitemShut {NoStop}%
\bibitem [{\citenamefont {Bulgac}(2019{\natexlab{a}})}]{Bulgac:2019}%
  \BibitemOpen
  \bibfield  {author} {\bibinfo {author} {\bibfnamefont {A.}~\bibnamefont
  {Bulgac}},\ }\bibfield  {title} {\enquote {\bibinfo {title} {{Time-Dependent
  Density Functional Theory for Fermionic Superfluids: from Cold Atomic gases,
  to Nuclei and Neutron Star Crust}},}\ }\href {\doibase
  10.1002/pssb.201800592} {\bibfield  {journal} {\bibinfo  {journal} {Physica
  Status Solidi B}\ }\textbf {\bibinfo {volume} {2019}},\ \bibinfo {pages}
  {1800592} (\bibinfo {year} {2019}{\natexlab{a}})}\BibitemShut {NoStop}%
\bibitem [{\citenamefont {Jin}\ \emph {et~al.}(2017)\citenamefont {Jin},
  \citenamefont {Bulgac}, \citenamefont {Roche},\ and\ \citenamefont
  {Wlaz\l{}owski}}]{Jin:2017}%
  \BibitemOpen
  \bibfield  {author} {\bibinfo {author} {\bibfnamefont {S.}~\bibnamefont
  {Jin}}, \bibinfo {author} {\bibfnamefont {A.}~\bibnamefont {Bulgac}},
  \bibinfo {author} {\bibfnamefont {K.}~\bibnamefont {Roche}}, \ and\ \bibinfo
  {author} {\bibfnamefont {G.}~\bibnamefont {Wlaz\l{}owski}},\ }\bibfield
  {title} {\enquote {\bibinfo {title} {Coordinate-space solver for superfluid
  many-fermion systems with the shifted conjugate-orthogonal conjugate-gradient
  method},}\ }\href {\doibase 10.1103/PhysRevC.95.044302} {\bibfield  {journal}
  {\bibinfo  {journal} {Phys. Rev. C}\ }\textbf {\bibinfo {volume} {95}},\
  \bibinfo {pages} {044302} (\bibinfo {year} {2017})}\BibitemShut {NoStop}%
\bibitem [{\citenamefont {Bertsch}\ \emph {et~al.}(2018)\citenamefont
  {Bertsch}, \citenamefont {Younes},\ and\ \citenamefont
  {Robledo}}]{Bertsch:2018}%
  \BibitemOpen
  \bibfield  {author} {\bibinfo {author} {\bibfnamefont {G.~F.}\ \bibnamefont
  {Bertsch}}, \bibinfo {author} {\bibfnamefont {W.}~\bibnamefont {Younes}}, \
  and\ \bibinfo {author} {\bibfnamefont {L.~M.}\ \bibnamefont {Robledo}},\
  }\bibfield  {title} {\enquote {\bibinfo {title} {Scission dynamics with $k$
  partitions},}\ }\href {\doibase 10.1103/PhysRevC.97.064619} {\bibfield
  {journal} {\bibinfo  {journal} {Phys. Rev. C}\ }\textbf {\bibinfo {volume}
  {97}},\ \bibinfo {pages} {064619} (\bibinfo {year} {2018})}\BibitemShut
  {NoStop}%
\bibitem [{\citenamefont {Bulgac}\ \emph {et~al.}(2016)\citenamefont {Bulgac},
  \citenamefont {Magierski}, \citenamefont {Roche},\ and\ \citenamefont
  {Stetcu}}]{Bulgac:2016}%
  \BibitemOpen
  \bibfield  {author} {\bibinfo {author} {\bibfnamefont {A.}~\bibnamefont
  {Bulgac}}, \bibinfo {author} {\bibfnamefont {P.}~\bibnamefont {Magierski}},
  \bibinfo {author} {\bibfnamefont {K.~J.}\ \bibnamefont {Roche}}, \ and\
  \bibinfo {author} {\bibfnamefont {I.}~\bibnamefont {Stetcu}},\ }\bibfield
  {title} {\enquote {\bibinfo {title} {{Induced Fission of $^{240}\mathrm{Pu}$
  within a Real-Time Microscopic Framework}},}\ }\href {\doibase
  10.1103/PhysRevLett.116.122504} {\bibfield  {journal} {\bibinfo  {journal}
  {Phys. Rev. Lett.}\ }\textbf {\bibinfo {volume} {116}},\ \bibinfo {pages}
  {122504} (\bibinfo {year} {2016})}\BibitemShut {NoStop}%
\bibitem [{\citenamefont {Weidenm\"uller}\ and\ \citenamefont
  {Zhang}(1984)}]{Weidenmuller:1984}%
  \BibitemOpen
  \bibfield  {author} {\bibinfo {author} {\bibfnamefont {H.~A.}\ \bibnamefont
  {Weidenm\"uller}}\ and\ \bibinfo {author} {\bibfnamefont {J.-S.}\
  \bibnamefont {Zhang}},\ }\bibfield  {title} {\enquote {\bibinfo {title}
  {Nuclear fission viewed as a diffusion process: Case of very large
  friction},}\ }\href {\doibase 10.1103/PhysRevC.29.879} {\bibfield  {journal}
  {\bibinfo  {journal} {Phys. Rev. C}\ }\textbf {\bibinfo {volume} {29}},\
  \bibinfo {pages} {879} (\bibinfo {year} {1984})}\BibitemShut {NoStop}%
\bibitem [{\citenamefont {Randrup}\ and\ \citenamefont
  {M\"oller}(2011)}]{Randrup:2011}%
  \BibitemOpen
  \bibfield  {author} {\bibinfo {author} {\bibfnamefont {J.}~\bibnamefont
  {Randrup}}\ and\ \bibinfo {author} {\bibfnamefont {P.}~\bibnamefont
  {M\"oller}},\ }\bibfield  {title} {\enquote {\bibinfo {title} {{Brownian
  Shape Motion on Five-Dimensional Potential-Energy Surfaces:Nuclear
  Fission-Fragment Mass Distributions}},}\ }\href {\doibase
  10.1103/PhysRevLett.106.132503} {\bibfield  {journal} {\bibinfo  {journal}
  {Phys. Rev. Lett.}\ }\textbf {\bibinfo {volume} {106}},\ \bibinfo {pages}
  {132503} (\bibinfo {year} {2011})}\BibitemShut {NoStop}%
\bibitem [{\citenamefont {Randrup}\ \emph {et~al.}(2011)\citenamefont
  {Randrup}, \citenamefont {M\"oller},\ and\ \citenamefont
  {Sierk}}]{Randrup:2011a}%
  \BibitemOpen
  \bibfield  {author} {\bibinfo {author} {\bibfnamefont {J.}~\bibnamefont
  {Randrup}}, \bibinfo {author} {\bibfnamefont {P.}~\bibnamefont {M\"oller}}, \
  and\ \bibinfo {author} {\bibfnamefont {A.~J.}\ \bibnamefont {Sierk}},\
  }\bibfield  {title} {\enquote {\bibinfo {title} {Fission-fragment mass
  distributions from strongly damped shape evolution},}\ }\href {\doibase
  10.1103/PhysRevC.84.034613} {\bibfield  {journal} {\bibinfo  {journal} {Phys.
  Rev. C}\ }\textbf {\bibinfo {volume} {84}},\ \bibinfo {pages} {034613}
  (\bibinfo {year} {2011})}\BibitemShut {NoStop}%
\bibitem [{\citenamefont {Randrup}\ and\ \citenamefont
  {M\"oller}(2013)}]{Randrup:2013}%
  \BibitemOpen
  \bibfield  {author} {\bibinfo {author} {\bibfnamefont {J.}~\bibnamefont
  {Randrup}}\ and\ \bibinfo {author} {\bibfnamefont {P.}~\bibnamefont
  {M\"oller}},\ }\bibfield  {title} {\enquote {\bibinfo {title} {Energy
  dependence of fission-fragment mass distributions from strongly damped shape
  evolution},}\ }\href {\doibase 10.1103/PhysRevC.88.064606} {\bibfield
  {journal} {\bibinfo  {journal} {Phys. Rev. C}\ }\textbf {\bibinfo {volume}
  {88}},\ \bibinfo {pages} {064606} (\bibinfo {year} {2013})}\BibitemShut
  {NoStop}%
\bibitem [{\citenamefont {Ward}\ \emph {et~al.}(2017)\citenamefont {Ward},
  \citenamefont {Carlsson}, \citenamefont {D\o{}ssing}, \citenamefont
  {M\"oller}, \citenamefont {Randrup},\ and\ \citenamefont
  {\AA{}berg}}]{Ward:2017}%
  \BibitemOpen
  \bibfield  {author} {\bibinfo {author} {\bibfnamefont {D.~E.}\ \bibnamefont
  {Ward}}, \bibinfo {author} {\bibfnamefont {B.~G.}\ \bibnamefont {Carlsson}},
  \bibinfo {author} {\bibfnamefont {T.}~\bibnamefont {D\o{}ssing}}, \bibinfo
  {author} {\bibfnamefont {P.}~\bibnamefont {M\"oller}}, \bibinfo {author}
  {\bibfnamefont {J.}~\bibnamefont {Randrup}}, \ and\ \bibinfo {author}
  {\bibfnamefont {S.}~\bibnamefont {\AA{}berg}},\ }\bibfield  {title} {\enquote
  {\bibinfo {title} {Nuclear shape evolution based on microscopic level
  densities},}\ }\href {\doibase 10.1103/PhysRevC.95.024618} {\bibfield
  {journal} {\bibinfo  {journal} {Phys. Rev. C}\ }\textbf {\bibinfo {volume}
  {95}},\ \bibinfo {pages} {024618} (\bibinfo {year} {2017})}\BibitemShut
  {NoStop}%
\bibitem [{\citenamefont {Albertsson}\ \emph {et~al.}(2020)\citenamefont
  {Albertsson}, \citenamefont {Carlsson}, \citenamefont {D{\o}ssing},
  \citenamefont {M{\o}ller}, \citenamefont {Randrup},\ and\ \citenamefont
  {{\AA}berg}}]{Albertsson:2020}%
  \BibitemOpen
  \bibfield  {author} {\bibinfo {author} {\bibfnamefont {M.}~\bibnamefont
  {Albertsson}}, \bibinfo {author} {\bibfnamefont {B.G.}\ \bibnamefont
  {Carlsson}}, \bibinfo {author} {\bibfnamefont {T.}~\bibnamefont
  {D{\o}ssing}}, \bibinfo {author} {\bibfnamefont {P.}~\bibnamefont
  {M{\o}ller}}, \bibinfo {author} {\bibfnamefont {J.}~\bibnamefont {Randrup}},
  \ and\ \bibinfo {author} {\bibfnamefont {S.}~\bibnamefont {{\AA}berg}},\
  }\bibfield  {title} {\enquote {\bibinfo {title} {{Excitation energy partition
  in fission}},}\ }\href {\doibase 10.1016/j.physletb.2020.135276} {\bibfield
  {journal} {\bibinfo  {journal} {Phys. Lett. B}\ }\textbf {\bibinfo {volume}
  {803}},\ \bibinfo {pages} {135276} (\bibinfo {year} {2020})}\BibitemShut
  {NoStop}%
\bibitem [{\citenamefont {Wilkins}\ and\ \citenamefont
  {Steinberg}(1972)}]{Wilkins:1972}%
  \BibitemOpen
  \bibfield  {author} {\bibinfo {author} {\bibfnamefont {B.D.}\ \bibnamefont
  {Wilkins}}\ and\ \bibinfo {author} {\bibfnamefont {E.P.}\ \bibnamefont
  {Steinberg}},\ }\bibfield  {title} {\enquote {\bibinfo {title}
  {Semi-empirical interpretation of nuclear fission based on deformed-shell
  effects},}\ }\href {\doibase 10.1016/0370-2693(72)90046-9} {\bibfield
  {journal} {\bibinfo  {journal} {Phys. Lett. B}\ }\textbf {\bibinfo {volume}
  {42}},\ \bibinfo {pages} {141} (\bibinfo {year} {1972})}\BibitemShut
  {NoStop}%
\bibitem [{\citenamefont {Wilkins}\ \emph {et~al.}(1976)\citenamefont
  {Wilkins}, \citenamefont {Steinberg},\ and\ \citenamefont
  {Chasman}}]{Wilkins:1976}%
  \BibitemOpen
  \bibfield  {author} {\bibinfo {author} {\bibfnamefont {B.~D.}\ \bibnamefont
  {Wilkins}}, \bibinfo {author} {\bibfnamefont {E.~P.}\ \bibnamefont
  {Steinberg}}, \ and\ \bibinfo {author} {\bibfnamefont {R.~R.}\ \bibnamefont
  {Chasman}},\ }\bibfield  {title} {\enquote {\bibinfo {title} {Scission-point
  model of nuclear fission based on deformed-shell effects},}\ }\href {\doibase
  10.1103/PhysRevC.14.1832} {\bibfield  {journal} {\bibinfo  {journal} {Phys.
  Rev. C}\ }\textbf {\bibinfo {volume} {14}},\ \bibinfo {pages} {1832}
  (\bibinfo {year} {1976})}\BibitemShut {NoStop}%
\bibitem [{\citenamefont {Lema\^{\i}tre}\ \emph {et~al.}(2015)\citenamefont
  {Lema\^{\i}tre}, \citenamefont {Panebianco}, \citenamefont {Sida},
  \citenamefont {Hilaire},\ and\ \citenamefont {Heinrich}}]{Lemaitre:2015}%
  \BibitemOpen
  \bibfield  {author} {\bibinfo {author} {\bibfnamefont {J.-F.}\ \bibnamefont
  {Lema\^{\i}tre}}, \bibinfo {author} {\bibfnamefont {S.}~\bibnamefont
  {Panebianco}}, \bibinfo {author} {\bibfnamefont {J.-L.}\ \bibnamefont
  {Sida}}, \bibinfo {author} {\bibfnamefont {S.}~\bibnamefont {Hilaire}}, \
  and\ \bibinfo {author} {\bibfnamefont {S.}~\bibnamefont {Heinrich}},\
  }\bibfield  {title} {\enquote {\bibinfo {title} {New statistical
  scission-point model to predict fission fragment observables},}\ }\href
  {\doibase 10.1103/PhysRevC.92.034617} {\bibfield  {journal} {\bibinfo
  {journal} {Phys. Rev. C}\ }\textbf {\bibinfo {volume} {92}},\ \bibinfo
  {pages} {034617} (\bibinfo {year} {2015})}\BibitemShut {NoStop}%
\bibitem [{\citenamefont {Lema\^{\i}tre}\ \emph {et~al.}(2019)\citenamefont
  {Lema\^{\i}tre}, \citenamefont {Goriely}, \citenamefont {Hilaire},\ and\
  \citenamefont {Sida}}]{Lemaitre:2019}%
  \BibitemOpen
  \bibfield  {author} {\bibinfo {author} {\bibfnamefont {J.-F.}\ \bibnamefont
  {Lema\^{\i}tre}}, \bibinfo {author} {\bibfnamefont {S.}~\bibnamefont
  {Goriely}}, \bibinfo {author} {\bibfnamefont {S.}~\bibnamefont {Hilaire}}, \
  and\ \bibinfo {author} {\bibfnamefont {J.-L.}\ \bibnamefont {Sida}},\
  }\bibfield  {title} {\enquote {\bibinfo {title} {Fully microscopic
  scission-point model to predict fission fragment observables},}\ }\href
  {\doibase 10.1103/PhysRevC.99.034612} {\bibfield  {journal} {\bibinfo
  {journal} {Phys. Rev. C}\ }\textbf {\bibinfo {volume} {99}},\ \bibinfo
  {pages} {034612} (\bibinfo {year} {2019})}\BibitemShut {NoStop}%
\bibitem [{\citenamefont {Bethe}(1936)}]{Bethe:1936}%
  \BibitemOpen
  \bibfield  {author} {\bibinfo {author} {\bibfnamefont {H.~A.}\ \bibnamefont
  {Bethe}},\ }\bibfield  {title} {\enquote {\bibinfo {title} {An attempt to
  calculate the number of energy levels of a heavy nucleus},}\ }\href {\doibase
  10.1103/PhysRev.50.332} {\bibfield  {journal} {\bibinfo  {journal} {Phys.
  Rev.}\ }\textbf {\bibinfo {volume} {50}},\ \bibinfo {pages} {332--341}
  (\bibinfo {year} {1936})}\BibitemShut {NoStop}%
\bibitem [{\citenamefont {Bohr}\ and\ \citenamefont
  {Mottelson}(1969)}]{Bohr:1969}%
  \BibitemOpen
  \bibfield  {author} {\bibinfo {author} {\bibfnamefont {A.}~\bibnamefont
  {Bohr}}\ and\ \bibinfo {author} {\bibfnamefont {B.R.}\ \bibnamefont
  {Mottelson}},\ }\href@noop {} {\emph {\bibinfo {title} {Nuclear Structure}}}\
  (\bibinfo  {publisher} {Benjamin Inc.},\ \bibinfo {address} {New York},\
  \bibinfo {year} {1969})\BibitemShut {NoStop}%
\bibitem [{\citenamefont {Bulgac}\ \emph {et~al.}(1996)\citenamefont {Bulgac},
  \citenamefont {Dang},\ and\ \citenamefont {Kusnezov}}]{Bulgac:1996}%
  \BibitemOpen
  \bibfield  {author} {\bibinfo {author} {\bibfnamefont {A.}~\bibnamefont
  {Bulgac}}, \bibinfo {author} {\bibfnamefont {G.~Do}\ \bibnamefont {Dang}}, \
  and\ \bibinfo {author} {\bibfnamefont {D.}~\bibnamefont {Kusnezov}},\
  }\bibfield  {title} {\enquote {\bibinfo {title} {Stochastic aspects of large
  amplitude collective motion},}\ }\href {\doibase
  10.1016/0370-1573(95)00028-3} {\bibfield  {journal} {\bibinfo  {journal}
  {Phys. Rep.}\ }\textbf {\bibinfo {volume} {264}},\ \bibinfo {pages} {67}
  (\bibinfo {year} {1996})}\BibitemShut {NoStop}%
\bibitem [{\citenamefont {Tully}\ and\ \citenamefont
  {Preston}(1971)}]{Tully:1971}%
  \BibitemOpen
  \bibfield  {author} {\bibinfo {author} {\bibfnamefont {J.~C.}\ \bibnamefont
  {Tully}}\ and\ \bibinfo {author} {\bibfnamefont {R.~K.}\ \bibnamefont
  {Preston}},\ }\bibfield  {title} {\enquote {\bibinfo {title} {{Trajectory
  Surface Hopping Approach to Nonadiabatic Molecular Collisions: The reaction
  of H$^+$ with D$_2$}},}\ }\href {\doibase 10.1063/1.1675788} {\bibfield
  {journal} {\bibinfo  {journal} {J. Chem. Phys.}\ }\textbf {\bibinfo {volume}
  {55}},\ \bibinfo {pages} {562} (\bibinfo {year} {1971})}\BibitemShut
  {NoStop}%
\bibitem [{\citenamefont {Tully}(1990)}]{Tully:1990}%
  \BibitemOpen
  \bibfield  {author} {\bibinfo {author} {\bibfnamefont {J.~C.}\ \bibnamefont
  {Tully}},\ }\bibfield  {title} {\enquote {\bibinfo {title} {Molecular
  dynamics with electronic transitions},}\ }\href {\doibase 10.1063/1.459170}
  {\bibfield  {journal} {\bibinfo  {journal} {J. Chem. Phys.}\ }\textbf
  {\bibinfo {volume} {93}},\ \bibinfo {pages} {1061} (\bibinfo {year}
  {1990})}\BibitemShut {NoStop}%
\bibitem [{\citenamefont {Hammes‐Schiffer}\ and\ \citenamefont
  {Tully}(1994)}]{Tully:1994}%
  \BibitemOpen
  \bibfield  {author} {\bibinfo {author} {\bibfnamefont {S.}~\bibnamefont
  {Hammes‐Schiffer}}\ and\ \bibinfo {author} {\bibfnamefont {J.~C.}\
  \bibnamefont {Tully}},\ }\bibfield  {title} {\enquote {\bibinfo {title}
  {{Proton transfer in solution: Molecular dynamics with quantum
  transitions}},}\ }\href {\doibase 10.1063/1.467455} {\bibfield  {journal}
  {\bibinfo  {journal} {J. Chem. Phys.}\ }\textbf {\bibinfo {volume} {101}},\
  \bibinfo {pages} {4657} (\bibinfo {year} {1994})}\BibitemShut {NoStop}%
\bibitem [{\citenamefont {Grang\'e}\ \emph {et~al.}(1983)\citenamefont
  {Grang\'e}, \citenamefont {Li},\ and\ \citenamefont
  {Weidenm\"uller}}]{Grange:1983}%
  \BibitemOpen
  \bibfield  {author} {\bibinfo {author} {\bibfnamefont {P.}~\bibnamefont
  {Grang\'e}}, \bibinfo {author} {\bibfnamefont {J.-Q.}\ \bibnamefont {Li}}, \
  and\ \bibinfo {author} {\bibfnamefont {H.~A.}\ \bibnamefont
  {Weidenm\"uller}},\ }\bibfield  {title} {\enquote {\bibinfo {title} {{Induced
  nuclear fission viewed as a diffusion process: Transients}},}\ }\href
  {\doibase 10.1103/PhysRevC.27.2063} {\bibfield  {journal} {\bibinfo
  {journal} {Phys. Rev. C}\ }\textbf {\bibinfo {volume} {27}},\ \bibinfo
  {pages} {2063} (\bibinfo {year} {1983})}\BibitemShut {NoStop}%
\bibitem [{\citenamefont {Fr{\"o}brich}\ and\ \citenamefont
  {Gontchar}(1998)}]{Frobrich:1998}%
  \BibitemOpen
  \bibfield  {author} {\bibinfo {author} {\bibfnamefont {P.}~\bibnamefont
  {Fr{\"o}brich}}\ and\ \bibinfo {author} {\bibfnamefont {I.I.}\ \bibnamefont
  {Gontchar}},\ }\bibfield  {title} {\enquote {\bibinfo {title} {Langevin
  description of fusion, deep-inelastic collisions and heavy-ion-induced
  fission},}\ }\href {\doibase https://doi.org/10.1016/S0370-1573(97)00042-2}
  {\bibfield  {journal} {\bibinfo  {journal} {Phys. Rep.}\ }\textbf {\bibinfo
  {volume} {292}},\ \bibinfo {pages} {131} (\bibinfo {year}
  {1998})}\BibitemShut {NoStop}%
\bibitem [{\citenamefont {Sierk}(2017)}]{Sierk:2017}%
  \BibitemOpen
  \bibfield  {author} {\bibinfo {author} {\bibfnamefont {A.~J.}\ \bibnamefont
  {Sierk}},\ }\bibfield  {title} {\enquote {\bibinfo {title} {Langevin model of
  low-energy fission},}\ }\href {\doibase 10.1103/PhysRevC.96.034603}
  {\bibfield  {journal} {\bibinfo  {journal} {Phys. Rev. C}\ }\textbf {\bibinfo
  {volume} {96}},\ \bibinfo {pages} {034603} (\bibinfo {year}
  {2017})}\BibitemShut {NoStop}%
\bibitem [{\citenamefont {Ishizuka}\ \emph {et~al.}(2017)\citenamefont
  {Ishizuka}, \citenamefont {Usang}, \citenamefont {Ivanyuk}, \citenamefont
  {Maruhn}, \citenamefont {Nishio},\ and\ \citenamefont
  {Chiba}}]{Ishizuka:2017}%
  \BibitemOpen
  \bibfield  {author} {\bibinfo {author} {\bibfnamefont {C.}~\bibnamefont
  {Ishizuka}}, \bibinfo {author} {\bibfnamefont {M.~D.}\ \bibnamefont {Usang}},
  \bibinfo {author} {\bibfnamefont {F.~A.}\ \bibnamefont {Ivanyuk}}, \bibinfo
  {author} {\bibfnamefont {J.~A.}\ \bibnamefont {Maruhn}}, \bibinfo {author}
  {\bibfnamefont {K.}~\bibnamefont {Nishio}}, \ and\ \bibinfo {author}
  {\bibfnamefont {S.}~\bibnamefont {Chiba}},\ }\bibfield  {title} {\enquote
  {\bibinfo {title} {{Four-dimensional Langevin approach to low-energy nuclear
  fission of $^{236}\mathbf{U}$}},}\ }\href {\doibase
  10.1103/PhysRevC.96.064616} {\bibfield  {journal} {\bibinfo  {journal} {Phys.
  Rev. C}\ }\textbf {\bibinfo {volume} {96}},\ \bibinfo {pages} {064616}
  (\bibinfo {year} {2017})}\BibitemShut {NoStop}%
\bibitem [{\citenamefont {Sadhukhan}\ \emph {et~al.}(2016)\citenamefont
  {Sadhukhan}, \citenamefont {Nazarewicz},\ and\ \citenamefont
  {Schunck}}]{Sadhukhan:2016}%
  \BibitemOpen
  \bibfield  {author} {\bibinfo {author} {\bibfnamefont {J.}~\bibnamefont
  {Sadhukhan}}, \bibinfo {author} {\bibfnamefont {W.}~\bibnamefont
  {Nazarewicz}}, \ and\ \bibinfo {author} {\bibfnamefont {N.}~\bibnamefont
  {Schunck}},\ }\bibfield  {title} {\enquote {\bibinfo {title} {{Microscopic
  modeling of mass and charge distributions in the spontaneous fission of
  $^{240}$Pu}},}\ }\href {\doibase 10.1103/PhysRevC.93.011304} {\bibfield
  {journal} {\bibinfo  {journal} {Phys. Rev. C}\ }\textbf {\bibinfo {volume}
  {93}},\ \bibinfo {pages} {011304} (\bibinfo {year} {2016})}\BibitemShut
  {NoStop}%
\bibitem [{\citenamefont {Sadhukhan}\ \emph {et~al.}(2017)\citenamefont
  {Sadhukhan}, \citenamefont {Zhang}, \citenamefont {Nazarewicz},\ and\
  \citenamefont {Schunck}}]{Sadhukhan:2017}%
  \BibitemOpen
  \bibfield  {author} {\bibinfo {author} {\bibfnamefont {J.}~\bibnamefont
  {Sadhukhan}}, \bibinfo {author} {\bibfnamefont {C.}~\bibnamefont {Zhang}},
  \bibinfo {author} {\bibfnamefont {W.}~\bibnamefont {Nazarewicz}}, \ and\
  \bibinfo {author} {\bibfnamefont {N.}~\bibnamefont {Schunck}},\ }\bibfield
  {title} {\enquote {\bibinfo {title} {{Formation and distribution of fragments
  in the spontaneous fission of ${}^{\mathbf{240}}\mathbf{Pu}$}},}\ }\href
  {\doibase 10.1103/PhysRevC.96.061301} {\bibfield  {journal} {\bibinfo
  {journal} {Phys. Rev. C}\ }\textbf {\bibinfo {volume} {96}},\ \bibinfo
  {pages} {061301} (\bibinfo {year} {2017})}\BibitemShut {NoStop}%
\bibitem [{\citenamefont {Regnier}\ \emph {et~al.}(2019)\citenamefont
  {Regnier}, \citenamefont {Dubray},\ and\ \citenamefont
  {Schunck}}]{Regnier:2019}%
  \BibitemOpen
  \bibfield  {author} {\bibinfo {author} {\bibfnamefont {D.}~\bibnamefont
  {Regnier}}, \bibinfo {author} {\bibfnamefont {N.}~\bibnamefont {Dubray}}, \
  and\ \bibinfo {author} {\bibfnamefont {N.}~\bibnamefont {Schunck}},\
  }\bibfield  {title} {\enquote {\bibinfo {title} {From asymmetric to symmetric
  fission in the fermium isotopes within the time-dependent
  generator-coordinate-method formalism},}\ }\href {\doibase
  10.1103/PhysRevC.99.024611} {\bibfield  {journal} {\bibinfo  {journal} {Phys.
  Rev. C}\ }\textbf {\bibinfo {volume} {99}},\ \bibinfo {pages} {024611}
  (\bibinfo {year} {2019})}\BibitemShut {NoStop}%
\bibitem [{\citenamefont {Scamps}\ and\ \citenamefont
  {Simenel}(2018)}]{Scamps:2018}%
  \BibitemOpen
  \bibfield  {author} {\bibinfo {author} {\bibfnamefont {G.}~\bibnamefont
  {Scamps}}\ and\ \bibinfo {author} {\bibfnamefont {C.}~\bibnamefont
  {Simenel}},\ }\bibfield  {title} {\enquote {\bibinfo {title} {Impact of
  pear-shaped fission fragments on mass-asymmetric fission in actinides},}\
  }\href {\doibase 10.1038/s41586-018-0780-0} {\bibfield  {journal} {\bibinfo
  {journal} {Nature}\ }\textbf {\bibinfo {volume} {564}},\ \bibinfo {pages}
  {382} (\bibinfo {year} {2018})}\BibitemShut {NoStop}%
\bibitem [{\citenamefont {Mustafa}\ \emph {et~al.}(1971)\citenamefont
  {Mustafa}, \citenamefont {Schmitt},\ and\ \citenamefont
  {Mosel}}]{Mustafa:1971}%
  \BibitemOpen
  \bibfield  {author} {\bibinfo {author} {\bibfnamefont {M.G.}\ \bibnamefont
  {Mustafa}}, \bibinfo {author} {\bibfnamefont {H.W.}\ \bibnamefont {Schmitt}},
  \ and\ \bibinfo {author} {\bibfnamefont {U.}~\bibnamefont {Mosel}},\
  }\bibfield  {title} {\enquote {\bibinfo {title} {Dipole excitations in
  fission fragments},}\ }\href {\doibase
  https://doi.org/10.1016/0375-9474(71)90180-1} {\bibfield  {journal} {\bibinfo
   {journal} {Nucl. Phys. A}\ }\textbf {\bibinfo {volume} {178}},\ \bibinfo
  {pages} {9} (\bibinfo {year} {1971})}\BibitemShut {NoStop}%
\bibitem [{\citenamefont {Simenel}\ and\ \citenamefont
  {Umar}(2014)}]{Simenel:2014}%
  \BibitemOpen
  \bibfield  {author} {\bibinfo {author} {\bibfnamefont {C.}~\bibnamefont
  {Simenel}}\ and\ \bibinfo {author} {\bibfnamefont {A.~S.}\ \bibnamefont
  {Umar}},\ }\bibfield  {title} {\enquote {\bibinfo {title} {Formation and
  dynamics of fission fragments},}\ }\href {\doibase
  10.1103/PhysRevC.89.031601} {\bibfield  {journal} {\bibinfo  {journal} {Phys.
  Rev. C}\ }\textbf {\bibinfo {volume} {89}},\ \bibinfo {pages} {031601}
  (\bibinfo {year} {2014})}\BibitemShut {NoStop}%
\bibitem [{\citenamefont {Randrup}\ and\ \citenamefont
  {Vogt}(2009)}]{Randrup:2009}%
  \BibitemOpen
  \bibfield  {author} {\bibinfo {author} {\bibfnamefont {J.}~\bibnamefont
  {Randrup}}\ and\ \bibinfo {author} {\bibfnamefont {R.}~\bibnamefont {Vogt}},\
  }\bibfield  {title} {\enquote {\bibinfo {title} {Calculation of fission
  observables through event-by-event simulation},}\ }\href {\doibase
  10.1103/PhysRevC.80.024601} {\bibfield  {journal} {\bibinfo  {journal} {Phys.
  Rev. C}\ }\textbf {\bibinfo {volume} {80}},\ \bibinfo {pages} {024601}
  (\bibinfo {year} {2009})}\BibitemShut {NoStop}%
\bibitem [{\citenamefont {Becker}\ \emph {et~al.}(2013)\citenamefont {Becker},
  \citenamefont {Talou}, \citenamefont {Kawano}, \citenamefont {Danon},\ and\
  \citenamefont {Stetcu}}]{Becker:2013}%
  \BibitemOpen
  \bibfield  {author} {\bibinfo {author} {\bibfnamefont {B.}~\bibnamefont
  {Becker}}, \bibinfo {author} {\bibfnamefont {P.}~\bibnamefont {Talou}},
  \bibinfo {author} {\bibfnamefont {T.}~\bibnamefont {Kawano}}, \bibinfo
  {author} {\bibfnamefont {Y.}~\bibnamefont {Danon}}, \ and\ \bibinfo {author}
  {\bibfnamefont {I.}~\bibnamefont {Stetcu}},\ }\bibfield  {title} {\enquote
  {\bibinfo {title} {{Monte Carlo Hauser-Feshbach predictions of prompt fission
  gamma rays: Application to n+235U, n+239Pu, and 252Cf (sf)}},}\ }\href
  {\doibase 10.1103/PhysRevC.87.014617} {\bibfield  {journal} {\bibinfo
  {journal} {Phys. Rev. C}\ }\textbf {\bibinfo {volume} {87}},\ \bibinfo
  {pages} {014617} (\bibinfo {year} {2013})}\BibitemShut {NoStop}%
\bibitem [{\citenamefont {Randrup}\ \emph {et~al.}(2019)\citenamefont
  {Randrup}, \citenamefont {Talou},\ and\ \citenamefont {Vogt}}]{Randrup:2019}%
  \BibitemOpen
  \bibfield  {author} {\bibinfo {author} {\bibfnamefont {J.}~\bibnamefont
  {Randrup}}, \bibinfo {author} {\bibfnamefont {P.}~\bibnamefont {Talou}}, \
  and\ \bibinfo {author} {\bibfnamefont {R.}~\bibnamefont {Vogt}},\ }\bibfield
  {title} {\enquote {\bibinfo {title} {Sensitivity of neutron observables to
  the model input in simulations of ${}^{252}\mathrm{Cf}(\mathrm{sf})$},}\
  }\href {\doibase 10.1103/PhysRevC.99.054619} {\bibfield  {journal} {\bibinfo
  {journal} {Phys. Rev. C}\ }\textbf {\bibinfo {volume} {99}},\ \bibinfo
  {pages} {054619} (\bibinfo {year} {2019})}\BibitemShut {NoStop}%
\bibitem [{\citenamefont {Bertsch}\ \emph {et~al.}(2019)\citenamefont
  {Bertsch}, \citenamefont {Kawano},\ and\ \citenamefont
  {Robledo}}]{Bertsch:2019}%
  \BibitemOpen
  \bibfield  {author} {\bibinfo {author} {\bibfnamefont {G.~F.}\ \bibnamefont
  {Bertsch}}, \bibinfo {author} {\bibfnamefont {T.}~\bibnamefont {Kawano}}, \
  and\ \bibinfo {author} {\bibfnamefont {L.~M.}\ \bibnamefont {Robledo}},\
  }\bibfield  {title} {\enquote {\bibinfo {title} {Angular momentum of fission
  fragments},}\ }\href {\doibase 10.1103/PhysRevC.99.034603} {\bibfield
  {journal} {\bibinfo  {journal} {Phys. Rev. C}\ }\textbf {\bibinfo {volume}
  {99}},\ \bibinfo {pages} {034603} (\bibinfo {year} {2019})}\BibitemShut
  {NoStop}%
\bibitem [{\citenamefont {Schmidt}\ and\ \citenamefont
  {Jurado}(2010)}]{Schmidt:2010}%
  \BibitemOpen
  \bibfield  {author} {\bibinfo {author} {\bibfnamefont {K.-H.}\ \bibnamefont
  {Schmidt}}\ and\ \bibinfo {author} {\bibfnamefont {B.}~\bibnamefont
  {Jurado}},\ }\bibfield  {title} {\enquote {\bibinfo {title} {Entropy driven
  excitation energy sorting in superfluid fission dynamics},}\ }\href {\doibase
  10.1103/PhysRevLett.104.212501} {\bibfield  {journal} {\bibinfo  {journal}
  {Phys. Rev. Lett.}\ }\textbf {\bibinfo {volume} {104}},\ \bibinfo {pages}
  {212501} (\bibinfo {year} {2010})}\BibitemShut {NoStop}%
\bibitem [{\citenamefont {Schmidt}\ and\ \citenamefont
  {Jurado}(2011)}]{Schmidt:2011}%
  \BibitemOpen
  \bibfield  {author} {\bibinfo {author} {\bibfnamefont {K.-H.}\ \bibnamefont
  {Schmidt}}\ and\ \bibinfo {author} {\bibfnamefont {B.}~\bibnamefont
  {Jurado}},\ }\bibfield  {title} {\enquote {\bibinfo {title} {Final excitation
  energy of fission fragments},}\ }\href {\doibase 10.1103/PhysRevC.83.061601}
  {\bibfield  {journal} {\bibinfo  {journal} {Phys. Rev. C}\ }\textbf {\bibinfo
  {volume} {83}},\ \bibinfo {pages} {061601} (\bibinfo {year}
  {2011})}\BibitemShut {NoStop}%
\bibitem [{\citenamefont {Gilbert}\ and\ \citenamefont
  {Cameron}(1965)}]{Gilbert:1965}%
  \BibitemOpen
  \bibfield  {author} {\bibinfo {author} {\bibfnamefont {A.}~\bibnamefont
  {Gilbert}}\ and\ \bibinfo {author} {\bibfnamefont {A.~G.~W.}\ \bibnamefont
  {Cameron}},\ }\bibfield  {title} {\enquote {\bibinfo {title} {A composite
  nuclear-level density formula with shell correction},}\ }\href {\doibase
  10.1139/p65-139} {\bibfield  {journal} {\bibinfo  {journal} {Can. J. Phys.}\
  }\textbf {\bibinfo {volume} {43}},\ \bibinfo {pages} {1446} (\bibinfo {year}
  {1965})}\BibitemShut {NoStop}%
\bibitem [{\citenamefont {M\"uller}\ \emph {et~al.}(1984)\citenamefont
  {M\"uller}, \citenamefont {Naqvi}, \citenamefont {K\"appeler},\ and\
  \citenamefont {Dickmann}}]{Muller:1984}%
  \BibitemOpen
  \bibfield  {author} {\bibinfo {author} {\bibfnamefont {R.}~\bibnamefont
  {M\"uller}}, \bibinfo {author} {\bibfnamefont {A.~A.}\ \bibnamefont {Naqvi}},
  \bibinfo {author} {\bibfnamefont {F.}~\bibnamefont {K\"appeler}}, \ and\
  \bibinfo {author} {\bibfnamefont {F.}~\bibnamefont {Dickmann}},\ }\bibfield
  {title} {\enquote {\bibinfo {title} {{Fragment velocities, energies, and
  masses from fast neutron induced fission of $^{235}\mathrm{U}$}},}\ }\href
  {\doibase 10.1103/PhysRevC.29.885} {\bibfield  {journal} {\bibinfo  {journal}
  {Phys. Rev. C}\ }\textbf {\bibinfo {volume} {29}},\ \bibinfo {pages} {885}
  (\bibinfo {year} {1984})}\BibitemShut {NoStop}%
\bibitem [{\citenamefont {Bulgac}\ \emph
  {et~al.}(2019{\natexlab{b}})\citenamefont {Bulgac}, \citenamefont {Jin},\
  and\ \citenamefont {Stetcu}}]{Bulgac:2019a}%
  \BibitemOpen
  \bibfield  {author} {\bibinfo {author} {\bibfnamefont {A.}~\bibnamefont
  {Bulgac}}, \bibinfo {author} {\bibfnamefont {S.}~\bibnamefont {Jin}}, \ and\
  \bibinfo {author} {\bibfnamefont {I.}~\bibnamefont {Stetcu}},\ }\bibfield
  {title} {\enquote {\bibinfo {title} {Unitary evolution with fluctuations and
  dissipation},}\ }\href {\doibase 10.1103/PhysRevC.100.014615} {\bibfield
  {journal} {\bibinfo  {journal} {Phys. Rev. C}\ }\textbf {\bibinfo {volume}
  {100}},\ \bibinfo {pages} {014615} (\bibinfo {year}
  {2019}{\natexlab{b}})}\BibitemShut {NoStop}%
\bibitem [{\citenamefont {Bertsch}\ \emph {et~al.}(2007)\citenamefont
  {Bertsch}, \citenamefont {Girod}, \citenamefont {Hilaire}, \citenamefont
  {Delaroche}, \citenamefont {Goutte},\ and\ \citenamefont
  {P\'eru}}]{Bertsch:2007}%
  \BibitemOpen
  \bibfield  {author} {\bibinfo {author} {\bibfnamefont {G.~F.}\ \bibnamefont
  {Bertsch}}, \bibinfo {author} {\bibfnamefont {M.}~\bibnamefont {Girod}},
  \bibinfo {author} {\bibfnamefont {S.}~\bibnamefont {Hilaire}}, \bibinfo
  {author} {\bibfnamefont {J.-P.}\ \bibnamefont {Delaroche}}, \bibinfo {author}
  {\bibfnamefont {H.}~\bibnamefont {Goutte}}, \ and\ \bibinfo {author}
  {\bibfnamefont {S.}~\bibnamefont {P\'eru}},\ }\bibfield  {title} {\enquote
  {\bibinfo {title} {Systematics of the first ${2}^{+}$ excitation with the
  gogny interaction},}\ }\href {\doibase 10.1103/PhysRevLett.99.032502}
  {\bibfield  {journal} {\bibinfo  {journal} {Phys. Rev. Lett.}\ }\textbf
  {\bibinfo {volume} {99}},\ \bibinfo {pages} {032502} (\bibinfo {year}
  {2007})}\BibitemShut {NoStop}%
\bibitem [{\citenamefont {Delaroche}\ \emph {et~al.}(2010)\citenamefont
  {Delaroche}, \citenamefont {Girod}, \citenamefont {Libert}, \citenamefont
  {Goutte}, \citenamefont {Hilaire}, \citenamefont {P\'eru}, \citenamefont
  {Pillet},\ and\ \citenamefont {Bertsch}}]{Delaroche:2010}%
  \BibitemOpen
  \bibfield  {author} {\bibinfo {author} {\bibfnamefont {J.~P.}\ \bibnamefont
  {Delaroche}}, \bibinfo {author} {\bibfnamefont {M.}~\bibnamefont {Girod}},
  \bibinfo {author} {\bibfnamefont {J.}~\bibnamefont {Libert}}, \bibinfo
  {author} {\bibfnamefont {H.}~\bibnamefont {Goutte}}, \bibinfo {author}
  {\bibfnamefont {S.}~\bibnamefont {Hilaire}}, \bibinfo {author} {\bibfnamefont
  {S.}~\bibnamefont {P\'eru}}, \bibinfo {author} {\bibfnamefont
  {N.}~\bibnamefont {Pillet}}, \ and\ \bibinfo {author} {\bibfnamefont {G.~F.}\
  \bibnamefont {Bertsch}},\ }\bibfield  {title} {\enquote {\bibinfo {title}
  {Structure of even-even nuclei using a mapped collective hamiltonian and the
  d1s gogny interaction},}\ }\href {\doibase 10.1103/PhysRevC.81.014303}
  {\bibfield  {journal} {\bibinfo  {journal} {Phys. Rev. C}\ }\textbf {\bibinfo
  {volume} {81}},\ \bibinfo {pages} {014303} (\bibinfo {year}
  {2010})}\BibitemShut {NoStop}%
\bibitem [{\citenamefont {Ryssens}\ \emph {et~al.}(2015)\citenamefont
  {Ryssens}, \citenamefont {Heenen},\ and\ \citenamefont
  {Bender}}]{Ryssens:2015}%
  \BibitemOpen
  \bibfield  {author} {\bibinfo {author} {\bibfnamefont {W.}~\bibnamefont
  {Ryssens}}, \bibinfo {author} {\bibfnamefont {P.-H.}\ \bibnamefont {Heenen}},
  \ and\ \bibinfo {author} {\bibfnamefont {M.}~\bibnamefont {Bender}},\
  }\bibfield  {title} {\enquote {\bibinfo {title} {Numerical accuracy of
  mean-field calculations in coordinate space},}\ }\href {\doibase
  10.1103/PhysRevC.92.064318} {\bibfield  {journal} {\bibinfo  {journal} {Phys.
  Rev. C}\ }\textbf {\bibinfo {volume} {92}},\ \bibinfo {pages} {064318}
  (\bibinfo {year} {2015})}\BibitemShut {NoStop}%
\bibitem [{\citenamefont {Moretto}\ \emph {et~al.}(1989)\citenamefont
  {Moretto}, \citenamefont {Peqaslee},\ and\ \citenamefont
  {Wozniak}}]{Moretto:1989}%
  \BibitemOpen
  \bibfield  {author} {\bibinfo {author} {\bibfnamefont {L.~G.}\ \bibnamefont
  {Moretto}}, \bibinfo {author} {\bibfnamefont {G.~F.}\ \bibnamefont
  {Peqaslee}}, \ and\ \bibinfo {author} {\bibfnamefont {G.~J.}\ \bibnamefont
  {Wozniak}},\ }\bibfield  {title} {\enquote {\bibinfo {title}
  {{Angular-Momentum-Bearing Modes in Fission}},}\ }\href {\doibase
  10.1016/0375-9474(89)90682-9} {\bibfield  {journal} {\bibinfo  {journal}
  {Nucl. Phys. A}\ }\textbf {\bibinfo {volume} {502}},\ \bibinfo {pages} {453c}
  (\bibinfo {year} {1989})}\BibitemShut {NoStop}%
\bibitem [{\citenamefont {Bulgac}(2019{\natexlab{b}})}]{Bulgac:2019x}%
  \BibitemOpen
  \bibfield  {author} {\bibinfo {author} {\bibfnamefont {A.}~\bibnamefont
  {Bulgac}},\ }\bibfield  {title} {\enquote {\bibinfo {title} {{Projection of
  Good Quantum Numbers for Reaction Fragments}},}\ }\href {\doibase
  10.1103/PhysRevC.100.034612} {\bibfield  {journal} {\bibinfo  {journal}
  {Phys. Rev. C}\ }\textbf {\bibinfo {volume} {100}},\ \bibinfo {pages}
  {034612} (\bibinfo {year} {2019}{\natexlab{b}})}\BibitemShut {NoStop}%
\bibitem [{\citenamefont {Haight}\ \emph {et~al.}(2015)\citenamefont {Haight},
  \citenamefont {Wu}, \citenamefont {Lee}, \citenamefont {Taddeucci},
  \citenamefont {Perdue}, \citenamefont {O'Donnell}, \citenamefont {Fotiades},
  \citenamefont {Devlin}, \citenamefont {Ullmann}, \citenamefont {Bredeweg},
  \citenamefont {Jandel}, \citenamefont {Nelson}, \citenamefont {Wender},
  \citenamefont {Neudecker}, \citenamefont {Rising}, \citenamefont {Mosby},
  \citenamefont {Sjue}, \citenamefont {White}, \citenamefont {Bucher},\ and\
  \citenamefont {Henderson}}]{Haight:2015aa}%
  \BibitemOpen
  \bibfield  {author} {\bibinfo {author} {\bibfnamefont {R.C.}\ \bibnamefont
  {Haight}}, \bibinfo {author} {\bibfnamefont {C.Y.}\ \bibnamefont {Wu}},
  \bibinfo {author} {\bibfnamefont {H.Y.}\ \bibnamefont {Lee}}, \bibinfo
  {author} {\bibfnamefont {T.N.}\ \bibnamefont {Taddeucci}}, \bibinfo {author}
  {\bibfnamefont {B.A.}\ \bibnamefont {Perdue}}, \bibinfo {author}
  {\bibfnamefont {J.M.}\ \bibnamefont {O'Donnell}}, \bibinfo {author}
  {\bibfnamefont {N.}~\bibnamefont {Fotiades}}, \bibinfo {author}
  {\bibfnamefont {M.}~\bibnamefont {Devlin}}, \bibinfo {author} {\bibfnamefont
  {J.L.}\ \bibnamefont {Ullmann}}, \bibinfo {author} {\bibfnamefont {T.A.}\
  \bibnamefont {Bredeweg}}, \bibinfo {author} {\bibfnamefont {M.}~\bibnamefont
  {Jandel}}, \bibinfo {author} {\bibfnamefont {R.O.}\ \bibnamefont {Nelson}},
  \bibinfo {author} {\bibfnamefont {S.A.}\ \bibnamefont {Wender}}, \bibinfo
  {author} {\bibfnamefont {D.}~\bibnamefont {Neudecker}}, \bibinfo {author}
  {\bibfnamefont {M.E.}\ \bibnamefont {Rising}}, \bibinfo {author}
  {\bibfnamefont {S.}~\bibnamefont {Mosby}}, \bibinfo {author} {\bibfnamefont
  {S.}~\bibnamefont {Sjue}}, \bibinfo {author} {\bibfnamefont {M.C.}\
  \bibnamefont {White}}, \bibinfo {author} {\bibfnamefont {B.}~\bibnamefont
  {Bucher}}, \ and\ \bibinfo {author} {\bibfnamefont {R.}~\bibnamefont
  {Henderson}},\ }\bibfield  {title} {\enquote {\bibinfo {title} {{The
  LANL/LLNL Prompt Fission Neutron Spectrum Program at LANSCE and Approach to
  Uncertainties}},}\ }\href {\doibase
  https://doi.org/10.1016/j.nds.2014.12.023} {\bibfield  {journal} {\bibinfo
  {journal} {Nuclear Data Sheets}\ }\textbf {\bibinfo {volume} {123}},\
  \bibinfo {pages} {130 -- 134} (\bibinfo {year} {2015})},\ \bibinfo {note}
  {special Issue on International Workshop on Nuclear Data Covariances April 28
  - May 1, 2014, Santa Fe, New Mexico, USA
  http://t2.lanl.gov/cw2014}\BibitemShut {NoStop}%
\bibitem [{\citenamefont {{Kelly, Keegan J.}}\ \emph
  {et~al.}(2018)\citenamefont {{Kelly, Keegan J.}}, \citenamefont {{Devlin,
  Matthew}}, \citenamefont {{Gomez, Jaime A.}}, \citenamefont {{O\'{}Donnell,
  John M.}}, \citenamefont {{Taddeucci, Terry N.}}, \citenamefont {{Haight,
  Robert C.}}, \citenamefont {{Lee, Hye Young}}, \citenamefont {{Mosby, Shea
  M.}}, \citenamefont {{Fotiades, Nikolaos}}, \citenamefont {{Neudecker,
  Denise}}, \citenamefont {{Talou, Patrick}}, \citenamefont {{Rising, Michael
  E.}}, \citenamefont {{White, Morgan C.}}, \citenamefont {{Solomon, Clell
  J.}}, \citenamefont {{Wu, Ching-Yen}}, \citenamefont {{Bucher, Brian}},
  \citenamefont {{Buckner, Matthew Q.}},\ and\ \citenamefont {{Henderson, Roger
  A.}}}]{Kelly-Keegan-J.:2018aa}%
  \BibitemOpen
  \bibfield  {author} {\bibinfo {author} {\bibnamefont {{Kelly, Keegan J.}}},
  \bibinfo {author} {\bibnamefont {{Devlin, Matthew}}}, \bibinfo {author}
  {\bibnamefont {{Gomez, Jaime A.}}}, \bibinfo {author} {\bibnamefont
  {{O\'{}Donnell, John M.}}}, \bibinfo {author} {\bibnamefont {{Taddeucci,
  Terry N.}}}, \bibinfo {author} {\bibnamefont {{Haight, Robert C.}}}, \bibinfo
  {author} {\bibnamefont {{Lee, Hye Young}}}, \bibinfo {author} {\bibnamefont
  {{Mosby, Shea M.}}}, \bibinfo {author} {\bibnamefont {{Fotiades, Nikolaos}}},
  \bibinfo {author} {\bibnamefont {{Neudecker, Denise}}}, \bibinfo {author}
  {\bibnamefont {{Talou, Patrick}}}, \bibinfo {author} {\bibnamefont {{Rising,
  Michael E.}}}, \bibinfo {author} {\bibnamefont {{White, Morgan C.}}},
  \bibinfo {author} {\bibnamefont {{Solomon, Clell J.}}}, \bibinfo {author}
  {\bibnamefont {{Wu, Ching-Yen}}}, \bibinfo {author} {\bibnamefont {{Bucher,
  Brian}}}, \bibinfo {author} {\bibnamefont {{Buckner, Matthew Q.}}}, \ and\
  \bibinfo {author} {\bibnamefont {{Henderson, Roger A.}}},\ }\bibfield
  {title} {\enquote {\bibinfo {title} {Measurements of the prompt fission
  neutron spectrum at lansce: The chi-nu experiment},}\ }\href {\doibase
  10.1051/epjconf/201819303003} {\bibfield  {journal} {\bibinfo  {journal} {EPJ
  Web Conf.}\ }\textbf {\bibinfo {volume} {193}},\ \bibinfo {pages} {03003}
  (\bibinfo {year} {2018})}\BibitemShut {NoStop}%
\bibitem [{\citenamefont {Maslin}\ \emph {et~al.}(1967)\citenamefont {Maslin},
  \citenamefont {Rodgers},\ and\ \citenamefont {Core}}]{Maslin:1967aa}%
  \BibitemOpen
  \bibfield  {author} {\bibinfo {author} {\bibfnamefont {E.~E.}\ \bibnamefont
  {Maslin}}, \bibinfo {author} {\bibfnamefont {A.~L.}\ \bibnamefont {Rodgers}},
  \ and\ \bibinfo {author} {\bibfnamefont {W.~G.~F.}\ \bibnamefont {Core}},\
  }\bibfield  {title} {\enquote {\bibinfo {title} {Prompt neutron emission from
  ${\mathrm{u}}^{235}$ fission fragments},}\ }\href {\doibase
  10.1103/PhysRev.164.1520} {\bibfield  {journal} {\bibinfo  {journal} {Phys.
  Rev.}\ }\textbf {\bibinfo {volume} {164}},\ \bibinfo {pages} {1520--1527}
  (\bibinfo {year} {1967})}\BibitemShut {NoStop}%
\bibitem [{\citenamefont {Nishio}\ \emph {et~al.}(1998)\citenamefont {Nishio},
  \citenamefont {Nakagome}, \citenamefont {Yamamoto},\ and\ \citenamefont
  {Kimura}}]{Nishio:1998aa}%
  \BibitemOpen
  \bibfield  {author} {\bibinfo {author} {\bibfnamefont {K.}~\bibnamefont
  {Nishio}}, \bibinfo {author} {\bibfnamefont {Y.}~\bibnamefont {Nakagome}},
  \bibinfo {author} {\bibfnamefont {H.}~\bibnamefont {Yamamoto}}, \ and\
  \bibinfo {author} {\bibfnamefont {I.}~\bibnamefont {Kimura}},\ }\bibfield
  {title} {\enquote {\bibinfo {title} {Multiplicity and energy of neutrons from
  235u(nth,f) fission fragments},}\ }\href {\doibase
  https://doi.org/10.1016/S0375-9474(98)00008-6} {\bibfield  {journal}
  {\bibinfo  {journal} {Nuclear Physics A}\ }\textbf {\bibinfo {volume}
  {632}},\ \bibinfo {pages} {540 -- 558} (\bibinfo {year} {1998})}\BibitemShut
  {NoStop}%
\bibitem [{\citenamefont {Tsuchiya}\ \emph {et~al.}(2000)\citenamefont
  {Tsuchiya}, \citenamefont {Nakagome}, \citenamefont {Yamana}, \citenamefont
  {Moriyama}, \citenamefont {Nishio}, \citenamefont {Kanno}, \citenamefont
  {Shin},\ and\ \citenamefont {Kimura}}]{Tsuchiya:2000}%
  \BibitemOpen
  \bibfield  {author} {\bibinfo {author} {\bibfnamefont {C.}~\bibnamefont
  {Tsuchiya}}, \bibinfo {author} {\bibfnamefont {Y.}~\bibnamefont {Nakagome}},
  \bibinfo {author} {\bibfnamefont {H.}~\bibnamefont {Yamana}}, \bibinfo
  {author} {\bibfnamefont {H.}~\bibnamefont {Moriyama}}, \bibinfo {author}
  {\bibfnamefont {K.}~\bibnamefont {Nishio}}, \bibinfo {author} {\bibfnamefont
  {I.}~\bibnamefont {Kanno}}, \bibinfo {author} {\bibfnamefont
  {K.}~\bibnamefont {Shin}}, \ and\ \bibinfo {author} {\bibfnamefont
  {I.}~\bibnamefont {Kimura}},\ }\bibfield  {title} {\enquote {\bibinfo {title}
  {{Simultaneous Measurement of Prompt Neutrons and Fission Fragments for
  239Pu(nth,f)}},}\ }\href {\doibase 10.1080/18811248.2000.9714976} {\bibfield
  {journal} {\bibinfo  {journal} {J. Nucl. Sci. Tech.}\ }\textbf {\bibinfo
  {volume} {37}},\ \bibinfo {pages} {941} (\bibinfo {year} {2000})}\BibitemShut
  {NoStop}%
\bibitem [{\citenamefont {Batenkov}\ \emph {et~al.}(2005)\citenamefont
  {Batenkov}, \citenamefont {Boykov}, \citenamefont {Hambsch}, \citenamefont
  {Hamilton}, \citenamefont {Jakovlev}, \citenamefont {Kalinin}, \citenamefont
  {Laptev}, \citenamefont {Sokolov},\ and\ \citenamefont
  {Vorobyev}}]{Batenkov:2005}%
  \BibitemOpen
  \bibfield  {author} {\bibinfo {author} {\bibfnamefont {O.~A.}\ \bibnamefont
  {Batenkov}}, \bibinfo {author} {\bibfnamefont {G.~A.}\ \bibnamefont
  {Boykov}}, \bibinfo {author} {\bibfnamefont {F.-J.}\ \bibnamefont {Hambsch}},
  \bibinfo {author} {\bibfnamefont {J.~H.}\ \bibnamefont {Hamilton}}, \bibinfo
  {author} {\bibfnamefont {V.~A.}\ \bibnamefont {Jakovlev}}, \bibinfo {author}
  {\bibfnamefont {V.~A.}\ \bibnamefont {Kalinin}}, \bibinfo {author}
  {\bibfnamefont {A.~B.}\ \bibnamefont {Laptev}}, \bibinfo {author}
  {\bibfnamefont {V.~E.}\ \bibnamefont {Sokolov}}, \ and\ \bibinfo {author}
  {\bibfnamefont {A.~S.}\ \bibnamefont {Vorobyev}},\ }\bibfield  {title}
  {\enquote {\bibinfo {title} {{Prompt Neutron Emmision in the Neutron-Induced
  Fission of $^{239}$Pu and $^{235}$U}},}\ }\href {\doibase 10.1063/1.1945175}
  {\bibfield  {journal} {\bibinfo  {journal} {AIP Conf. Proc.}\ }\textbf
  {\bibinfo {volume} {769}},\ \bibinfo {pages} {1003} (\bibinfo {year}
  {2005})}\BibitemShut {NoStop}%
\bibitem [{\citenamefont {G\"o\"ok}\ \emph {et~al.}(2014)\citenamefont
  {G\"o\"ok}, \citenamefont {Hambsch},\ and\ \citenamefont
  {Vidali}}]{Gook:2014aa}%
  \BibitemOpen
  \bibfield  {author} {\bibinfo {author} {\bibfnamefont {A.}~\bibnamefont
  {G\"o\"ok}}, \bibinfo {author} {\bibfnamefont {F.-J.}\ \bibnamefont
  {Hambsch}}, \ and\ \bibinfo {author} {\bibfnamefont {M.}~\bibnamefont
  {Vidali}},\ }\bibfield  {title} {\enquote {\bibinfo {title} {Prompt neutron
  multiplicity in correlation with fragments from spontaneous fission of
  $^{252}\mathrm{Cf}$},}\ }\href {\doibase 10.1103/PhysRevC.90.064611}
  {\bibfield  {journal} {\bibinfo  {journal} {Phys. Rev. C}\ }\textbf {\bibinfo
  {volume} {90}},\ \bibinfo {pages} {064611} (\bibinfo {year}
  {2014})}\BibitemShut {NoStop}%
\bibitem [{\citenamefont {G\"o\"ok}\ \emph {et~al.}(2018)\citenamefont
  {G\"o\"ok}, \citenamefont {Hambsch}, \citenamefont {Oberstedt},\ and\
  \citenamefont {Vidali}}]{Gook:2018aa}%
  \BibitemOpen
  \bibfield  {author} {\bibinfo {author} {\bibfnamefont {Alf}\ \bibnamefont
  {G\"o\"ok}}, \bibinfo {author} {\bibfnamefont {Franz-Josef}\ \bibnamefont
  {Hambsch}}, \bibinfo {author} {\bibfnamefont {Stephan}\ \bibnamefont
  {Oberstedt}}, \ and\ \bibinfo {author} {\bibfnamefont {Marzio}\ \bibnamefont
  {Vidali}},\ }\bibfield  {title} {\enquote {\bibinfo {title} {Prompt neutrons
  in correlation with fission fragments from $^{235}\mathrm{U}(n,f)$},}\ }\href
  {\doibase 10.1103/PhysRevC.98.044615} {\bibfield  {journal} {\bibinfo
  {journal} {Phys. Rev. C}\ }\textbf {\bibinfo {volume} {98}},\ \bibinfo
  {pages} {044615} (\bibinfo {year} {2018})}\BibitemShut {NoStop}%
\bibitem [{\citenamefont {Akindele}\ \emph {et~al.}(2019)\citenamefont
  {Akindele}, \citenamefont {Alan}, \citenamefont {Burke}, \citenamefont
  {Casperson}, \citenamefont {Hughes}, \citenamefont {Koglin}, \citenamefont
  {Kolos}, \citenamefont {Norman}, \citenamefont {Ota},\ and\ \citenamefont
  {Saastamoinen}}]{Akindele:2019aa}%
  \BibitemOpen
  \bibfield  {author} {\bibinfo {author} {\bibfnamefont {O.~A.}\ \bibnamefont
  {Akindele}}, \bibinfo {author} {\bibfnamefont {B.~S.}\ \bibnamefont {Alan}},
  \bibinfo {author} {\bibfnamefont {J.~T.}\ \bibnamefont {Burke}}, \bibinfo
  {author} {\bibfnamefont {R.~J.}\ \bibnamefont {Casperson}}, \bibinfo {author}
  {\bibfnamefont {R.~O.}\ \bibnamefont {Hughes}}, \bibinfo {author}
  {\bibfnamefont {J.~D.}\ \bibnamefont {Koglin}}, \bibinfo {author}
  {\bibfnamefont {K.}~\bibnamefont {Kolos}}, \bibinfo {author} {\bibfnamefont
  {E.~B.}\ \bibnamefont {Norman}}, \bibinfo {author} {\bibfnamefont
  {S.}~\bibnamefont {Ota}}, \ and\ \bibinfo {author} {\bibfnamefont
  {A.}~\bibnamefont {Saastamoinen}},\ }\bibfield  {title} {\enquote {\bibinfo
  {title} {Expansion of the surrogate method to measure the prompt fission
  neutron multiplicity for $^{241}\mathrm{Pu}$},}\ }\href {\doibase
  10.1103/PhysRevC.99.054601} {\bibfield  {journal} {\bibinfo  {journal} {Phys.
  Rev. C}\ }\textbf {\bibinfo {volume} {99}},\ \bibinfo {pages} {054601}
  (\bibinfo {year} {2019})}\BibitemShut {NoStop}%
\bibitem [{\citenamefont {Wang}\ \emph {et~al.}(2019)\citenamefont {Wang},
  \citenamefont {Harke}, \citenamefont {Akindele}, \citenamefont {Casperson},
  \citenamefont {Hughes}, \citenamefont {Koglin}, \citenamefont {Kolos},
  \citenamefont {Norman}, \citenamefont {Ota},\ and\ \citenamefont
  {Saastamoinen}}]{Wang:2019aa}%
  \BibitemOpen
  \bibfield  {author} {\bibinfo {author} {\bibfnamefont {B.~S.}\ \bibnamefont
  {Wang}}, \bibinfo {author} {\bibfnamefont {J.~T.}\ \bibnamefont {Harke}},
  \bibinfo {author} {\bibfnamefont {O.~A.}\ \bibnamefont {Akindele}}, \bibinfo
  {author} {\bibfnamefont {R.~J.}\ \bibnamefont {Casperson}}, \bibinfo {author}
  {\bibfnamefont {R.~O.}\ \bibnamefont {Hughes}}, \bibinfo {author}
  {\bibfnamefont {J.~D.}\ \bibnamefont {Koglin}}, \bibinfo {author}
  {\bibfnamefont {K.}~\bibnamefont {Kolos}}, \bibinfo {author} {\bibfnamefont
  {E.~B.}\ \bibnamefont {Norman}}, \bibinfo {author} {\bibfnamefont
  {S.}~\bibnamefont {Ota}}, \ and\ \bibinfo {author} {\bibfnamefont
  {A.}~\bibnamefont {Saastamoinen}},\ }\bibfield  {title} {\enquote {\bibinfo
  {title} {Determining the average prompt-fission-neutron multiplicity for
  $^{239}\mathrm{Pu}(n,f)$ via a
  $^{240}\mathrm{Pu}(\ensuremath{\alpha},{\ensuremath{\alpha}}^{\ensuremath{'}}f)$
  surrogate reaction},}\ }\href {\doibase 10.1103/PhysRevC.100.064609}
  {\bibfield  {journal} {\bibinfo  {journal} {Phys. Rev. C}\ }\textbf {\bibinfo
  {volume} {100}},\ \bibinfo {pages} {064609} (\bibinfo {year}
  {2019})}\BibitemShut {NoStop}%
\bibitem [{\citenamefont {Ullmann}\ \emph {et~al.}(2013)\citenamefont
  {Ullmann}, \citenamefont {Bond}, \citenamefont {Bredeweg}, \citenamefont
  {Couture}, \citenamefont {Haight}, \citenamefont {Jandel}, \citenamefont
  {Kawano}, \citenamefont {Lee}, \citenamefont {O'Donnell}, \citenamefont
  {Hayes}, \citenamefont {Stetcu}, \citenamefont {Taddeucci}, \citenamefont
  {Talou}, \citenamefont {Vieira}, \citenamefont {Wilhelmy}, \citenamefont
  {Becker}, \citenamefont {Chyzh}, \citenamefont {Gostic}, \citenamefont
  {Henderson}, \citenamefont {Kwan},\ and\ \citenamefont {Wu}}]{Ullmann:2013}%
  \BibitemOpen
  \bibfield  {author} {\bibinfo {author} {\bibfnamefont {J.~L.}\ \bibnamefont
  {Ullmann}}, \bibinfo {author} {\bibfnamefont {E.~M.}\ \bibnamefont {Bond}},
  \bibinfo {author} {\bibfnamefont {T.~A.}\ \bibnamefont {Bredeweg}}, \bibinfo
  {author} {\bibfnamefont {A.}~\bibnamefont {Couture}}, \bibinfo {author}
  {\bibfnamefont {R.~C.}\ \bibnamefont {Haight}}, \bibinfo {author}
  {\bibfnamefont {M.}~\bibnamefont {Jandel}}, \bibinfo {author} {\bibfnamefont
  {T.}~\bibnamefont {Kawano}}, \bibinfo {author} {\bibfnamefont {H.~Y.}\
  \bibnamefont {Lee}}, \bibinfo {author} {\bibfnamefont {J.~M.}\ \bibnamefont
  {O'Donnell}}, \bibinfo {author} {\bibfnamefont {A.~C.}\ \bibnamefont
  {Hayes}}, \bibinfo {author} {\bibfnamefont {I.}~\bibnamefont {Stetcu}},
  \bibinfo {author} {\bibfnamefont {T.~N.}\ \bibnamefont {Taddeucci}}, \bibinfo
  {author} {\bibfnamefont {P.}~\bibnamefont {Talou}}, \bibinfo {author}
  {\bibfnamefont {D.~J.}\ \bibnamefont {Vieira}}, \bibinfo {author}
  {\bibfnamefont {J.~B.}\ \bibnamefont {Wilhelmy}}, \bibinfo {author}
  {\bibfnamefont {J.~A.}\ \bibnamefont {Becker}}, \bibinfo {author}
  {\bibfnamefont {A.}~\bibnamefont {Chyzh}}, \bibinfo {author} {\bibfnamefont
  {J.}~\bibnamefont {Gostic}}, \bibinfo {author} {\bibfnamefont
  {R.}~\bibnamefont {Henderson}}, \bibinfo {author} {\bibfnamefont
  {E.}~\bibnamefont {Kwan}}, \ and\ \bibinfo {author} {\bibfnamefont {C.~Y.}\
  \bibnamefont {Wu}},\ }\bibfield  {title} {\enquote {\bibinfo {title} {Prompt
  $\ensuremath{\gamma}$-ray production in neutron-induced fission of
  ${}^{239}${P}u},}\ }\href {\doibase 10.1103/PhysRevC.87.044607} {\bibfield
  {journal} {\bibinfo  {journal} {Phys. Rev. C}\ }\textbf {\bibinfo {volume}
  {87}},\ \bibinfo {pages} {044607} (\bibinfo {year} {2013})}\BibitemShut
  {NoStop}%
\bibitem [{\citenamefont {Chyzh}\ \emph {et~al.}(2013)\citenamefont {Chyzh},
  \citenamefont {Wu}, \citenamefont {Kwan}, \citenamefont {Henderson},
  \citenamefont {Gostic}, \citenamefont {Bredeweg}, \citenamefont {Couture},
  \citenamefont {Haight}, \citenamefont {Hayes-Sterbenz}, \citenamefont
  {Jandel}, \citenamefont {Lee}, \citenamefont {O'Donnell},\ and\ \citenamefont
  {Ullmann}}]{Chyzh:2013}%
  \BibitemOpen
  \bibfield  {author} {\bibinfo {author} {\bibfnamefont {A.}~\bibnamefont
  {Chyzh}}, \bibinfo {author} {\bibfnamefont {C.~Y.}\ \bibnamefont {Wu}},
  \bibinfo {author} {\bibfnamefont {E.}~\bibnamefont {Kwan}}, \bibinfo {author}
  {\bibfnamefont {R.~A.}\ \bibnamefont {Henderson}}, \bibinfo {author}
  {\bibfnamefont {J.~M.}\ \bibnamefont {Gostic}}, \bibinfo {author}
  {\bibfnamefont {T.~A.}\ \bibnamefont {Bredeweg}}, \bibinfo {author}
  {\bibfnamefont {A.}~\bibnamefont {Couture}}, \bibinfo {author} {\bibfnamefont
  {R.~C.}\ \bibnamefont {Haight}}, \bibinfo {author} {\bibfnamefont {A.~C.}\
  \bibnamefont {Hayes-Sterbenz}}, \bibinfo {author} {\bibfnamefont
  {M.}~\bibnamefont {Jandel}}, \bibinfo {author} {\bibfnamefont {H.~Y.}\
  \bibnamefont {Lee}}, \bibinfo {author} {\bibfnamefont {J.~M.}\ \bibnamefont
  {O'Donnell}}, \ and\ \bibinfo {author} {\bibfnamefont {J.~L.}\ \bibnamefont
  {Ullmann}},\ }\bibfield  {title} {\enquote {\bibinfo {title} {Systematics of
  prompt $\ensuremath{\gamma}$-ray emission in fission},}\ }\href {\doibase
  10.1103/PhysRevC.87.034620} {\bibfield  {journal} {\bibinfo  {journal} {Phys.
  Rev. C}\ }\textbf {\bibinfo {volume} {87}},\ \bibinfo {pages} {034620}
  (\bibinfo {year} {2013})}\BibitemShut {NoStop}%
\bibitem [{\citenamefont {Chyzh}\ \emph {et~al.}(2014)\citenamefont {Chyzh},
  \citenamefont {Wu}, \citenamefont {Kwan}, \citenamefont {Henderson},
  \citenamefont {Bredeweg}, \citenamefont {Haight}, \citenamefont
  {Hayes-Sterbenz}, \citenamefont {Lee}, \citenamefont {O'Donnell},\ and\
  \citenamefont {Ullmann}}]{Chyzh:2014}%
  \BibitemOpen
  \bibfield  {author} {\bibinfo {author} {\bibfnamefont {A.}~\bibnamefont
  {Chyzh}}, \bibinfo {author} {\bibfnamefont {C.~Y.}\ \bibnamefont {Wu}},
  \bibinfo {author} {\bibfnamefont {E.}~\bibnamefont {Kwan}}, \bibinfo {author}
  {\bibfnamefont {R.~A.}\ \bibnamefont {Henderson}}, \bibinfo {author}
  {\bibfnamefont {T.~A.}\ \bibnamefont {Bredeweg}}, \bibinfo {author}
  {\bibfnamefont {R.~C.}\ \bibnamefont {Haight}}, \bibinfo {author}
  {\bibfnamefont {A.~C.}\ \bibnamefont {Hayes-Sterbenz}}, \bibinfo {author}
  {\bibfnamefont {H.~Y.}\ \bibnamefont {Lee}}, \bibinfo {author} {\bibfnamefont
  {J.~M.}\ \bibnamefont {O'Donnell}}, \ and\ \bibinfo {author} {\bibfnamefont
  {J.~L.}\ \bibnamefont {Ullmann}},\ }\bibfield  {title} {\enquote {\bibinfo
  {title} {Total prompt $\ensuremath{\gamma}$-ray emission in fission of
  $^{235}\mathrm{U}$, $^{239,241}\mathrm{Pu}$, and $^{252}\mathrm{Cf}$},}\
  }\href {\doibase 10.1103/PhysRevC.90.014602} {\bibfield  {journal} {\bibinfo
  {journal} {Phys. Rev. C}\ }\textbf {\bibinfo {volume} {90}},\ \bibinfo
  {pages} {014602} (\bibinfo {year} {2014})}\BibitemShut {NoStop}%
\bibitem [{\citenamefont {Jandel}\ \emph {et~al.}(2014)\citenamefont {Jandel},
  \citenamefont {Rusev}, \citenamefont {Bond}, \citenamefont {Bredeweg},
  \citenamefont {Chadwick}, \citenamefont {Couture}, \citenamefont {Fowler},
  \citenamefont {Haight}, \citenamefont {Kawano}, \citenamefont {Keksis},
  \citenamefont {Mosby}, \citenamefont {O'Donnell}, \citenamefont {Rundberg},
  \citenamefont {Stetcu}, \citenamefont {Talou}, \citenamefont {Ullmann},
  \citenamefont {Vieira}, \citenamefont {Wilhelmy}, \citenamefont {Stoyer},
  \citenamefont {Haslett}, \citenamefont {Henderson}, \citenamefont {Becker},\
  and\ \citenamefont {Wu}}]{Jandel:2014aa}%
  \BibitemOpen
  \bibfield  {author} {\bibinfo {author} {\bibfnamefont {M.}~\bibnamefont
  {Jandel}}, \bibinfo {author} {\bibfnamefont {G.}~\bibnamefont {Rusev}},
  \bibinfo {author} {\bibfnamefont {E.M.}\ \bibnamefont {Bond}}, \bibinfo
  {author} {\bibfnamefont {T.A.}\ \bibnamefont {Bredeweg}}, \bibinfo {author}
  {\bibfnamefont {M.B.}\ \bibnamefont {Chadwick}}, \bibinfo {author}
  {\bibfnamefont {A.}~\bibnamefont {Couture}}, \bibinfo {author} {\bibfnamefont
  {M.M.}\ \bibnamefont {Fowler}}, \bibinfo {author} {\bibfnamefont {R.C.}\
  \bibnamefont {Haight}}, \bibinfo {author} {\bibfnamefont {T.}~\bibnamefont
  {Kawano}}, \bibinfo {author} {\bibfnamefont {A.L.}\ \bibnamefont {Keksis}},
  \bibinfo {author} {\bibfnamefont {S.M.}\ \bibnamefont {Mosby}}, \bibinfo
  {author} {\bibfnamefont {J.M.}\ \bibnamefont {O'Donnell}}, \bibinfo {author}
  {\bibfnamefont {R.S.}\ \bibnamefont {Rundberg}}, \bibinfo {author}
  {\bibfnamefont {I.}~\bibnamefont {Stetcu}}, \bibinfo {author} {\bibfnamefont
  {P.}~\bibnamefont {Talou}}, \bibinfo {author} {\bibfnamefont {J.L.}\
  \bibnamefont {Ullmann}}, \bibinfo {author} {\bibfnamefont {D.J.}\
  \bibnamefont {Vieira}}, \bibinfo {author} {\bibfnamefont {J.B.}\ \bibnamefont
  {Wilhelmy}}, \bibinfo {author} {\bibfnamefont {M.A.}\ \bibnamefont {Stoyer}},
  \bibinfo {author} {\bibfnamefont {R.J.}\ \bibnamefont {Haslett}}, \bibinfo
  {author} {\bibfnamefont {R.A.}\ \bibnamefont {Henderson}}, \bibinfo {author}
  {\bibfnamefont {J.A.}\ \bibnamefont {Becker}}, \ and\ \bibinfo {author}
  {\bibfnamefont {C.Y.}\ \bibnamefont {Wu}},\ }\bibfield  {title} {\enquote
  {\bibinfo {title} {Prompt fission gamma-ray studies at dance},}\ }\href
  {\doibase https://doi.org/10.1016/j.phpro.2014.10.016} {\bibfield  {journal}
  {\bibinfo  {journal} {Physics Procedia}\ }\textbf {\bibinfo {volume} {59}},\
  \bibinfo {pages} {101 -- 106} (\bibinfo {year} {2014})},\ \bibinfo {note}
  {gAMMA-2 Scientific Workshop on the Emission of Prompt Gamma-Rays in Fission
  and Related Topics}\BibitemShut {NoStop}%
\bibitem [{\citenamefont {Billnert}\ \emph {et~al.}(2013)\citenamefont
  {Billnert}, \citenamefont {Hambsch}, \citenamefont {Oberstedt},\ and\
  \citenamefont {Oberstedt}}]{Billnert:2013}%
  \BibitemOpen
  \bibfield  {author} {\bibinfo {author} {\bibfnamefont {R.}~\bibnamefont
  {Billnert}}, \bibinfo {author} {\bibfnamefont {F.-J.}\ \bibnamefont
  {Hambsch}}, \bibinfo {author} {\bibfnamefont {A.}~\bibnamefont {Oberstedt}},
  \ and\ \bibinfo {author} {\bibfnamefont {S.}~\bibnamefont {Oberstedt}},\
  }\bibfield  {title} {\enquote {\bibinfo {title} {New prompt spectral
  $\gamma$-ray data from the reaction ${}^{252}${C}f(sf) and its implication on
  present evaluated nuclear data files},}\ }\href {\doibase
  10.1103/PhysRevC.87.024601} {\bibfield  {journal} {\bibinfo  {journal} {Phys.
  Rev. C}\ }\textbf {\bibinfo {volume} {87}},\ \bibinfo {pages} {024601}
  (\bibinfo {year} {2013})}\BibitemShut {NoStop}%
\bibitem [{\citenamefont {Oberstedt}\ \emph {et~al.}(2013)\citenamefont
  {Oberstedt}, \citenamefont {Belgya}, \citenamefont {Billnert}, \citenamefont
  {Borcea}, \citenamefont {Bry\ifmmode~\acute{s}\else \'{s}\fi{}},
  \citenamefont {Geerts}, \citenamefont {G\"o\"ok}, \citenamefont {Hambsch},
  \citenamefont {Kis}, \citenamefont {Martinez}, \citenamefont {Oberstedt},
  \citenamefont {Szentmiklosi}, \citenamefont {Tak\`acs},\ and\ \citenamefont
  {Vidali}}]{Oberstedt:2013}%
  \BibitemOpen
  \bibfield  {author} {\bibinfo {author} {\bibfnamefont {A.}~\bibnamefont
  {Oberstedt}}, \bibinfo {author} {\bibfnamefont {T.}~\bibnamefont {Belgya}},
  \bibinfo {author} {\bibfnamefont {R.}~\bibnamefont {Billnert}}, \bibinfo
  {author} {\bibfnamefont {R.}~\bibnamefont {Borcea}}, \bibinfo {author}
  {\bibfnamefont {T.}~\bibnamefont {Bry\ifmmode~\acute{s}\else \'{s}\fi{}}},
  \bibinfo {author} {\bibfnamefont {W.}~\bibnamefont {Geerts}}, \bibinfo
  {author} {\bibfnamefont {A.}~\bibnamefont {G\"o\"ok}}, \bibinfo {author}
  {\bibfnamefont {F.-J.}\ \bibnamefont {Hambsch}}, \bibinfo {author}
  {\bibfnamefont {Z.}~\bibnamefont {Kis}}, \bibinfo {author} {\bibfnamefont
  {T.}~\bibnamefont {Martinez}}, \bibinfo {author} {\bibfnamefont
  {S.}~\bibnamefont {Oberstedt}}, \bibinfo {author} {\bibfnamefont
  {L.}~\bibnamefont {Szentmiklosi}}, \bibinfo {author} {\bibfnamefont
  {K.}~\bibnamefont {Tak\`acs}}, \ and\ \bibinfo {author} {\bibfnamefont
  {M.}~\bibnamefont {Vidali}},\ }\bibfield  {title} {\enquote {\bibinfo {title}
  {Improved values for the characteristics of prompt-fission $\gamma$-ray
  spectra from the reaction ${}^{235}${U}(${n}_{\mathrm{th}}$,$f$)},}\ }\href
  {\doibase 10.1103/PhysRevC.87.051602} {\bibfield  {journal} {\bibinfo
  {journal} {Phys. Rev. C}\ }\textbf {\bibinfo {volume} {87}},\ \bibinfo
  {pages} {051602} (\bibinfo {year} {2013})}\BibitemShut {NoStop}%
\bibitem [{\citenamefont {Lebois}\ \emph {et~al.}(2015)\citenamefont {Lebois},
  \citenamefont {Wilson}, \citenamefont {Halipr\'e}, \citenamefont {Oberstedt},
  \citenamefont {Oberstedt}, \citenamefont {Marini}, \citenamefont {Schmitt},
  \citenamefont {Rose}, \citenamefont {Siem}, \citenamefont {Fallot},
  \citenamefont {Porta},\ and\ \citenamefont {Zakari}}]{Lebois:2015aa}%
  \BibitemOpen
  \bibfield  {author} {\bibinfo {author} {\bibfnamefont {M.}~\bibnamefont
  {Lebois}}, \bibinfo {author} {\bibfnamefont {J.~N.}\ \bibnamefont {Wilson}},
  \bibinfo {author} {\bibfnamefont {P.}~\bibnamefont {Halipr\'e}}, \bibinfo
  {author} {\bibfnamefont {A.}~\bibnamefont {Oberstedt}}, \bibinfo {author}
  {\bibfnamefont {S.}~\bibnamefont {Oberstedt}}, \bibinfo {author}
  {\bibfnamefont {P.}~\bibnamefont {Marini}}, \bibinfo {author} {\bibfnamefont
  {C.}~\bibnamefont {Schmitt}}, \bibinfo {author} {\bibfnamefont {S.~J.}\
  \bibnamefont {Rose}}, \bibinfo {author} {\bibfnamefont {S.}~\bibnamefont
  {Siem}}, \bibinfo {author} {\bibfnamefont {M.}~\bibnamefont {Fallot}},
  \bibinfo {author} {\bibfnamefont {A.}~\bibnamefont {Porta}}, \ and\ \bibinfo
  {author} {\bibfnamefont {A.-A.}\ \bibnamefont {Zakari}},\ }\bibfield  {title}
  {\enquote {\bibinfo {title} {Comparative measurement of prompt fission
  $\ensuremath{\gamma}$-ray emission from fast-neutron-induced fission of
  $^{235}\mathrm{U}$ and $^{238}\mathrm{U}$},}\ }\href {\doibase
  10.1103/PhysRevC.92.034618} {\bibfield  {journal} {\bibinfo  {journal} {Phys.
  Rev. C}\ }\textbf {\bibinfo {volume} {92}},\ \bibinfo {pages} {034618}
  (\bibinfo {year} {2015})}\BibitemShut {NoStop}%
\bibitem [{\citenamefont {Gatera}\ \emph {et~al.}(2017)\citenamefont {Gatera},
  \citenamefont {Belgya}, \citenamefont {Geerts}, \citenamefont {G\"o\"ok},
  \citenamefont {Hambsch}, \citenamefont {Lebois}, \citenamefont {Mar\'oti},
  \citenamefont {Moens}, \citenamefont {Oberstedt}, \citenamefont {Oberstedt},
  \citenamefont {Postelt}, \citenamefont {Qi}, \citenamefont {Szentmikl\'osi},
  \citenamefont {Sibbens}, \citenamefont {Vanleeuw}, \citenamefont {Vidali},\
  and\ \citenamefont {Zeiser}}]{Gatera:2017}%
  \BibitemOpen
  \bibfield  {author} {\bibinfo {author} {\bibfnamefont {A.}~\bibnamefont
  {Gatera}}, \bibinfo {author} {\bibfnamefont {T.}~\bibnamefont {Belgya}},
  \bibinfo {author} {\bibfnamefont {W.}~\bibnamefont {Geerts}}, \bibinfo
  {author} {\bibfnamefont {A.}~\bibnamefont {G\"o\"ok}}, \bibinfo {author}
  {\bibfnamefont {F.-J.}\ \bibnamefont {Hambsch}}, \bibinfo {author}
  {\bibfnamefont {M.}~\bibnamefont {Lebois}}, \bibinfo {author} {\bibfnamefont
  {B.}~\bibnamefont {Mar\'oti}}, \bibinfo {author} {\bibfnamefont
  {A.}~\bibnamefont {Moens}}, \bibinfo {author} {\bibfnamefont
  {A.}~\bibnamefont {Oberstedt}}, \bibinfo {author} {\bibfnamefont
  {S.}~\bibnamefont {Oberstedt}}, \bibinfo {author} {\bibfnamefont
  {F.}~\bibnamefont {Postelt}}, \bibinfo {author} {\bibfnamefont
  {L.}~\bibnamefont {Qi}}, \bibinfo {author} {\bibfnamefont {L.}~\bibnamefont
  {Szentmikl\'osi}}, \bibinfo {author} {\bibfnamefont {G.}~\bibnamefont
  {Sibbens}}, \bibinfo {author} {\bibfnamefont {D.}~\bibnamefont {Vanleeuw}},
  \bibinfo {author} {\bibfnamefont {M.}~\bibnamefont {Vidali}}, \ and\ \bibinfo
  {author} {\bibfnamefont {F.}~\bibnamefont {Zeiser}},\ }\bibfield  {title}
  {\enquote {\bibinfo {title} {Prompt-fission $\ensuremath{\gamma}$-ray
  spectral characteristics from $^{239}\mathrm{Pu}({n}_{\mathrm{th}},f)$},}\
  }\href {\doibase 10.1103/PhysRevC.95.064609} {\bibfield  {journal} {\bibinfo
  {journal} {Phys. Rev. C}\ }\textbf {\bibinfo {volume} {95}},\ \bibinfo
  {pages} {064609} (\bibinfo {year} {2017})}\BibitemShut {NoStop}%
\bibitem [{\citenamefont {Makii}\ \emph {et~al.}(2019)\citenamefont {Makii},
  \citenamefont {Nishio}, \citenamefont {Hirose}, \citenamefont {Orlandi},
  \citenamefont {L\'eguillon}, \citenamefont {Ogawa}, \citenamefont {Soldner},
  \citenamefont {K\"oster}, \citenamefont {Pollitt}, \citenamefont {Hambsch},
  \citenamefont {Tsekhanovich}, \citenamefont {A\"{\i}che}, \citenamefont
  {Czajkowski}, \citenamefont {Mathieu}, \citenamefont {Petrache},
  \citenamefont {Astier}, \citenamefont {Guo}, \citenamefont {Ohtsuki},
  \citenamefont {Sekimoto}, \citenamefont {Takamiya}, \citenamefont {Frost},\
  and\ \citenamefont {Kawano}}]{Makii:2019aa}%
  \BibitemOpen
  \bibfield  {author} {\bibinfo {author} {\bibfnamefont {H.}~\bibnamefont
  {Makii}}, \bibinfo {author} {\bibfnamefont {K.}~\bibnamefont {Nishio}},
  \bibinfo {author} {\bibfnamefont {K.}~\bibnamefont {Hirose}}, \bibinfo
  {author} {\bibfnamefont {R.}~\bibnamefont {Orlandi}}, \bibinfo {author}
  {\bibfnamefont {R.}~\bibnamefont {L\'eguillon}}, \bibinfo {author}
  {\bibfnamefont {T.}~\bibnamefont {Ogawa}}, \bibinfo {author} {\bibfnamefont
  {T.}~\bibnamefont {Soldner}}, \bibinfo {author} {\bibfnamefont
  {U.}~\bibnamefont {K\"oster}}, \bibinfo {author} {\bibfnamefont
  {A.}~\bibnamefont {Pollitt}}, \bibinfo {author} {\bibfnamefont {F.-J.}\
  \bibnamefont {Hambsch}}, \bibinfo {author} {\bibfnamefont {I.}~\bibnamefont
  {Tsekhanovich}}, \bibinfo {author} {\bibfnamefont {M.}~\bibnamefont
  {A\"{\i}che}}, \bibinfo {author} {\bibfnamefont {S.}~\bibnamefont
  {Czajkowski}}, \bibinfo {author} {\bibfnamefont {L.}~\bibnamefont {Mathieu}},
  \bibinfo {author} {\bibfnamefont {C.~M.}\ \bibnamefont {Petrache}}, \bibinfo
  {author} {\bibfnamefont {A.}~\bibnamefont {Astier}}, \bibinfo {author}
  {\bibfnamefont {S.}~\bibnamefont {Guo}}, \bibinfo {author} {\bibfnamefont
  {T.}~\bibnamefont {Ohtsuki}}, \bibinfo {author} {\bibfnamefont
  {S.}~\bibnamefont {Sekimoto}}, \bibinfo {author} {\bibfnamefont
  {K.}~\bibnamefont {Takamiya}}, \bibinfo {author} {\bibfnamefont {R.~J.~W.}\
  \bibnamefont {Frost}}, \ and\ \bibinfo {author} {\bibfnamefont
  {T.}~\bibnamefont {Kawano}},\ }\bibfield  {title} {\enquote {\bibinfo {title}
  {Effects of the nuclear structure of fission fragments on the high-energy
  prompt fission $\ensuremath{\gamma}$-ray spectrum in
  $^{235}\mathrm{U}({n}_{\mathrm{th}},f)$},}\ }\href {\doibase
  10.1103/PhysRevC.100.044610} {\bibfield  {journal} {\bibinfo  {journal}
  {Phys. Rev. C}\ }\textbf {\bibinfo {volume} {100}},\ \bibinfo {pages}
  {044610} (\bibinfo {year} {2019})}\BibitemShut {NoStop}%
\bibitem [{\citenamefont {Verbinski}\ \emph {et~al.}(1973)\citenamefont
  {Verbinski}, \citenamefont {Weber},\ and\ \citenamefont
  {Sund}}]{Verbinski:1973}%
  \BibitemOpen
  \bibfield  {author} {\bibinfo {author} {\bibfnamefont {V.~V.}\ \bibnamefont
  {Verbinski}}, \bibinfo {author} {\bibfnamefont {Hans}\ \bibnamefont {Weber}},
  \ and\ \bibinfo {author} {\bibfnamefont {R.~E.}\ \bibnamefont {Sund}},\
  }\bibfield  {title} {\enquote {\bibinfo {title} {Prompt gamma rays from
  $^{235}\mathrm{U}(n,f)$, $^{239}\mathrm{Pu}(n,f)$, and spontaneous fission of
  $^{252}\mathrm{Cf}$},}\ }\href {\doibase 10.1103/PhysRevC.7.1173} {\bibfield
  {journal} {\bibinfo  {journal} {Phys. Rev. C}\ }\textbf {\bibinfo {volume}
  {7}},\ \bibinfo {pages} {1173--1185} (\bibinfo {year} {1973})}\BibitemShut
  {NoStop}%
\bibitem [{\citenamefont {Peelle}\ and\ \citenamefont
  {Maienschein}(1971)}]{Peelle1971}%
  \BibitemOpen
  \bibfield  {author} {\bibinfo {author} {\bibfnamefont {R.~W.}\ \bibnamefont
  {Peelle}}\ and\ \bibinfo {author} {\bibfnamefont {F.~C.}\ \bibnamefont
  {Maienschein}},\ }\bibfield  {title} {\enquote {\bibinfo {title} {Spectrum of
  photons emitted in coincidence with fission of $^{235}\mathrm{U}$ by thermal
  neutrons},}\ }\href {\doibase 10.1103/PhysRevC.3.373} {\bibfield  {journal}
  {\bibinfo  {journal} {Phys. Rev. C}\ }\textbf {\bibinfo {volume} {3}},\
  \bibinfo {pages} {373--390} (\bibinfo {year} {1971})}\BibitemShut {NoStop}%
\bibitem [{\citenamefont {Pleasonton}\ \emph {et~al.}(1972)\citenamefont
  {Pleasonton}, \citenamefont {Ferguson},\ and\ \citenamefont
  {Schmitt}}]{Pleasonton:1972}%
  \BibitemOpen
  \bibfield  {author} {\bibinfo {author} {\bibfnamefont {Frances}\ \bibnamefont
  {Pleasonton}}, \bibinfo {author} {\bibfnamefont {Robert~L.}\ \bibnamefont
  {Ferguson}}, \ and\ \bibinfo {author} {\bibfnamefont {H.~W.}\ \bibnamefont
  {Schmitt}},\ }\bibfield  {title} {\enquote {\bibinfo {title} {Prompt gamma
  rays emitted in the thermal-neutron-induced fission of $^{235}\mathrm{U}$},}\
  }\href {\doibase 10.1103/PhysRevC.6.1023} {\bibfield  {journal} {\bibinfo
  {journal} {Phys. Rev. C}\ }\textbf {\bibinfo {volume} {6}},\ \bibinfo {pages}
  {1023--1039} (\bibinfo {year} {1972})}\BibitemShut {NoStop}%
\bibitem [{\citenamefont {Pleasonton}(1973)}]{Pleasonton:1973}%
  \BibitemOpen
  \bibfield  {author} {\bibinfo {author} {\bibfnamefont {Frances}\ \bibnamefont
  {Pleasonton}},\ }\bibfield  {title} {\enquote {\bibinfo {title} {Prompt
  $\gamma$-rays emitted in the thermal-neutron induced fission of $^{233}${U}
  and $^{239}${P}u},}\ }\href {\doibase
  http://dx.doi.org/10.1016/0375-9474(73)90161-9} {\bibfield  {journal}
  {\bibinfo  {journal} {Nucl. Phys. A}\ }\textbf {\bibinfo {volume} {213}},\
  \bibinfo {pages} {413 -- 425} (\bibinfo {year} {1973})}\BibitemShut {NoStop}%
\bibitem [{\citenamefont {Stetcu}\ \emph {et~al.}(2020)\citenamefont {Stetcu},
  \citenamefont {Chadwick}, \citenamefont {Kawano}, \citenamefont {Talou},
  \citenamefont {Capote},\ and\ \citenamefont {Trkov}}]{Stetcu:2020aa}%
  \BibitemOpen
  \bibfield  {author} {\bibinfo {author} {\bibfnamefont {I.}~\bibnamefont
  {Stetcu}}, \bibinfo {author} {\bibfnamefont {M.~B.}\ \bibnamefont
  {Chadwick}}, \bibinfo {author} {\bibfnamefont {T.}~\bibnamefont {Kawano}},
  \bibinfo {author} {\bibfnamefont {P.}~\bibnamefont {Talou}}, \bibinfo
  {author} {\bibfnamefont {R.}~\bibnamefont {Capote}}, \ and\ \bibinfo {author}
  {\bibfnamefont {A.}~\bibnamefont {Trkov}},\ }\bibfield  {title} {\enquote
  {\bibinfo {title} {{Evaluation of the Prompt Fission Gamma Properties for
  Neutron Induced Fission of $^{235;238}$U and $^{239}$Pu}},}\ }\href@noop {}
  {\bibfield  {journal} {\bibinfo  {journal} {Nuclear Data Sheets}\ }\textbf
  {\bibinfo {volume} {163}},\ \bibinfo {pages} {261} (\bibinfo {year}
  {2020})}\BibitemShut {NoStop}%
\bibitem [{\citenamefont {Straede}\ \emph {et~al.}(1987)\citenamefont
  {Straede}, \citenamefont {Budtz-J{\o}rgensen},\ and\ \citenamefont
  {Knitter}}]{Straede:1987aa}%
  \BibitemOpen
  \bibfield  {author} {\bibinfo {author} {\bibfnamefont {Ch.}\ \bibnamefont
  {Straede}}, \bibinfo {author} {\bibfnamefont {C.}~\bibnamefont
  {Budtz-J{\o}rgensen}}, \ and\ \bibinfo {author} {\bibfnamefont {H.-H.}\
  \bibnamefont {Knitter}},\ }\bibfield  {title} {\enquote {\bibinfo {title}
  {235u(n, f) fragment mass-, kinetic energy- and angular distributions for
  incident neutron energies between thermal and 6 mev},}\ }\href {\doibase
  https://doi.org/10.1016/0375-9474(87)90381-2} {\bibfield  {journal} {\bibinfo
   {journal} {Nuclear Physics A}\ }\textbf {\bibinfo {volume} {462}},\ \bibinfo
  {pages} {85 -- 108} (\bibinfo {year} {1987})}\BibitemShut {NoStop}%
\bibitem [{\citenamefont {Viv\`es}\ \emph {et~al.}(2000)\citenamefont
  {Viv\`es}, \citenamefont {Hambsch}, \citenamefont {Bax},\ and\ \citenamefont
  {Oberstedt}}]{Vives:2000aa}%
  \BibitemOpen
  \bibfield  {author} {\bibinfo {author} {\bibfnamefont {F.}~\bibnamefont
  {Viv\`es}}, \bibinfo {author} {\bibfnamefont {F.-J.}\ \bibnamefont
  {Hambsch}}, \bibinfo {author} {\bibfnamefont {H.}~\bibnamefont {Bax}}, \ and\
  \bibinfo {author} {\bibfnamefont {S.}~\bibnamefont {Oberstedt}},\ }\bibfield
  {title} {\enquote {\bibinfo {title} {Investigation of the fission fragment
  properties of the reaction 238u(n,f) at incident neutron energies up to 5.8
  mev},}\ }\href {\doibase http://dx.doi.org/10.1016/S0375-9474(99)00413-3}
  {\bibfield  {journal} {\bibinfo  {journal} {Nuclear Physics A}\ }\textbf
  {\bibinfo {volume} {662}},\ \bibinfo {pages} {63 -- 92} (\bibinfo {year}
  {2000})}\BibitemShut {NoStop}%
\bibitem [{\citenamefont {Birgersson}\ \emph {et~al.}(2009)\citenamefont
  {Birgersson}, \citenamefont {Oberstedt}, \citenamefont {Oberstedt},\ and\
  \citenamefont {Hambsch}}]{Birgersson:2009aa}%
  \BibitemOpen
  \bibfield  {author} {\bibinfo {author} {\bibfnamefont {E.}~\bibnamefont
  {Birgersson}}, \bibinfo {author} {\bibfnamefont {A.}~\bibnamefont
  {Oberstedt}}, \bibinfo {author} {\bibfnamefont {S.}~\bibnamefont
  {Oberstedt}}, \ and\ \bibinfo {author} {\bibfnamefont {F.-J.}\ \bibnamefont
  {Hambsch}},\ }\bibfield  {title} {\enquote {\bibinfo {title} {Properties of
  the reaction 238u at the vibrational resonances},}\ }\href {\doibase
  http://dx.doi.org/10.1016/j.nuclphysa.2008.12.001} {\bibfield  {journal}
  {\bibinfo  {journal} {Nuclear Physics A}\ }\textbf {\bibinfo {volume}
  {817}},\ \bibinfo {pages} {1 -- 34} (\bibinfo {year} {2009})}\BibitemShut
  {NoStop}%
\bibitem [{\citenamefont {Duke}\ \emph {et~al.}(2016)\citenamefont {Duke},
  \citenamefont {Tovesson}, \citenamefont {Laptev}, \citenamefont {Mosby},
  \citenamefont {Hambsch}, \citenamefont {Bry\ifmmode~\acute{s}\else
  \'{s}\fi{}},\ and\ \citenamefont {Vidali}}]{Duke:2016}%
  \BibitemOpen
  \bibfield  {author} {\bibinfo {author} {\bibfnamefont {D.~L.}\ \bibnamefont
  {Duke}}, \bibinfo {author} {\bibfnamefont {F.}~\bibnamefont {Tovesson}},
  \bibinfo {author} {\bibfnamefont {A.~B.}\ \bibnamefont {Laptev}}, \bibinfo
  {author} {\bibfnamefont {S.}~\bibnamefont {Mosby}}, \bibinfo {author}
  {\bibfnamefont {F.-J.}\ \bibnamefont {Hambsch}}, \bibinfo {author}
  {\bibfnamefont {T.}~\bibnamefont {Bry\ifmmode~\acute{s}\else \'{s}\fi{}}}, \
  and\ \bibinfo {author} {\bibfnamefont {M.}~\bibnamefont {Vidali}},\
  }\bibfield  {title} {\enquote {\bibinfo {title} {Fission-fragment properties
  in $^{238}\mathrm{U}(n,f)$ between 1 and 30 mev},}\ }\href {\doibase
  10.1103/PhysRevC.94.054604} {\bibfield  {journal} {\bibinfo  {journal} {Phys.
  Rev. C}\ }\textbf {\bibinfo {volume} {94}},\ \bibinfo {pages} {054604}
  (\bibinfo {year} {2016})}\BibitemShut {NoStop}%
\bibitem [{\citenamefont {Tovesson}\ \emph {et~al.}(2013)\citenamefont
  {Tovesson}, \citenamefont {Arnold}, \citenamefont {Bredeweg}, \citenamefont
  {Jandel}, \citenamefont {Laptev}, \citenamefont {Meierbachtol}, \citenamefont
  {Sierk}, \citenamefont {M.}, \citenamefont {Hecht}, \citenamefont {Mader},
  \citenamefont {R.}, \citenamefont {Greife}, \citenamefont {Moore},
  \citenamefont {Shields},\ and\ \citenamefont {Snyder}}]{Tovesson:2013}%
  \BibitemOpen
  \bibfield  {author} {\bibinfo {author} {\bibfnamefont {F.}~\bibnamefont
  {Tovesson}}, \bibinfo {author} {\bibfnamefont {C.~W.}\ \bibnamefont
  {Arnold}}, \bibinfo {author} {\bibfnamefont {T.}~\bibnamefont {Bredeweg}},
  \bibinfo {author} {\bibfnamefont {M.}~\bibnamefont {Jandel}}, \bibinfo
  {author} {\bibfnamefont {A.~B.}\ \bibnamefont {Laptev}}, \bibinfo {author}
  {\bibfnamefont {K.}~\bibnamefont {Meierbachtol}}, \bibinfo {author}
  {\bibfnamefont {A.}~\bibnamefont {Sierk}}, \bibinfo {author} {\bibfnamefont
  {White}\ \bibnamefont {M.}}, \bibinfo {author} {\bibfnamefont {A.~A.}\
  \bibnamefont {Hecht}}, \bibinfo {author} {\bibfnamefont {D.}~\bibnamefont
  {Mader}}, \bibinfo {author} {\bibfnamefont {Blakeley}\ \bibnamefont {R.}},
  \bibinfo {author} {\bibfnamefont {U.}~\bibnamefont {Greife}}, \bibinfo
  {author} {\bibfnamefont {B.}~\bibnamefont {Moore}}, \bibinfo {author}
  {\bibfnamefont {D.}~\bibnamefont {Shields}}, \ and\ \bibinfo {author}
  {\bibfnamefont {L.}~\bibnamefont {Snyder}},\ }\bibfield  {title} {\enquote
  {\bibinfo {title} {Spider: A new instrument for fission yield measurement},}\
  }in\ \href@noop {} {\emph {\bibinfo {booktitle} {Proceedings of the Fifth
  International Conference on ICFN5}}},\ \bibinfo {editor} {edited by\ \bibinfo
  {editor} {\bibfnamefont {J.~H.}\ \bibnamefont {Hamilton}}\ and\ \bibinfo
  {editor} {\bibfnamefont {A.~V.}\ \bibnamefont {Ramayya}}}\ (\bibinfo
  {publisher} {World Scientific},\ \bibinfo {year} {2013})\ p.\ \bibinfo
  {pages} {361}\BibitemShut {NoStop}%
\bibitem [{\citenamefont {Meierbachtol}\ \emph {et~al.}(2016)\citenamefont
  {Meierbachtol}, \citenamefont {Tovesson}, \citenamefont {Duke}, \citenamefont
  {Geppert-Kleinrath}, \citenamefont {Manning}, \citenamefont {Meharchand},
  \citenamefont {Mosby},\ and\ \citenamefont {Shields}}]{Meierbachtol:2016}%
  \BibitemOpen
  \bibfield  {author} {\bibinfo {author} {\bibfnamefont {K.}~\bibnamefont
  {Meierbachtol}}, \bibinfo {author} {\bibfnamefont {F.}~\bibnamefont
  {Tovesson}}, \bibinfo {author} {\bibfnamefont {D.~L.}\ \bibnamefont {Duke}},
  \bibinfo {author} {\bibfnamefont {V.}~\bibnamefont {Geppert-Kleinrath}},
  \bibinfo {author} {\bibfnamefont {B.}~\bibnamefont {Manning}}, \bibinfo
  {author} {\bibfnamefont {R.}~\bibnamefont {Meharchand}}, \bibinfo {author}
  {\bibfnamefont {S.}~\bibnamefont {Mosby}}, \ and\ \bibinfo {author}
  {\bibfnamefont {D.}~\bibnamefont {Shields}},\ }\bibfield  {title} {\enquote
  {\bibinfo {title} {Total kinetic energy release in $^{239}\mathrm{Pu}(n,f)$
  post-neutron emission from 0.5 to 50 mev incident neutron energy},}\ }\href
  {\doibase 10.1103/PhysRevC.94.034611} {\bibfield  {journal} {\bibinfo
  {journal} {Phys. Rev. C}\ }\textbf {\bibinfo {volume} {94}},\ \bibinfo
  {pages} {034611} (\bibinfo {year} {2016})}\BibitemShut {NoStop}%
\bibitem [{\citenamefont {Pellereau}\ \emph {et~al.}(2017)\citenamefont
  {Pellereau}, \citenamefont {Ta\"{\i}eb}, \citenamefont {Chatillon},
  \citenamefont {Alvarez-Pol}, \citenamefont {Audouin}, \citenamefont {Ayyad},
  \citenamefont {B\'elier}, \citenamefont {Benlliure}, \citenamefont {Boutoux},
  \citenamefont {Caama\~no}, \citenamefont {Casarejos}, \citenamefont
  {Cortina-Gil}, \citenamefont {Ebran}, \citenamefont {Farget}, \citenamefont
  {Fern\'andez-Dom\'{\i}nguez}, \citenamefont {Gorbinet}, \citenamefont
  {Grente}, \citenamefont {Heinz}, \citenamefont {Johansson}, \citenamefont
  {Jurado}, \citenamefont {Keli\ifmmode \acute{c}\else~\'{c}\fi{} Heil},
  \citenamefont {Kurz}, \citenamefont {Laurent}, \citenamefont {Martin},
  \citenamefont {Nociforo}, \citenamefont {Paradela}, \citenamefont {Pietri},
  \citenamefont {Rodr\'{\i}guez-S\'anchez}, \citenamefont {Schmidt},
  \citenamefont {Simon}, \citenamefont {Tassan-Got}, \citenamefont {Vargas},
  \citenamefont {Voss},\ and\ \citenamefont {Weick}}]{Pellereau:2017aa}%
  \BibitemOpen
  \bibfield  {author} {\bibinfo {author} {\bibfnamefont {E.}~\bibnamefont
  {Pellereau}}, \bibinfo {author} {\bibfnamefont {J.}~\bibnamefont
  {Ta\"{\i}eb}}, \bibinfo {author} {\bibfnamefont {A.}~\bibnamefont
  {Chatillon}}, \bibinfo {author} {\bibfnamefont {H.}~\bibnamefont
  {Alvarez-Pol}}, \bibinfo {author} {\bibfnamefont {L.}~\bibnamefont
  {Audouin}}, \bibinfo {author} {\bibfnamefont {Y.}~\bibnamefont {Ayyad}},
  \bibinfo {author} {\bibfnamefont {G.}~\bibnamefont {B\'elier}}, \bibinfo
  {author} {\bibfnamefont {J.}~\bibnamefont {Benlliure}}, \bibinfo {author}
  {\bibfnamefont {G.}~\bibnamefont {Boutoux}}, \bibinfo {author} {\bibfnamefont
  {M.}~\bibnamefont {Caama\~no}}, \bibinfo {author} {\bibfnamefont
  {E.}~\bibnamefont {Casarejos}}, \bibinfo {author} {\bibfnamefont
  {D.}~\bibnamefont {Cortina-Gil}}, \bibinfo {author} {\bibfnamefont
  {A.}~\bibnamefont {Ebran}}, \bibinfo {author} {\bibfnamefont
  {F.}~\bibnamefont {Farget}}, \bibinfo {author} {\bibfnamefont
  {B.}~\bibnamefont {Fern\'andez-Dom\'{\i}nguez}}, \bibinfo {author}
  {\bibfnamefont {T.}~\bibnamefont {Gorbinet}}, \bibinfo {author}
  {\bibfnamefont {L.}~\bibnamefont {Grente}}, \bibinfo {author} {\bibfnamefont
  {A.}~\bibnamefont {Heinz}}, \bibinfo {author} {\bibfnamefont
  {H.}~\bibnamefont {Johansson}}, \bibinfo {author} {\bibfnamefont
  {B.}~\bibnamefont {Jurado}}, \bibinfo {author} {\bibfnamefont
  {A.}~\bibnamefont {Keli\ifmmode \acute{c}\else~\'{c}\fi{} Heil}}, \bibinfo
  {author} {\bibfnamefont {N.}~\bibnamefont {Kurz}}, \bibinfo {author}
  {\bibfnamefont {B.}~\bibnamefont {Laurent}}, \bibinfo {author} {\bibfnamefont
  {J.-F.}\ \bibnamefont {Martin}}, \bibinfo {author} {\bibfnamefont
  {C.}~\bibnamefont {Nociforo}}, \bibinfo {author} {\bibfnamefont
  {C.}~\bibnamefont {Paradela}}, \bibinfo {author} {\bibfnamefont
  {S.}~\bibnamefont {Pietri}}, \bibinfo {author} {\bibfnamefont {J.~L.}\
  \bibnamefont {Rodr\'{\i}guez-S\'anchez}}, \bibinfo {author} {\bibfnamefont
  {K.-H.}\ \bibnamefont {Schmidt}}, \bibinfo {author} {\bibfnamefont
  {H.}~\bibnamefont {Simon}}, \bibinfo {author} {\bibfnamefont
  {L.}~\bibnamefont {Tassan-Got}}, \bibinfo {author} {\bibfnamefont
  {J.}~\bibnamefont {Vargas}}, \bibinfo {author} {\bibfnamefont
  {B.}~\bibnamefont {Voss}}, \ and\ \bibinfo {author} {\bibfnamefont
  {H.}~\bibnamefont {Weick}},\ }\bibfield  {title} {\enquote {\bibinfo {title}
  {Accurate isotopic fission yields of electromagnetically induced fission of
  $^{238}\mathrm{U}$ measured in inverse kinematics at relativistic
  energies},}\ }\href {\doibase 10.1103/PhysRevC.95.054603} {\bibfield
  {journal} {\bibinfo  {journal} {Phys. Rev. C}\ }\textbf {\bibinfo {volume}
  {95}},\ \bibinfo {pages} {054603} (\bibinfo {year} {2017})}\BibitemShut
  {NoStop}%
\bibitem [{\citenamefont {Gooden}\ \emph {et~al.}(2016)\citenamefont {Gooden},
  \citenamefont {Arnold}, \citenamefont {Becker}, \citenamefont {Bhatia},
  \citenamefont {Bhike}, \citenamefont {Bond}, \citenamefont {Bredeweg},
  \citenamefont {Fallin}, \citenamefont {Fowler}, \citenamefont {Howell},
  \citenamefont {Kelley}, \citenamefont {Krishichayan}, \citenamefont {Macri},
  \citenamefont {Rusev}, \citenamefont {Ryan}, \citenamefont {Sheets},
  \citenamefont {Stoyer}, \citenamefont {Tonchev}, \citenamefont {Tornow},
  \citenamefont {Vieira},\ and\ \citenamefont {Wilhelmy}}]{Gooden:2016aa}%
  \BibitemOpen
  \bibfield  {author} {\bibinfo {author} {\bibfnamefont {M.E.}\ \bibnamefont
  {Gooden}}, \bibinfo {author} {\bibfnamefont {C.W.}\ \bibnamefont {Arnold}},
  \bibinfo {author} {\bibfnamefont {J.A.}\ \bibnamefont {Becker}}, \bibinfo
  {author} {\bibfnamefont {C.}~\bibnamefont {Bhatia}}, \bibinfo {author}
  {\bibfnamefont {M.}~\bibnamefont {Bhike}}, \bibinfo {author} {\bibfnamefont
  {E.M.}\ \bibnamefont {Bond}}, \bibinfo {author} {\bibfnamefont {T.A.}\
  \bibnamefont {Bredeweg}}, \bibinfo {author} {\bibfnamefont {B.}~\bibnamefont
  {Fallin}}, \bibinfo {author} {\bibfnamefont {M.M.}\ \bibnamefont {Fowler}},
  \bibinfo {author} {\bibfnamefont {C.R.}\ \bibnamefont {Howell}}, \bibinfo
  {author} {\bibfnamefont {J.H.}\ \bibnamefont {Kelley}}, \bibinfo {author}
  {\bibnamefont {Krishichayan}}, \bibinfo {author} {\bibfnamefont
  {R.}~\bibnamefont {Macri}}, \bibinfo {author} {\bibfnamefont
  {G.}~\bibnamefont {Rusev}}, \bibinfo {author} {\bibfnamefont
  {C.}~\bibnamefont {Ryan}}, \bibinfo {author} {\bibfnamefont {S.A.}\
  \bibnamefont {Sheets}}, \bibinfo {author} {\bibfnamefont {M.A.}\ \bibnamefont
  {Stoyer}}, \bibinfo {author} {\bibfnamefont {A.P.}\ \bibnamefont {Tonchev}},
  \bibinfo {author} {\bibfnamefont {W.}~\bibnamefont {Tornow}}, \bibinfo
  {author} {\bibfnamefont {D.J.}\ \bibnamefont {Vieira}}, \ and\ \bibinfo
  {author} {\bibfnamefont {J.B.}\ \bibnamefont {Wilhelmy}},\ }\bibfield
  {title} {\enquote {\bibinfo {title} {Energy dependence of fission product
  yields from 235u, 238u and 239pu for incident neutron energies between 0.5
  and 14.8 mev},}\ }\href {\doibase https://doi.org/10.1016/j.nds.2015.12.006}
  {\bibfield  {journal} {\bibinfo  {journal} {Nuclear Data Sheets}\ }\textbf
  {\bibinfo {volume} {131}},\ \bibinfo {pages} {319 -- 356} (\bibinfo {year}
  {2016})},\ \bibinfo {note} {special Issue on Nuclear Reaction
  Data}\BibitemShut {NoStop}%
\bibitem [{\citenamefont {Silano}\ \emph {et~al.}(2019)\citenamefont {Silano},
  \citenamefont {Tonchev}, \citenamefont {Henderson}, \citenamefont
  {{Schunck}}, \citenamefont {Tornow}, \citenamefont {Howell}, \citenamefont
  {Krishichayan}, \citenamefont {Finch},\ and\ \citenamefont
  {Gooden}}]{Silano:2020aa}%
  \BibitemOpen
  \bibfield  {author} {\bibinfo {author} {\bibfnamefont {J.}~\bibnamefont
  {Silano}}, \bibinfo {author} {\bibfnamefont {A.}~\bibnamefont {Tonchev}},
  \bibinfo {author} {\bibfnamefont {R.}~\bibnamefont {Henderson}}, \bibinfo
  {author} {\bibfnamefont {N.}~\bibnamefont {{Schunck}}}, \bibinfo {author}
  {\bibfnamefont {W.}~\bibnamefont {Tornow}}, \bibinfo {author} {\bibfnamefont
  {C.}~\bibnamefont {Howell}}, \bibinfo {author} {\bibfnamefont
  {F.}~\bibnamefont {Krishichayan}}, \bibinfo {author} {\bibfnamefont
  {S.}~\bibnamefont {Finch}}, \ and\ \bibinfo {author} {\bibfnamefont
  {M.}~\bibnamefont {Gooden}},\ }\href@noop {} {\enquote {\bibinfo {title}
  {Comparing fission-product yields from photon-induced fission of $^{240}${Pu}
  and neutron-induced fission of $^{239}${Pu} as a test of the bohr hypothesis
  in nuclear fission},}\ } (\bibinfo {year} {2019}),\ \bibinfo {note}
  {subbmited to the Proceedings of the Santa Fe workshop on fission product
  yields}\BibitemShut {NoStop}%
\bibitem [{\citenamefont {Savard}(2019)}]{Savard:2019aa}%
  \BibitemOpen
  \bibfield  {author} {\bibinfo {author} {\bibfnamefont {G.}~\bibnamefont
  {Savard}},\ }\href@noop {} {\enquote {\bibinfo {title} {Fission product yield
  measurements using $^{252}$cf spontaneous fission and neutron-induced fission
  on actinide targets at {CARIBU}},}\ } (\bibinfo {year} {2019}),\ \bibinfo
  {note} {talk at the International Workshop on Fission Product
  Yields}\BibitemShut {NoStop}%
\bibitem [{HRS()}]{HRS}%
  \BibitemOpen
  \href@noop {} {\enquote {\bibinfo {title} {{A High Rigidity Spectrometer fro
  FRIB, {hrs.lbl.gov/documents/HRS-WhitePaper122017.pdf}}},}\ }\BibitemShut
  {NoStop}%
\bibitem [{FRI()}]{FRIB}%
  \BibitemOpen
  \href@noop {} {\enquote {\bibinfo {title} {{Facility for Rare Isotopes Beams,
  www.frib.msu.edu}},}\ }\BibitemShut {NoStop}%
\bibitem [{\citenamefont {Carjan}\ and\ \citenamefont
  {Rizea}(2010)}]{Carjan:2010}%
  \BibitemOpen
  \bibfield  {author} {\bibinfo {author} {\bibfnamefont {N.}~\bibnamefont
  {Carjan}}\ and\ \bibinfo {author} {\bibfnamefont {M.}~\bibnamefont {Rizea}},\
  }\bibfield  {title} {\enquote {\bibinfo {title} {{Scission neutrons and other
  scission properties as function of mass asymmetry in
  $^{235}\mathrm{U}$(${n}_{\mathrm{th}}$,$f$)}},}\ }\href {\doibase
  10.1103/PhysRevC.82.014617} {\bibfield  {journal} {\bibinfo  {journal} {Phys.
  Rev. C}\ }\textbf {\bibinfo {volume} {82}},\ \bibinfo {pages} {014617}
  (\bibinfo {year} {2010})}\BibitemShut {NoStop}%
\bibitem [{\citenamefont {Carjan}\ \emph {et~al.}(2012)\citenamefont {Carjan},
  \citenamefont {Hambsch}, \citenamefont {Rizea},\ and\ \citenamefont
  {Serot}}]{Carjan:2012}%
  \BibitemOpen
  \bibfield  {author} {\bibinfo {author} {\bibfnamefont {N.}~\bibnamefont
  {Carjan}}, \bibinfo {author} {\bibfnamefont {F.-J.}\ \bibnamefont {Hambsch}},
  \bibinfo {author} {\bibfnamefont {M.}~\bibnamefont {Rizea}}, \ and\ \bibinfo
  {author} {\bibfnamefont {O.}~\bibnamefont {Serot}},\ }\bibfield  {title}
  {\enquote {\bibinfo {title} {Partition between the fission fragments of the
  excitation energy and of the neutron multiplicity at scission in low-energy
  fission},}\ }\href {\doibase 10.1103/PhysRevC.85.044601} {\bibfield
  {journal} {\bibinfo  {journal} {Phys. Rev. C}\ }\textbf {\bibinfo {volume}
  {85}},\ \bibinfo {pages} {044601} (\bibinfo {year} {2012})}\BibitemShut
  {NoStop}%
\bibitem [{\citenamefont {Rizea}\ and\ \citenamefont
  {Carjan}(2013)}]{Rizea:2013}%
  \BibitemOpen
  \bibfield  {author} {\bibinfo {author} {\bibfnamefont {M.}~\bibnamefont
  {Rizea}}\ and\ \bibinfo {author} {\bibfnamefont {N.}~\bibnamefont {Carjan}},\
  }\bibfield  {title} {\enquote {\bibinfo {title} {Dynamical scission model},}\
  }\href {\doibase Rizea, M. and Carjan, N.} {\bibfield  {journal} {\bibinfo
  {journal} {Nucl. Phys. A}\ }\textbf {\bibinfo {volume} {909}},\ \bibinfo
  {pages} {50} (\bibinfo {year} {2013})}\BibitemShut {NoStop}%
\bibitem [{\citenamefont {Carjan}\ and\ \citenamefont
  {Rizea}(2015)}]{Carjan:2015}%
  \BibitemOpen
  \bibfield  {author} {\bibinfo {author} {\bibfnamefont {N.}~\bibnamefont
  {Carjan}}\ and\ \bibinfo {author} {\bibfnamefont {M.}~\bibnamefont {Rizea}},\
  }\bibfield  {title} {\enquote {\bibinfo {title} {Similarities between
  calculated scission-neutron properties and experimental data on prompt
  fission neutrons},}\ }\href {\doibase 10.1016/j.physletb.2015.05.050}
  {\bibfield  {journal} {\bibinfo  {journal} {Phys. Lett. B}\ }\textbf
  {\bibinfo {volume} {747}},\ \bibinfo {pages} {178} (\bibinfo {year}
  {2015})}\BibitemShut {NoStop}%
\bibitem [{\citenamefont {Capote}\ \emph {et~al.}(2016)\citenamefont {Capote},
  \citenamefont {Carjan},\ and\ \citenamefont {Chiba}}]{Capote:2016a}%
  \BibitemOpen
  \bibfield  {author} {\bibinfo {author} {\bibfnamefont {R.}~\bibnamefont
  {Capote}}, \bibinfo {author} {\bibfnamefont {N.}~\bibnamefont {Carjan}}, \
  and\ \bibinfo {author} {\bibfnamefont {S.}~\bibnamefont {Chiba}},\ }\bibfield
   {title} {\enquote {\bibinfo {title} {{Scission neutrons for U, Pu, Cm, and
  Cf isotopes: Relative multiplicities calculated in the sudden limit}},}\
  }\href {\doibase 10.1103/PhysRevC.93.024609} {\bibfield  {journal} {\bibinfo
  {journal} {Phys. Rev. C}\ }\textbf {\bibinfo {volume} {93}},\ \bibinfo
  {pages} {024609} (\bibinfo {year} {2016})}\BibitemShut {NoStop}%
\bibitem [{\citenamefont {Carjan}\ and\ \citenamefont
  {Rizea}(2019)}]{Carjan:2019}%
  \BibitemOpen
  \bibfield  {author} {\bibinfo {author} {\bibfnamefont {N.}~\bibnamefont
  {Carjan}}\ and\ \bibinfo {author} {\bibfnamefont {M.}~\bibnamefont {Rizea}},\
  }\bibfield  {title} {\enquote {\bibinfo {title} {Structures in the energy
  distribution of the scission neutrons: Finite neutron-number effect},}\
  }\href {\doibase 10.1103/PhysRevC.99.034613} {\bibfield  {journal} {\bibinfo
  {journal} {Phys. Rev. C}\ }\textbf {\bibinfo {volume} {99}},\ \bibinfo
  {pages} {034613} (\bibinfo {year} {2019})}\BibitemShut {NoStop}%
\bibitem [{\citenamefont {Vogt}\ and\ \citenamefont
  {Randrup}(2013)}]{Vogt:2013}%
  \BibitemOpen
  \bibfield  {author} {\bibinfo {author} {\bibfnamefont {R.}~\bibnamefont
  {Vogt}}\ and\ \bibinfo {author} {\bibfnamefont {J.}~\bibnamefont {Randrup}},\
  }\bibfield  {title} {\enquote {\bibinfo {title} {{Event-by-event Modeling of
  Prompt Neutrons and Photons from Neutron-induced and Spontaneous Fission with
  FREYA}},}\ }\href {\doibase 10.1016/j.phpro.2013.06.013} {\bibfield
  {journal} {\bibinfo  {journal} {Physics Procedia}\ }\textbf {\bibinfo
  {volume} {47}},\ \bibinfo {pages} {82} (\bibinfo {year} {2013})}\BibitemShut
  {NoStop}%
\bibitem [{\citenamefont {Litaize}\ \emph {et~al.}(2015)\citenamefont
  {Litaize}, \citenamefont {Serot},\ and\ \citenamefont
  {Berge}}]{Litaize:2015}%
  \BibitemOpen
  \bibfield  {author} {\bibinfo {author} {\bibfnamefont {O.}~\bibnamefont
  {Litaize}}, \bibinfo {author} {\bibfnamefont {O.}~\bibnamefont {Serot}}, \
  and\ \bibinfo {author} {\bibfnamefont {L.}~\bibnamefont {Berge}},\ }\bibfield
   {title} {\enquote {\bibinfo {title} {Fission modelling with {FIFRELIN}},}\
  }\href {\doibase 10.1140/epja/i2015-15177-9} {\bibfield  {journal} {\bibinfo
  {journal} {Eur. Phys. Jour. A}\ }\textbf {\bibinfo {volume} {51}},\ \bibinfo
  {pages} {177} (\bibinfo {year} {2015})}\BibitemShut {NoStop}%
\bibitem [{\citenamefont {Schmidt}\ \emph {et~al.}(2016)\citenamefont
  {Schmidt}, \citenamefont {Jurado}, \citenamefont {Amouroux},\ and\
  \citenamefont {Schmitt}}]{Schmidt:2016}%
  \BibitemOpen
  \bibfield  {author} {\bibinfo {author} {\bibfnamefont {K.-H.}\ \bibnamefont
  {Schmidt}}, \bibinfo {author} {\bibfnamefont {B.}~\bibnamefont {Jurado}},
  \bibinfo {author} {\bibfnamefont {C.}~\bibnamefont {Amouroux}}, \ and\
  \bibinfo {author} {\bibfnamefont {C.}~\bibnamefont {Schmitt}},\ }\bibfield
  {title} {\enquote {\bibinfo {title} {General description of fission
  observables: Gef model code},}\ }\href {\doibase
  http://dx.doi.org/10.1016/j.nds.2015.12.009} {\bibfield  {journal} {\bibinfo
  {journal} {Nuclear Data Sheets}\ }\textbf {\bibinfo {volume} {131}},\
  \bibinfo {pages} {107 -- 221} (\bibinfo {year} {2016})}\BibitemShut {NoStop}%
\bibitem [{\citenamefont {Schmidt}\ and\ \citenamefont
  {Jurado}(2018)}]{Schmidt:2018}%
  \BibitemOpen
  \bibfield  {author} {\bibinfo {author} {\bibfnamefont {K.-H.}\ \bibnamefont
  {Schmidt}}\ and\ \bibinfo {author} {\bibfnamefont {B.}~\bibnamefont
  {Jurado}},\ }\bibfield  {title} {\enquote {\bibinfo {title} {Review on the
  progress in nuclear fission - experimental methods and theoretical
  descriptions},}\ }\href {\doibase 10.1088/1361-6633/aacfa7} {\bibfield
  {journal} {\bibinfo  {journal} {Rep. Prog. Phys.}\ }\textbf {\bibinfo
  {volume} {81}},\ \bibinfo {pages} {106301} (\bibinfo {year}
  {2018})}\BibitemShut {NoStop}%
\bibitem [{\citenamefont {Talou}\ \emph {et~al.}(2018)\citenamefont {Talou},
  \citenamefont {Vogt}, \citenamefont {Randrup}, \citenamefont {Rising},
  \citenamefont {Pozzi}, \citenamefont {Verbeke}, \citenamefont {Andrews},
  \citenamefont {Clarke}, \citenamefont {Jaffke}, \citenamefont {Jandel},
  \citenamefont {Kawano}, \citenamefont {Marcath}, \citenamefont
  {Meierbachtol}, \citenamefont {Nakae}, \citenamefont {Rusev}, \citenamefont
  {Sood}, \citenamefont {Stetcu},\ and\ \citenamefont {Walker}}]{Talou:2018}%
  \BibitemOpen
  \bibfield  {author} {\bibinfo {author} {\bibfnamefont {P.}~\bibnamefont
  {Talou}}, \bibinfo {author} {\bibfnamefont {R.}~\bibnamefont {Vogt}},
  \bibinfo {author} {\bibfnamefont {J.}~\bibnamefont {Randrup}}, \bibinfo
  {author} {\bibfnamefont {M.~E.}\ \bibnamefont {Rising}}, \bibinfo {author}
  {\bibfnamefont {S.~A.}\ \bibnamefont {Pozzi}}, \bibinfo {author}
  {\bibfnamefont {J.}~\bibnamefont {Verbeke}}, \bibinfo {author} {\bibfnamefont
  {M.~T.}\ \bibnamefont {Andrews}}, \bibinfo {author} {\bibfnamefont {S.~D.}\
  \bibnamefont {Clarke}}, \bibinfo {author} {\bibfnamefont {P.}~\bibnamefont
  {Jaffke}}, \bibinfo {author} {\bibfnamefont {M.}~\bibnamefont {Jandel}},
  \bibinfo {author} {\bibfnamefont {T.}~\bibnamefont {Kawano}}, \bibinfo
  {author} {\bibfnamefont {M.~J.}\ \bibnamefont {Marcath}}, \bibinfo {author}
  {\bibfnamefont {K.}~\bibnamefont {Meierbachtol}}, \bibinfo {author}
  {\bibfnamefont {L.}~\bibnamefont {Nakae}}, \bibinfo {author} {\bibfnamefont
  {G.}~\bibnamefont {Rusev}}, \bibinfo {author} {\bibfnamefont
  {A.}~\bibnamefont {Sood}}, \bibinfo {author} {\bibfnamefont {I.}~\bibnamefont
  {Stetcu}}, \ and\ \bibinfo {author} {\bibfnamefont {C.}~\bibnamefont
  {Walker}},\ }\bibfield  {title} {\enquote {\bibinfo {title} {Correlated
  prompt fission data in transport simulations},}\ }\href {\doibase
  10.1140/epja/i2018-12455-0} {\bibfield  {journal} {\bibinfo  {journal} {Eur.
  Phys. Jour. A}\ }\textbf {\bibinfo {volume} {54}},\ \bibinfo {pages} {9}
  (\bibinfo {year} {2018})}\BibitemShut {NoStop}%
\bibitem [{\citenamefont {Stetcu}\ \emph {et~al.}(2014)\citenamefont {Stetcu},
  \citenamefont {Talou}, \citenamefont {Kawano},\ and\ \citenamefont
  {Jandel}}]{Stetcu:2014a}%
  \BibitemOpen
  \bibfield  {author} {\bibinfo {author} {\bibfnamefont {I.}~\bibnamefont
  {Stetcu}}, \bibinfo {author} {\bibfnamefont {P.}~\bibnamefont {Talou}},
  \bibinfo {author} {\bibfnamefont {T.}~\bibnamefont {Kawano}}, \ and\ \bibinfo
  {author} {\bibfnamefont {M.}~\bibnamefont {Jandel}},\ }\bibfield  {title}
  {\enquote {\bibinfo {title} {Properties of prompt-fission
  $\ensuremath{\gamma}$ rays},}\ }\href {\doibase 10.1103/PhysRevC.90.024617}
  {\bibfield  {journal} {\bibinfo  {journal} {Phys. Rev. C}\ }\textbf {\bibinfo
  {volume} {90}},\ \bibinfo {pages} {024617} (\bibinfo {year}
  {2014})}\BibitemShut {NoStop}%
\bibitem [{\citenamefont {Stetcu}\ \emph {et~al.}(2013)\citenamefont {Stetcu},
  \citenamefont {Talou}, \citenamefont {Kawano},\ and\ \citenamefont
  {Jandel}}]{Stetcu:2013}%
  \BibitemOpen
  \bibfield  {author} {\bibinfo {author} {\bibfnamefont {I.}~\bibnamefont
  {Stetcu}}, \bibinfo {author} {\bibfnamefont {P.}~\bibnamefont {Talou}},
  \bibinfo {author} {\bibfnamefont {T.}~\bibnamefont {Kawano}}, \ and\ \bibinfo
  {author} {\bibfnamefont {M.}~\bibnamefont {Jandel}},\ }\bibfield  {title}
  {\enquote {\bibinfo {title} {Isomer production ratios and the angular
  momentum distribution of fission fragments},}\ }\href {\doibase
  10.1103/PhysRevC.88.044603} {\bibfield  {journal} {\bibinfo  {journal} {Phys.
  Rev. C}\ }\textbf {\bibinfo {volume} {88}},\ \bibinfo {pages} {044603}
  (\bibinfo {year} {2013})}\BibitemShut {NoStop}%
\bibitem [{\citenamefont {Fr\'ehaut}\ \emph {et~al.}(1983)\citenamefont
  {Fr\'ehaut}, \citenamefont {Bertin},\ and\ \citenamefont
  {Bois}}]{Frehaut:1983}%
  \BibitemOpen
  \bibfield  {author} {\bibinfo {author} {\bibfnamefont {J.}~\bibnamefont
  {Fr\'ehaut}}, \bibinfo {author} {\bibfnamefont {A.}~\bibnamefont {Bertin}}, \
  and\ \bibinfo {author} {\bibfnamefont {R.}~\bibnamefont {Bois}},\ }\bibfield
  {title} {\enquote {\bibinfo {title} {Mesure de $\bar\nu_p$ pour la fission de
  $^{232}${T}h, $^{235}${U} et $^{237}${N}p induite par des neutrons d'energie
  comprise entre 1 et 15 {MeV}},}\ }in\ \href@noop {} {\emph {\bibinfo
  {booktitle} {International Conference on Nuclear Data for Science and
  Technology, Antwerp, Belgium}}}\ (\bibinfo  {publisher} {Reidel, Dordrech,
  Holland},\ \bibinfo {year} {1983})\ p.~\bibinfo {pages} {78}\BibitemShut
  {NoStop}%
\bibitem [{\citenamefont {Goddard}\ \emph {et~al.}(2015)\citenamefont
  {Goddard}, \citenamefont {Stevenson},\ and\ \citenamefont
  {Rios}}]{Goddard:2015}%
  \BibitemOpen
  \bibfield  {author} {\bibinfo {author} {\bibfnamefont {P.}~\bibnamefont
  {Goddard}}, \bibinfo {author} {\bibfnamefont {P.}~\bibnamefont {Stevenson}},
  \ and\ \bibinfo {author} {\bibfnamefont {A.}~\bibnamefont {Rios}},\
  }\bibfield  {title} {\enquote {\bibinfo {title} {{Fission dynamics within
  time-dependent Hartree-Fock: Deformation-induced fission}},}\ }\href
  {\doibase 10.1103/PhysRevC.92.054610} {\bibfield  {journal} {\bibinfo
  {journal} {Phys. Rev. C}\ }\textbf {\bibinfo {volume} {92}},\ \bibinfo
  {pages} {054610} (\bibinfo {year} {2015})}\BibitemShut {NoStop}%
\bibitem [{\citenamefont {Goddard}\ \emph {et~al.}(2016)\citenamefont
  {Goddard}, \citenamefont {Stevenson},\ and\ \citenamefont
  {Rios}}]{Goddard:2016}%
  \BibitemOpen
  \bibfield  {author} {\bibinfo {author} {\bibfnamefont {P.}~\bibnamefont
  {Goddard}}, \bibinfo {author} {\bibfnamefont {P.}~\bibnamefont {Stevenson}},
  \ and\ \bibinfo {author} {\bibfnamefont {A.}~\bibnamefont {Rios}},\
  }\bibfield  {title} {\enquote {\bibinfo {title} {{Fission dynamics within
  time-dependent Hartree-Fock. II. Boost-induced fission}},}\ }\href {\doibase
  10.1103/PhysRevC.93.014620} {\bibfield  {journal} {\bibinfo  {journal} {Phys.
  Rev. C}\ }\textbf {\bibinfo {volume} {93}},\ \bibinfo {pages} {014620}
  (\bibinfo {year} {2016})}\BibitemShut {NoStop}%
\bibitem [{\citenamefont {Tanimura}\ \emph {et~al.}(2015)\citenamefont
  {Tanimura}, \citenamefont {Lacroix},\ and\ \citenamefont
  {Scamps}}]{Tanimura:2015}%
  \BibitemOpen
  \bibfield  {author} {\bibinfo {author} {\bibfnamefont {Y.}~\bibnamefont
  {Tanimura}}, \bibinfo {author} {\bibfnamefont {D.}~\bibnamefont {Lacroix}}, \
  and\ \bibinfo {author} {\bibfnamefont {G.}~\bibnamefont {Scamps}},\
  }\bibfield  {title} {\enquote {\bibinfo {title} {{Collective aspects deduced
  from time-dependent microscopic mean-field with pairing: Application to the
  fission process}},}\ }\href {\doibase 10.1103/PhysRevC.92.034601} {\bibfield
  {journal} {\bibinfo  {journal} {Phys. Rev. C}\ }\textbf {\bibinfo {volume}
  {92}},\ \bibinfo {pages} {034601} (\bibinfo {year} {2015})}\BibitemShut
  {NoStop}%
\bibitem [{\citenamefont {Bulgac}\ \emph {et~al.}(2017)\citenamefont {Bulgac},
  \citenamefont {Jin}, \citenamefont {Magierski}, \citenamefont {Roche},
  \citenamefont {Schunck},\ and\ \citenamefont {Stetcu}}]{Bulgac:2017a}%
  \BibitemOpen
  \bibfield  {author} {\bibinfo {author} {\bibfnamefont {A.}~\bibnamefont
  {Bulgac}}, \bibinfo {author} {\bibfnamefont {S.}~\bibnamefont {Jin}},
  \bibinfo {author} {\bibfnamefont {P.}~\bibnamefont {Magierski}}, \bibinfo
  {author} {\bibfnamefont {K.~J}\ \bibnamefont {Roche}}, \bibinfo {author}
  {\bibfnamefont {N.}~\bibnamefont {Schunck}}, \ and\ \bibinfo {author}
  {\bibfnamefont {I.}~\bibnamefont {Stetcu}},\ }\bibfield  {title} {\enquote
  {\bibinfo {title} {{Nuclear Fission: from more phenomenology and adjusted
  parameters to more fundamental theory and increased predictive power}},}\
  }\href {\doibase 10.1051/epjconf/2017/16300007} {\bibfield  {journal}
  {\bibinfo  {journal} {EPJ Web of Conferences}\ }\textbf {\bibinfo {volume}
  {163}},\ \bibinfo {pages} {00007} (\bibinfo {year} {2017})}\BibitemShut
  {NoStop}%
\bibitem [{\citenamefont {Meitner}\ and\ \citenamefont
  {Frisch}(1939{\natexlab{b}})}]{Meitner:1939a}%
  \BibitemOpen
  \bibfield  {author} {\bibinfo {author} {\bibfnamefont {L.}~\bibnamefont
  {Meitner}}\ and\ \bibinfo {author} {\bibfnamefont {O.~R.}\ \bibnamefont
  {Frisch}},\ }\bibfield  {title} {\enquote {\bibinfo {title} {{Products of the
  Fission of the Uranium Nucleus}},}\ }\href {\doibase 10.1038/143471a0}
  {\bibfield  {journal} {\bibinfo  {journal} {Nature}\ }\textbf {\bibinfo
  {volume} {143}},\ \bibinfo {pages} {471} (\bibinfo {year}
  {1939}{\natexlab{b}})}\BibitemShut {NoStop}%
\bibitem [{\citenamefont {Andreyev}\ \emph {et~al.}(2018)\citenamefont
  {Andreyev}, \citenamefont {Nishio},\ and\ \citenamefont
  {Schmidt}}]{Andreyev:2018}%
  \BibitemOpen
  \bibfield  {author} {\bibinfo {author} {\bibfnamefont {A.~N.}\ \bibnamefont
  {Andreyev}}, \bibinfo {author} {\bibfnamefont {K.}~\bibnamefont {Nishio}}, \
  and\ \bibinfo {author} {\bibfnamefont {K.-H.}\ \bibnamefont {Schmidt}},\
  }\bibfield  {title} {\enquote {\bibinfo {title} {{Nuclear fission: a review
  of experimental advances and phenomenology}},}\ }\href {\doibase
  10.1088/1361-6633/aa82eb} {\bibfield  {journal} {\bibinfo  {journal} {Rep.
  Prog. Phys.}\ }\textbf {\bibinfo {volume} {81}},\ \bibinfo {pages} {016301}
  (\bibinfo {year} {2018})}\BibitemShut {NoStop}%
\bibitem [{\citenamefont {Bulgac}\ \emph
  {et~al.}(2019{\natexlab{c}})\citenamefont {Bulgac}, \citenamefont {Jin},\
  and\ \citenamefont {Stetcu}}]{Bulgac:2020}%
  \BibitemOpen
  \bibfield  {author} {\bibinfo {author} {\bibfnamefont {A.}~\bibnamefont
  {Bulgac}}, \bibinfo {author} {\bibfnamefont {S.}~\bibnamefont {Jin}}, \ and\
  \bibinfo {author} {\bibfnamefont {I.}~\bibnamefont {Stetcu}},\ }\bibfield
  {title} {\enquote {\bibinfo {title} {{Nuclear Fission Dynamics: Past,
  Present, Needs, and Future}},}\ }\href@noop {} {\  (\bibinfo {year}
  {2019}{\natexlab{c}})},\ \Eprint {http://arxiv.org/abs/1912:00287v1}
  {arXiv:1912:00287v1} \BibitemShut {NoStop}%
\end{thebibliography}%

\end{document}